\documentclass[preprint,3p,12pt]{elsarticle} % Options: preprint, review, or final
\biboptions{sort&compress}

\usepackage{amsmath,amssymb} % for mathematical symbols
\usepackage{graphicx} % for including figures
\usepackage{hyperref} % jump to citation
\usepackage{subfigure}
\usepackage{booktabs}

\begin{document}

\begin{frontmatter}

\title{An efficient solution algorithm for force-driven continuum and rarefied flows}

%% Group authors per affiliation:
\author[imech,ucas]{Shuangqing Liu}
\author[imech,ucas]{Zuoxu Li}
\author[imech,ucas]{Yonghao Zhang}
\author[imech,ucas]{Tianbai Xiao\corref{cor1}}
\ead{txiao@imech.ac.cn}

\cortext[cor1]{Corresponding author}

\address[imech]{Centre for Interdisciplinary Research in Fluids, Institute of Mechanics, Chinese Academy of Sciences, Beijing, China}
\address[ucas]{School of Engineering Science, University of Chinese Academy of Sciences, Beijing, China}

\begin{abstract}
Gaseous flows under an external force are intrinsically defined by their multi-scale nature due to the large variation of densities along the forcing direction.
Devising a numerical method capable of accurately and efficiently solving force-driven cross-scale flow dynamics, encompassing both continuum and rarefied regimes, continues to pose a formidable and enduring challenge.
In this work, a novel solution algorithm for multi-scale and non-equilibrium flow transport under an external force is developed based on the Boltzmann-BGK equation.
The core innovation lies in the fusion of the Hermite spectral method (employed to characterize non-equilibrium particle distributions) with a multi-scale evolution model (sourced from the unified gas-kinetic scheme), achieving a seamless connection between computational methods and physical models.
To accommodate the properties of the spectral-collocation method, a series of collocation points and weights are adapted based on the Gauss-Hermite quadrature.
As a result, the computational efficiency of the solution algorithm is significantly improved (up to 50 times) while maintaining comparable accuracy as the classical discrete velocity method.
It is demonstrated that the solution algorithm effectively preserves the key structural features of gas-dynamic systems subjected to an external force, e.g., the well-balanced property.
Extensive numerical experiments have been performed to verify the accuracy and efficiency of the proposed method, including the one-dimensional hydrostatic equilibrium problem, the Sod shock tube, the Fourier flow, the Poiseuille flow, and the Rayleigh-Taylor instability problem.
The proposed methodology can provide substantive theoretical insights into a wide range of engineering challenges involving force-driven multi-scale flows.
\end{abstract}


\end{frontmatter}

%\newpage
% Nomenclature
\begin{table}
    \centering
    \caption{Nomenclature.}
    \begin{tabular*}{16cm}{lll}
        \hline
        %\hline
        $t$ & time variable \\
        $\mathbf x$ & space variable $(x,y,z)$ \\
        $\mathbf v$ & velocity variable $(u,v,w)$ \\
        $\mathbf a$ & external force acceleration \\
        $f$ & particle distribution function \\
        $\mathcal Q$ & collision operator in kinetic equation \\
        $\mathcal M$ & Maxwellian distribution function \\
        $\nu$ & relaxation frequency \\
        $\tau$ & relaxation time $(\tau=1/\nu)$ \\
        $m$ & molecular mass \\
        $k$ & Boltzmann constant \\
        $\rho$ & macroscopic density \\
        $\mathbf V$ & macroscopic velocity \\
        $T$ & macroscopic temperature \\
        $E$ & macroscopic energy \\
        $\psi$ & collision invariants \\
        $\mathbf W$ & macroscopic conservative variables $(\rho,\rho \mathbf V,\rho E)$ \\
        $\mathbf P$ & stress tensor \\
        $\mathbf q$ & heat flux \\
        $\mathbf \Omega$ & computational domain \\
        $V$ & volume of domain \\
        $\mathbf F^f$ & numerical flux for distribution function \\
        $\mathbf F^W$ & numerical flux for conservative variables \\
        $\mathbf S$ & area vector of cell face \\
        $G$ & source term for distribution function \\
        $\mathbf G$ & source term for conservative variables \\
        $\boldsymbol\sigma$ & reconstructed spatial slope of distribution function \\
        $\boldsymbol\theta$ & spatial slope of equilibrium distribution function \\
        $\phi$ & temporal slope of equilibrium distribution function \\
        $\tilde m$ & equilibrium flux contribution \\
        $\tilde f$ & initial flux contribution \\
        $H$ & generalized Heaviside function \\
        $\mathcal H$ & Hermite polynomials \\
        $\mathbf u$ & rescaled particle velocity \\
        $\mathbf d$ & expansion coefficients \\
        $\omega$ & weight function in the Hermite expansion \\
        $\boldsymbol\delta$ & identity tensor \\
        $\mathbf T$ & modified stress tensor \\
        $\mathbf Q$ & modified heat flux tensor \\
        \hline
        %\hline
    \end{tabular*}
    \label{table:nomenclature}
\end{table}
\newpage

%\linenumbers

\section{Introduction}

Classical gas dynamics is established based on corresponding characteristic scales.
At the spatial and temporal scales of molecular mean free path and collision time, the Boltzmann equation is established to statistically describe the transport of a vast ensemble of particles, thereby depicting fluid motions at the kinetic scale.
At a macroscopic level, the Navier-Stokes equations are used to characterize the collective behaviors of particle transport, e.g., phenomena involving wave propagation and interaction.
Scale separation provides the foundation for the applicability of different models, accordingly, especially for shifting from intricate molecular descriptions to manageable continuum models.
Numerical methods for computational fluid dynamics have been developed for solving the governing equations at different scales, e.g., the discrete velocity method for the Boltzmann equation \cite{aristov2001direct} and the Riemann-problem-based flux solver for compressible Navier-Stokes equations \cite{toro2013riemann}.

The principle of scale separation is not universally applicable.
An easily overlooked example is gaseous flows under external force.
It can be theoretically demonstrated that the density and pressure of gases are exponentially distributed under a unidirectional constant external force \cite{chapman1990mathematical}.
The exponential factor manifests as $ah/RT$, with $a$ being the magnitude of forcing acceleration, $h$ the length of the domain, $T$ the temperature, and $R$ the gas constant.
In an ordinary laboratory environment, the factor is moderate, and thus, classical gas dynamics routinely ignores the effects of external force.
As we focus on large-scale flows, e.g., atmospheric circulation, the influence exerted by the external force cannot be disregarded.
Comparable scenarios may occur at small scales accompanied by large accelerations, e.g., Vlasov-type forces in plasmas and dense gases \cite{benilov2019enskog}.
Note that the acceleration of a non-inertial frame also gives rise to fictitious forces, e.g., centrifugal and Coriolis forces. 
The study of gas dynamics under an external force is both a fascinating research pursuit and of notable engineering significance.

Differences in gas density along the forcing direction result in varying characteristic scales of flows.
Employing the Knudsen number (i.e., the ratio of the molecular mean free path to a length scale of interest) as the criterion, flows within spatially distinct regions can be categorized into different regimes \cite{tsien1946superaerodynamics}.
The Navier-Stokes equations, formulated in the continuum regime, prove inadequate for the precise description of rarefied flows.
This inapplicability is even more pronounced under an external force.
In the Navier-Stokes equations, the external force appears only as a source term for the velocity field.
I.e., it contributes sorely to the kinematics and does not alter the dynamics of the fluid system.
Asymptotic analysis of the Boltzmann equation shows that external force can induce the generation of non-equilibrium stress and heat flux  \cite{tij1994perturbation,xiao2018investigation}.
This phenomenon arises due to the asymmetric tendency of particle transport caused by acceleration and is therefore observable under a non-vanishing mean free path (e.g., in the Navier-Stokes regime).
Fine-grained mesoscopic modeling and simulation are essential to accurately capture non-equilibrium transport phenomena under external forces.

A wealth of numerical methods has been advanced to address non-equilibrium flows, with the stochastic particle method and the discrete velocity method for the Boltzmann equation standing as two of the most prominent approaches.
By integrating coupled transport and collision processes in the particle evolution model and developing implicit and semi-implicit update algorithms, these methodologies can be extended to encompass cross-scale flow simulations.
%\cite{jin2010asymptotic,dimarco2013asymptotic,xiao2021flux,chacon2017multiscale,xu2010unified,guo2013discrete,xiao2020velocity,zhu2021general,fei2020unified,liu2020unified}.
Among others, exemplary achievements include asymptotic-preserving methods \cite{jin2010asymptotic,dimarco2013asymptotic,xiao2021flux}, high-order/low-order algorithms \cite{chacon2017multiscale}, gas-kinetic schemes \cite{xu2010unified,guo2013discrete,xiao2020velocity}, iterative schemes \cite{zhu2021general}, multi-scale particle methods \cite{fei2020unified,liu2020unified,liu2020simplified,yang2023discrete}.
However, the presence of external forces places additional demands on the performance of numerical methods.
Under a unidirectional constant force, a gas dynamic system will evolve toward and persist with a steady state with zero macroscopic velocity from an arbitrary initial state.
Such hydrostatic equilibrium should be maintained by realizing an exact balance between the force term and the inhomogeneous flux function in the numerical method.
The lack of such a well-balanced property leads to spurious
numerical solutions, especially in long-time self-evolving gravitational systems.

Current research on well-balanced methods predominantly centers on hyperbolic conservation laws \cite{li2021simple,navas2023family,mantri2021well,ghosh2016well}, while limited attention has been devoted to rarefied and multi-scale flows.
Developing well-balanced methods for macroscopic and kinetic equations fundamentally constitutes different challenges.
In the continuum limit, it can be deduced that in hydrostatic equilibrium the gas dynamic system has the same temperature everywhere \cite{luo2011well}.
However, as the local Knudsen number increases, the force-induced heat fluxes lead to a heterogeneous temperature distribution.
At this juncture, the temperature can no longer be routinely treated as a constant multiplier.
Instead, it must be incorporated into the evolving framework of the gas dynamic model.
The unified gas-kinetic scheme (UGKS) provides such a model, where the characteristic lines of the Bhatnagar-Gross-Krook model \cite{bhatnagar1954model} (as a simplified model of the Boltzmann equation) that contain particle trajectory information, can be introduced in the construction of numerical fluxes.
Based on the UGKS, well-balanced methods have been developed for single- and multi-component gas transport \cite{xiao2017well,xiao2022well}.
Other pioneering work on multi-scale methods for the flows under an external force can be found in the literature \cite{li2020gas,premnath2009steady}.

Among other discrete velocity methods, the UGKS employs a uniform nodal description in the physical-velocity phase space.
The well-balanced property is satisfied in an asymptotic basis as the number of collocation points $N_c \rightarrow 0$.
However, in real numerical simulations, the number of discrete meshes is always finite.
The preservation of physical structure only holds practical significance when discussed in conjunction with aliasing, interpolation, and integral errors.
A prime illustration lies in the fact that particle methods, while fundamentally grounded in first-principles modeling, are rendered impractical for the development of well-balanced methods due to statistical errors.
The existence of an external force field necessitates solving the derivative of particle distributions in the velocity space.
A straightforward finite difference approximation leads to an error of $O(\Delta \mathbf v)$ where $\mathbf v\in (-\infty,\infty)^3$.
Since the number of velocity collocation points is usually smaller than the number of elements in the physical space, such an error is considerably substantial.

Spectral methods offer an accurate approach for approximating solutions to differential equations.
By approximating the solution as a sum of basis functions, the error can decrease exponentially with the number of bases, and fewer degrees of freedom are usually required to obtain comparable solutions from the finite difference method.
Hermite polynomials prove exceptionally advantageous for approximating distribution functions in the Boltzmann equation, owing to the orthogonality with respect to Gaussian weighting functions, which corresponds to the Maxwellian equilibrium distribution.
Such orthogonality warrants that the deviation from equilibrium can be accurately represented and efficiently decomposed into physically meaningful moment variables.
It has been shown that the Hermite polynomials can asymptotically converge to non-equilibrium distribution functions \cite{shan2006kinetic}.
In this study, we present a unified Galerkin-collocation method for multi-scale flows under external force by synergistically integrating the Hermite spectral method with the multiscale evolutionary model, i.e., the UGKS.
By employing the spectral and collocation methods to compute the forcing and advection terms of the Boltzmann model equation, respectively, the solution algorithm is anticipated to harmonize accuracy with the capability to provide highly non-equilibrium and discontinuous solutions.
A series of collocation points and weights applicable to the solution algorithm are adapted based on the Gauss-Hermite quadrature.

The rest of this paper is organized as follows.
Section \ref{sec:theory} briefly introduces the kinetic theory of gases under an external force.
Section \ref{sec:algorithm} provides the implementation of the solution algorithm and the corresponding numerical analysis.
Section \ref{sec:experiment} includes numerical experiments to demonstrate the performance of the proposed approach,
followed by the conclusions.

%It is more important to capture the non-equilibrium flow solution as accurately as possible at a discrete mesh level.
%highly dissipative region.
%exclude the use of particle methods
%fine-grained inscription at the mesoscopic scale.
%patterns 
%at the intersection of 
%tackle
%Unlike the moment method, 

\section{Kinetic Theory}\label{sec:theory}

The Boltzmann equation in the presence of an external force writes
\begin{equation}
    \frac{\partial f}{\partial t} + \mathbf v\cdot \nabla_{\mathbf x}f+\mathbf a\cdot \nabla_{\mathbf v}f=\mathcal Q(f),
    \label{eq:boltzmann}
\end{equation}
where $f(t,\mathbf x,\mathbf v)$ is the particle distribution function in the phase space, $\mathbf v$ is the particle velocity, $\mathbf a$ is the forcing acceleration, and $\mathcal Q$ is the intermolecular collision term.
The full Boltzmann collision term is a nonlinear fivefold integral.
Simplified models, e.g., the BGK model, are often adopted in engineering analysis and simulation to reduce the computational overhead.
In the BGK model, the collision term is replaced by a relaxation operator, i.e.,
\begin{equation}
    \mathcal Q(f)=\nu (\mathcal M - f),
    \label{eq:bgk}
\end{equation}
where $\nu$ is the relaxation frequency, and $\mathcal M$ denotes the Maxwellian distribution,
\begin{equation}
    \mathcal M = \rho\left(\frac{m}{2\pi k T}\right)^{3/2} \exp(-\frac{m}{2kT} \left(\mathbf v - \mathbf V)^2 \right),
\end{equation}
where $m$ is the molecular mass, $k$ is the Boltzmann constant, $\{\rho,\mathbf V,T\}$ are the macroscopic density, velocity, and temperature, respectively.

Microscopic and macroscopic states are interconnected.
The macroscopic conservative variables in fluid mechanics can be obtained by taking moments of the particle distribution function, i.e.,
\begin{equation}
    \mathbf{W}(t, \mathbf{x}):=\left(\begin{array}{c}
    \rho \\
    \rho \mathbf{V} \\
    \rho E
    \end{array}\right) = \int_{\mathbb R^3} f \psi d \mathbf{v},
\end{equation}
where $\psi=(1,\mathbf v,\mathbf v^2/2)^T$ is the vector of collision invariants that satisfies
\begin{equation}
    \int_{\mathbb R^3} \mathcal Q(f) \psi d\mathbf v = 0.
\end{equation}
The transport equations of conservative variables can thus be obtained, i.e.,
\begin{equation}
    \begin{aligned}
        &\frac{\partial \rho}{\partial t} + \nabla_{\mathbf x} \cdot (\rho \mathbf V)=0, \\
        &\frac{\partial \rho \mathbf V}{\partial t} + \nabla_{\mathbf x} \cdot (\rho \mathbf V \otimes \mathbf V)+\nabla_{\mathbf x} \cdot \mathbf P=\rho \mathbf a, \\
        &\frac{\partial \rho E}{\partial t} + \nabla_{\mathbf x} \cdot (\rho E\mathbf V)+\nabla_{\mathbf x} \cdot (\mathbf P \cdot \mathbf V)+\nabla_{\mathbf x} \cdot \mathbf q=\rho \mathbf V \cdot \mathbf a,
    \end{aligned}
\label{eq:conservation laws}
\end{equation}
where $\otimes$ denotes the dyadic product operator, and the stress tensor $\mathbf P$ and heat flux $\mathbf q$ are defined as
\begin{equation}
    \mathbf P=\int_{\mathbb R^3} (\mathbf v-\mathbf V)\otimes(\mathbf v-\mathbf V)fd\mathbf v, \quad \mathbf q=\int_{\mathbb R^3} \frac{1}{2}(\mathbf v-\mathbf V)(\mathbf v-\mathbf V)^2fd\mathbf v.
\end{equation}
Note that the derivation from Eq.(\ref{eq:boltzmann}) to (\ref{eq:conservation laws}) introduces no assumptions before incorporating a closure model, e.g., the Chapman-Enskog expansion.

\section{Solution Algorithm}\label{sec:algorithm}

\subsection{Update algorithm}

The solution algorithm is built upon the finite volume method.
The computational domain is divided into $N_x$ non-overlapping cells, i.e.,
\begin{equation}
    \mathbf{\Omega} = \bigcup_{i=1}^{N_{x}} \mathbf \Omega_{i}, \quad \bigcap_{i=1}^{N_{x}} \mathbf \Omega_{i} = \emptyset.
\end{equation}
In each cell, the particle distribution function is approximated using $N_v$ collocation points.
The update algorithm of cell-averaged particle distribution function can be written as
\begin{equation}
    \begin{aligned}
        \frac{\partial f_{i,j}}{\partial t} &= -\frac{1}{V_{i}} \oint_{\partial \mathbf \Omega_i} \mathbf F^f_j(t,\mathbf x) \cdot d\mathbf S + G(f_{i,j}) + \mathcal Q(f_{i,j}) \\
        &= -\frac{1}{V_{i}} \sum_{k=1}^{N_f} \mathbf F_{k,j}^f \cdot \Delta \mathbf S_k - \mathbf a \cdot \nabla_{\mathbf v_j} f_{i,j} +\nu_i ({\mathcal M}_{i,j} - f_{i,j}),
    \end{aligned}
\label{eq:fvm micro}
\end{equation}
where $f_{i,j}$ represents the averaged distribution function at the $j$-th collocation point in the $i$-th cell,
$V_i$ denotes the cell volume, $\mathbf F^f$ is the numerical flux of distribution function, $\mathbf S=\mathbf n \Delta S$ is the area vector pointing out of the cell, $N_f$ is the number of faces, and $G$ is the external force term.

Following the derivation of Eq.(\ref{eq:conservation laws}), taking moments of Eq.(\ref{eq:fvm micro}) over the velocity space yields the semi-discrete conservation laws, i.e.,
\begin{equation}
\begin{aligned}
    \frac{\partial {\mathbf W}_{i}}{\partial t} &= -\frac{1}{V_{i}} \oint_{\partial \mathbf \Omega_i} \mathbf F^W \cdot d\mathbf S + \mathbf G_i = -\frac{1}{V_{i}} \sum_{k=1}^{N_f} \mathbf F_{k}^W \cdot \Delta \mathbf S_k + \mathbf G_i,
\end{aligned}
\label{eq:fvm macro}
\end{equation}
where the source term related to external force is $\mathbf G=(0,\rho\mathbf a,\rho\mathbf V\cdot a)^T$.
The cell-averaged conservative variables in $\mathbf \Omega_{i}$ can be approximated by numerical quadrature at the discrete level, i.e.,
\begin{equation}
    {\mathbf W}_{i} := \int_{\mathbb R^3} f_{i} \psi d \mathbf{v} \simeq \sum_{j=1}^{N_{v}} w_j f_{i,j} \psi_j,
\end{equation}
where $w_j$ denotes the quadrature weights.
Within the allowable range of integral errors, the update of macroscopic flow variables is equivalent to that of particle distribution function.
However, due to the absence of source term in Eq.(\ref{eq:fvm macro}), nested updating of macroscopic variables and distribution function leads to gains in numerical solution.
As an example, the updated macroscopic variables can be used to evaluate the equilibrium distribution function in Eq.(\ref{eq:bgk}), and subsequently solve Eq.(\ref{eq:fvm micro}).
Such an implicit-explicit (IMEX) scheme is capable of mitigating the stiffness of the kinetic equation in the continuum limit.
In this paper, the macroscopic conservative variables and particle distribution function are simultaneously updated.

\subsection{Numerical flux}\label{sec:flux}

The core task of the finite volume method is the computation of numerical fluxes.
In this study, we employ the integral solution of the BGK equation to formulate the flux function.
This practice has gained widespread acceptance in a series of gas-kinetic schemes to construct multi-scale numerical methods \cite{xu2014direct}.
If we assume the relaxation frequency to be a local constant, Eq.(\ref{eq:bgk}) yields an integral solution in the form
\begin{equation}
    f(t,\mathbf x,\mathbf v)=\frac{1}{\tau}\int_{t^{0}}^{t} \mathcal M (t^{\prime},\mathbf x^{\prime},\mathbf v^{\prime})e^{-(t-t^{\prime})/\tau}dt^{\prime}+e^{-(t-t^{n})/\tau}f(t^{0},\mathbf x^{0},\mathbf v^{0}),
    \label{eq:integral}
\end{equation}
where $\tau=1/\nu$ is the the relaxation time, $\mathbf x'=\mathbf x-\mathbf v'(t-t')-\frac{1}{2}\mathbf a(t-t')^2$ and $\mathbf v'=\mathbf v-\mathbf a(t-t')$ are the particle trajectories in physical and velocity space, and $\{\mathbf x^0=\mathbf x-\mathbf v^0(t-t^0)-\frac{1}{2}\mathbf a(t-t^0)^2,\mathbf v^0=\mathbf v-\mathbf a(t-t^0)\}$ are the initial locations in the phase space.

We then construct the numerical fluxes by approximating the equilibrium and initial distributions in Eq.(\ref{eq:integral}), respectively.
For brevity, we denote a face located between cell $i$ and $i+1$ as $\mathbf x_{i+1/2}=0$, and the $n$-th time step as $t^n=0$.
Eq.(\ref{eq:integral}) can be rewritten in the discrete velocity coordinates as 
\begin{equation}
    f(t,0,\mathbf v_j)=\frac{1}{\tau}\int_{0}^{t} \mathcal M (t^{\prime},\mathbf x^{\prime},\mathbf v^{\prime}_j)e^{-(t-t^{\prime})/\tau}dt^{\prime}+e^{-t/\tau}f(0,\mathbf x^{0},\mathbf v^0_{k}).
    \label{eq:integral1}
\end{equation}

The initial particle distribution function is constructed as
\begin{equation}
    f(0,\mathbf x,\mathbf v_{k})=f_0(\mathbf x,\mathbf v_k)=\begin{cases}f_{i+1/2,k}^I+\boldsymbol\sigma_{i,k} \cdot \mathbf x,\quad &\mathbf x \in \mathbf\Omega_i,\\
    f_{i+1/2,k}^O+\boldsymbol\sigma_{i+1,k} \cdot \mathbf x,\quad &\mathbf x\in\mathbf\Omega_{i+1},\end{cases}
\end{equation}
where $\{f^I_{i+1/2},f^0_{i+1/2}\}$ are the distribution functions at the face $\mathbf x_{i+1/2}$ reconstructed inside and outside the cell $\mathbf \Omega_i$, and $\{\boldsymbol\sigma_{i},\boldsymbol\sigma_{i+1}\}$ are the spatial slopes that can be obtained with the help of various limiters.
Note that under external force, in Eq.(\ref{eq:integral1}), $\mathbf v_k^0$ may not coincide with any node of the pre-established Eulerian mesh.
The present study adopts central difference schemes to interpolate the distribution function value at this point.

The equilibrium distribution function is expanded as
\begin{equation}
    \mathcal M(t,\mathbf x,\mathbf v)=\begin{cases}
    \mathcal M_{0}(1+\boldsymbol\theta_{i,k}\cdot \mathbf x+\phi t),\quad &\mathbf x \in \mathbf\Omega_i,\\
    \mathcal M_{0}(1+\boldsymbol\theta_{i+1,k}\cdot \mathbf x+\phi t),\quad &\mathbf x\in\mathbf\Omega_{i+1},\end{cases}
\end{equation}
where $\mathcal M_0$ is the initial Maxwellian distribution function at $\{t=0,\mathbf x=0\}$, which can be determined using the compatibility condition of the kinetic equation, i.e.,
\begin{equation}
\begin{aligned}
    \int_{\mathbb R^3}\mathcal M_0 \psi d\mathbf v&=\int_{\mathbf v\cdot \mathbf n\ge 0} f^I_{i+1/2} \psi d\mathbf v+\int_{\mathbf v\cdot \mathbf n < 0} f^O_{i+1/2} \psi d\mathbf v \\
    &\simeq \sum_{\mathbf v_j\cdot \mathbf n\ge 0} w_j f^I_{i+1/2,j} \psi_j + \sum_{\mathbf v_j\cdot \mathbf n< 0} w_j f^O_{i+1/2,j} \psi_j.
\end{aligned}
\end{equation}

The spatial and temporal slopes of the equilibrium distribution function are related to the slopes of macroscopic variables, i.e.,
\begin{equation}
    \nabla_{\mathbf x} \mathbf W = \int_{\mathbb R^3} \boldsymbol\theta f\psi d\mathbf v, \quad \frac{\partial \mathbf W}{\partial t}=\int_{\mathbb R^3} \phi f\psi d\mathbf v.
\end{equation}
If we expand the slopes with respect to collision invariants, i.e.,
\begin{equation}
    \boldsymbol \theta = \hat{\boldsymbol\theta} \cdot \psi, \quad \phi=\hat{\phi}\cdot \psi,
\end{equation}
then the coefficients can be determined by solving the corresponding linear systems \cite{xu2001gas}.
Note that the form of temporal slope can be transformed with the help of Euler equations, i.e.,
\begin{equation}
    -\int \mathbf v \cdot\nabla_{\mathbf x}{\mathcal M}\psi d\mathbf v-\int \mathbf a \cdot\nabla_{\mathbf v}{\mathcal M}\psi d\mathbf v=\frac{\partial W}{\partial t}=\int \phi \mathcal M\psi d\mathbf v,
\end{equation}
and then solved in the same way as spatial slopes.

After all the physical quantities and coefficients are determined, the integral solution in Eq.(\ref{eq:integral1}) becomes
\begin{equation}
    \begin{aligned}
    f(t,0,\mathbf v_{j})&=\left(1-e^{-t/\tau}\right)\mathcal M_{0,j}+\left(\tau(-1+e^{-t/\tau})+te^{-t/\tau}\right)\boldsymbol\theta_{i,i+1}\cdot \mathbf v_{j}\mathcal M_{0,j}\\&-\left[\tau\left(\tau(-1+e^{-t/\tau})+te^{-t/\tau}\right)+\frac{1}{2}t^2e^{-t/\tau}\right]\boldsymbol\theta_{i,i+1}\cdot\mathbf a \mathcal M_{0,j}+\tau\left(t/\tau-1+e^{-t/\tau}\right)\phi \mathcal M_{0,j}\\
    &+e^{-t/\tau}\left[\left(f_{i+1/2,j^0}^I-\left(\mathbf v_j t-\frac{1}{2}\mathbf a t^2\right)\cdot\boldsymbol\sigma_{i,j^0}\right)H\left(-\mathbf v_j t+\frac{1}{2}\mathbf at^2\right)\right.\\
    &\left.+\left(f_{i+1/2,j^0}^O-\left(\mathbf v_j t-\frac{1}{2}\mathbf a t^2\right)\cdot\boldsymbol\sigma_{i+1,j^0}\right)\left(1-H\left(-\mathbf v_j t+\frac{1}{2}\mathbf a t^2\right)\right)\right]\\
    &=\widetilde{m}_{j}+\widetilde{f}_{j},
    \end{aligned}
    \label{eq:integral2}
\end{equation}
which can be decomposed into the contributions from equilibrium and initial distribution functions.
The generalized Heaviside function is defined as
\begin{equation}
    H(\mathbf x)=\begin{cases}
    1,\quad &\mathbf x \in \mathbf\Omega_i,\\
    0,\quad &\mathbf x\in\mathbf\Omega_{i+1}.\end{cases}
\end{equation}
Finally, the numerical fluxes are determined by the time-dependent distribution function at the face, i.e.,
\begin{equation}
\begin{aligned}
    &\mathbf F^f=\mathbf v f(t,0,\mathbf v), \\
    &\mathbf F^W=\int_{\mathbb R^3} \mathbf v f(t,0,\mathbf v) \psi d\mathbf v.
\end{aligned}
\end{equation}

\subsection{Hermite approximation}\label{sec:hermite}

For non-equilibrium flows under external force, a core task is to compute the derivatives of the distribution function in the velocity space.
This study employs the Hermite polynomial expansion method proposed by Grad \cite{grad1949kinetic}. 
The distribution function is expanded as an $N_m$-th order Hermite polynomial series, i.e.,
\begin{equation}
    f(t,\mathbf{x},\mathbf{v})\simeq f_{N_m} =\omega(\mathbf{u})
\sum_{n=0}^{N_m} \frac{1}{n!}\mathbf{d}^{(n)}(t,\mathbf{x}) : \mathcal{H}^{(n)}(\mathbf{u}),
\label{eq:ex_f}
\end{equation}
where $\mathbf u=(\mathbf v-\mathbf V)/\sqrt{RT}$ is the rescaled particle velocity based on local bulk velocity and temperature, $R$ is the gas constant, and $\omega(\mathbf u)\equiv \frac{1}{\sqrt{2\pi}^3} \exp(-\mathbf u^2)$ is the weight function. 
The tensor contraction is denoted as $:$.
The $n$-th order tensor Hermite polynomial is defined by
\begin{equation}
    \mathcal{H}^{(n)}(\mathbf{u})
= \frac{1}{\omega(\mathbf{u})}\nabla_{\mathbf u}^{(n)}\omega(\mathbf{u}),
\end{equation}
which satisfies the recurrence relation, i.e.,
\begin{equation}
    \mathcal{H}^{(n+1)}(\mathbf{u})
= \mathbf{u}\mathcal{H}^{(n)}(\mathbf{u})
- n\boldsymbol{\delta}_i\mathcal{H}^{(n-1)}(\mathbf{u}),
\end{equation}
with $\mathcal{H}^{(0)}=1$, $\mathcal{H}^{(1)}=\mathbf{u}$, and $\boldsymbol{\delta}$ the identity tensor. The expansion coefficients can be determined by
\begin{equation}
    \mathbf{d}^{(n)}(t,\mathbf{x})
= \int_{\mathbb R^3} f\mathcal{H}^{(n)}(\mathbf{u})d\mathbf{u} \simeq \sum_{j=1}^{N_{v}} w_j f_{j} \mathcal H^{(n)}_j.
\end{equation}
Notably, truncating the expansion in Eq.(\ref{eq:ex_f}) at the zeroth-order (\( N_m=0 \)) recovers the Maxwellian distribution.

Substituting $f_{N_m}$ into the forcing term of the Boltzmann equation yields
\begin{align}
    \mathbf{a}\cdot\nabla_{\mathbf v} f \simeq &\frac{1}{\sqrt{RT}}\mathbf{a}\cdot\nabla_{\mathbf u}\left( \omega(\mathbf u) \sum_{n=0}^{N_m} \frac{1}{n!} \mathbf d^{(n)}(t,\mathbf{x}) : \mathcal H^{(n)}(\mathbf u)\right) \nonumber \\
    =& \frac{1}{\sqrt{RT}} \mathbf{a}\cdot\sum_{n=0}^{N_m} \frac{1}{n!} \mathbf d^{(n)}(t,\mathbf{x}) : \mathcal \nabla_{\mathbf u} \left(  \omega(\mathbf u) H^{(n)}(\mathbf u)\right) \nonumber \\
    =& -\frac{1}{\sqrt{RT}}\omega(\mathbf u) \sum_{n=0}^{N} \frac{1}{n!} \mathbf {a\otimes d}^{(n)}(t,\mathbf{x}) : \mathcal H^{(n+1)}(\mathbf u).
\end{align}
By substituting the Hermite polynomials and moments, the above equation can be further simplified as
\begin{equation}
    \mathbf{a}\cdot\nabla_{\mathbf{v}}f
    \approx \frac{1}{\sqrt{RT}}\,\mathcal M
    \Bigl(
    \mathbf{a}\cdot\mathbf{u}
    - \frac{\mathbf{a\otimes\mathbf T}}{2\rho RT} : \mathcal{H}^{(3)}(\mathbf{u})
    + \frac{\mathbf{a\otimes Q}}{6\rho(RT)^{3/2}} : \mathcal{H}^{(4)}(\mathbf{u})
    +\dots\Bigr),
\label{eq:hermite1}
\end{equation}
where the modified stress and heat transfer tensors are defined as
\begin{equation}
    \mathbf T=\int_{\mathbb R^3} f(\mathbf u \otimes \mathbf u-\boldsymbol \delta)d\mathbf v, \quad \mathbf Q=\int_{\mathbb R^3} f(\mathbf u \otimes \mathbf u\otimes \mathbf u)d\mathbf v.
\end{equation}
It can be seen that the leading term in Eq.(\ref{eq:hermite1}) describes the equilibrium effects of external force, while the rest higher-order terms quantify the non-equilibrium effects.
The Hermite expansion conserves the moments of external force up to $(N_m+1)$-th order.

The spectral convergence of Hermite expansions undoubtedly provides an efficient tool to characterize the non-equilibrium distribution function and compute its derivatives.
The other remaining question to address is its universality.
As analyzed in \cite{grad1949kinetic}, provided that $\omega^{-1/2}f$ is square integrable, the series in Eq.(\ref{eq:ex_f}) converges spectrally. The necessary condition is
\begin{align}
    \lim\limits_{|\mathbf v| \to \infty} f \exp\left(\frac{|\mathbf{v}-\mathbf{V}|^2}{4RT}\right) = 0.
\end{align}
This implies that the velocity distribution to be approximated must decay at a faster rate than the exponential factor \( \exp\left(-{|\mathbf{v}-\mathbf{V}|^2}/{4RT}\right) \), which can be regarded as an equilibrium distribution with the temperature \( 2T \).
For most non-equilibrium distributions, this condition can be naturally satisfied.
Occasionally, in highly dissipative regions, e.g., in the vicinity of a shock wave with a large temperature gradient, the scaled temperature may be chosen (to surpass no less than half of the domain's peak temperature) as a surrogate of the local temperature $T$ to guarantee the convergence of the series \cite{Cai2019,Yan2024}.

\subsection{Numerical quadrature}

The numerical integration based on a particle distribution function can be written as
\begin{equation}
    \int_{\mathbb R^3} g(\mathbf v) f(\mathbf v) d\mathbf v \simeq \sum_j^{N_v} w_j g(\mathbf v_j) f(\mathbf v_j),
\end{equation}
where $g$ denotes an arbitrary function of $\mathbf v$, and $w_j$ is the quadrature weight at the collocation point $\mathbf v_j$.
Two sets of numerical quadratures are considered in this study to solve the BGK equation.
For brevity, we describe their use in the one-dimensional case below, while the extension to the multi-dimensional case is straightforward via the tensor product rule.

The first numerical quadrature follows the Newton-Cotes (NC) rule.
This method samples the integrand at equally spaced points.
The quadrature weights are determined by integrating the Lagrange basis polynomials over the velocity space.
If the integrand is sufficiently smooth, 
it can be proved that the error scaling of the Newton-Cotes rule follows
\begin{equation}
    \epsilon \propto (\Delta u)^{N_p} \max_{u\in[-U,U]} I^{(N_p)}(u),
\end{equation}
where $I$ denotes the integrand, and $N_p=N_v+1$ for a $N_v$-panel Newton–Cotes rule.

The second numerical quadrature is the Gauss-Hermite (GH) quadrature.
This technique is tailored to approximate integrals of the form
\begin{equation}
    \int_{-\infty}^\infty e^{-u^2} I(u) du \simeq \sum_{j=1}^{N_v} w_j I(u_j).
\end{equation}
The quadrature nodes are placed at the $N_v$ real roots of $N_v$-th Hermite polynomial (as discussed in Section \ref{sec:hermite}), and the weights are given by
\begin{equation}
    w_j=\frac{2^{N_v-1}N_v!\sqrt{\pi}}{N_v^2 [\mathcal H^{(N_v-1)}(u_i)]^2}.
\end{equation}
The Gauss-Hermite quadrature is especially efficient for those quadrature problems in which the integrand is multiplied by $e^{-u^2}$, e.g., the Maxwellian distribution function,
It can be applied to an arbitrary integrand by refactoring it to match the Hermite–weight form, i.e.,
\begin{equation}
    \int_{-\infty}^\infty I(u) du = \int_{-\infty}^\infty e^{-u^2}(e^{u^2}I(u))=\int_{-\infty}^\infty e^{-u^2}I'(u) \simeq \sum_{j=1}^{N_v} w_j I'(u_j).
\end{equation}
The quadrature is exact when the integrand is a polynomial of degree up to 
$2N_v-1$.
Note that besides the classic Gauss-Hermite quadrature, the quadrature nodes can also be shifted and rescaled 
to adapt peculiar non-equilibrium distributions.

\subsection{Well-balanced property}

The well-balanced property is achieved due to the precise offset of the external forcing terms and inhomogeneous fluxes in Eq.(\ref{eq:fvm micro}) (and Eq.(\ref{eq:fvm macro}), correspondingly).
Here, we analyze the numerical behavior of the solution algorithm in the asymptotic limit of the BGK equation.
For brevity, a local coordinate system aligned with the direction of the external force is established, thereby simplifying the flow equations to one dimension.
In the continuum limit, i.e., the Euler equations, the hydrostatic equilibrium solution takes the form
\begin{equation}
    \rho=\rho_0 \exp\left(\frac{ax}{RT}\right), \quad \mathbf V=0, \quad p=p_0 \exp\left(\frac{ax}{RT}\right),
    \label{eq:hydrostatic}
\end{equation}
where $\{\rho_0,p_0\}$ are the density and pressure at an arbitrarily selected reference state.
The temperature $T$ is equal everywhere \cite{chapman1990mathematical}.
With infinitely frequent intermolecular collisions, the particle distribution function is everywhere in equilibrium.

We first consider the source term part.
Provided that the particle distribution function is Maxwellian, the Hermite expansion in Eq.(\ref{eq:hermite1}) degenerates to
\begin{equation}
    G=-\mathbf{a}\cdot\nabla_{\mathbf{v}}f
    = -\frac{au}{{RT}}\,\mathcal M,
\end{equation}
and the source terms related to the external force can be obtained via
\begin{equation}
    \mathbf{G}=\int G\psi d\mathbf v=\left(\begin{array}{c}
    0 \\
    \rho {a} \\
    \rho Ua
    \end{array}\right)=\left(\begin{array}{c}
    0 \\
    \rho {a} \\
    0
    \end{array}\right).
\end{equation}
Since the zeroth-order truncation in the Hermite expansion corresponds to the Maxwellian distribution, the two source terms above do not introduce any numerical errors.

We then turn to the numerical fluxes.
In the Euler limit with $\tau\rightarrow 0$, the integral solution in Eq.(\ref{eq:integral2}) reduces to
\begin{equation}
    f=\mathcal M_0(1+t\phi).
\end{equation}
As analyzed in Section \ref{sec:flux}, the temporal slope of the distribution function can be obtained via
\begin{equation}
    -\int u \partial_{x}{\mathcal M}\psi du-\int a \partial_{v}{\mathcal M}\psi du=\frac{\partial W}{\partial t}=\int \phi \mathcal M\psi du.
\end{equation}
The derivatives of the Maxwellian distribution can be obtained using the chain rule, i.e.,
\begin{equation}
\begin{aligned}
    &\frac{\partial \mathcal M}{\partial x}=\frac{1}{\rho}\frac{\partial \rho}{\partial x}\mathcal M+\frac{1}{2\lambda}\frac{\partial \lambda}{\partial x}\mathcal M-(u^2-2uU+U^2)\frac{\partial \lambda}{\partial x}\mathcal M+2\lambda(u-U)\frac{\partial U}{\partial x}\mathcal M, \\
    &\frac{\partial \mathcal M}{\partial u}=-2\lambda(u-U)\mathcal M,
\end{aligned}
\end{equation}
where $\lambda=1/(2RT)$.
Inserting the hydrostatic solution in Eq.(\ref{eq:hydrostatic}) yields
\begin{equation}
    \phi=0.
\end{equation}

Therefore, the numerical fluxes depend only on the Maxwellian distribution at the cell face, which are of the form
\begin{equation}
\begin{aligned}
    &\mathbf F^f=u\mathcal M_0, \\
    &\mathbf F^W = \int_{-\infty}^\infty u \mathcal M_0 \psi du=\left(\begin{array}{c}
    0 \\
    \rho RT \\
    0
    \end{array}\right).
\end{aligned}
\end{equation}
It is clear that density and energy are well balanced.
For the momentum equation, we can obtain the following relation between the source term and numerical flux, i.e.,
\begin{equation}
    \int_{x_{i-1/2}}^{x_{i+1/2}} \mathbf G_i^{\rho U}dx=\int_{x_{i-1/2}}^{x_{i+1/2}} \rho a dx = RT(\rho_{x_{i+1/2}}-\rho_{x_{i-1/2}})=-(\mathbf F^{\rho U}_{i-1/2}-\mathbf F^{\rho U}_{i+1/2}).
\end{equation}
from which the well balance in a control volume is clearly identified.
%$O(\Delta x^2)$

The above analysis is performed only in the asymptotic limit with vanishing molecular mean free path.
When the mean free path is not zero (i.e., the flow regimes of Navier-Stokes equations and beyond), it can be shown that there exist non-equilibrium stress and heat flux aligned with the direction of the external force, and thus the hydrostatic solution is no longer isothermal \cite{xiao2018investigation}.
It is difficult to prove the well-balanced property by numerical analysis anymore.
It should be noted, however, that it is the non-equilibrium effects described above where the role of external forces becomes exceptionally important.
By introducing particle trajectories under external forces into the numerical flux function (from the integral solution of the BGK equation) and solving exactly for the derivatives in the phase space (using the Hermite spectral method), the current solution algorithm is consistent with the physical mechanisms of multi-scale flow transport, and should consequently demonstrate superior performance in non-equilibrium flow simulations.
In the next section, we will conduct numerical experiments to validate the efficacy of the solution algorithm.

\section{Numerical Experiments}\label{sec:experiment}

Numerical experiments are presented in this section to validate the current solution algorithm.
Dimensionless variables are introduced as
\begin{equation}
\begin{aligned}
    &\tilde t = \frac{t}{L_0/V_0}, \ \tilde{\mathbf x} = \frac{\mathbf x}{L_0}, \ \tilde{\rho} = \frac{\rho}{\rho_0}, \ \tilde{\mathbf V} = \frac{\mathbf V}{V_0}, \ \tilde{T} = \frac{T}{T_0}, \ \tilde{E} = \frac{E}{V_0^2}, \ \\
    &\tilde{\mathbf P} = \frac{\mathbf P}{\rho_0 V_0^2}, \ \tilde{\mathbf q} = \frac{\mathbf q}{\rho_0 V_0^{3}}, \ \tilde{\mathbf v} = \frac{\mathbf v}{V_0}, \ \tilde f = \frac{f}{\rho_0 /V_0^3}, \ 
\end{aligned}
\end{equation}
where a physical quantity with a subscript of 0 indicates its value in the reference state, $V_0=\sqrt{2kT_0/m}$ is the most probable molecular speed.
The referenced Knudsen number is defined as
\begin{equation}
    \mathrm{Kn}=\frac{\ell_0}{L_0}=\frac{V_0}{L_0 \nu_0},
\end{equation}
where $\ell_0=V_0/\nu_0$ denotes the molecular mean free path and $\nu_0$ is the mean collision frequency.
For brevity, we drop the tilde notation to denote dimensionless variables henceforth.

\subsection{Convergence of Hermite approximation}

The first numerical experiment examines the convergence properties of the Hermite approximation when applied to particle distributions in both near-equilibrium and non-equilibrium states.
Four forms of the particle distribution function are considered, i.e.,
\begin{enumerate}
  \item $f = \exp(-\frac{1}{4}u^2)$,
  \item $f = \exp(-\frac{1}{4}u^2)(1+0.5\sin(u))$,
  \item $f=\exp(-\frac{1}{4}(u-2)^2)+\frac{1}{2}(-\frac{1}{4}(u+2)^2)$,
  \item $f = \exp(-\frac{1}{4}u^4)$.
\end{enumerate}

\begin{figure}[htb!]
    \centering
    \subfigure[$f = \exp(-\frac{1}{4}u^2)$]{
        \includegraphics[width=0.47\textwidth]{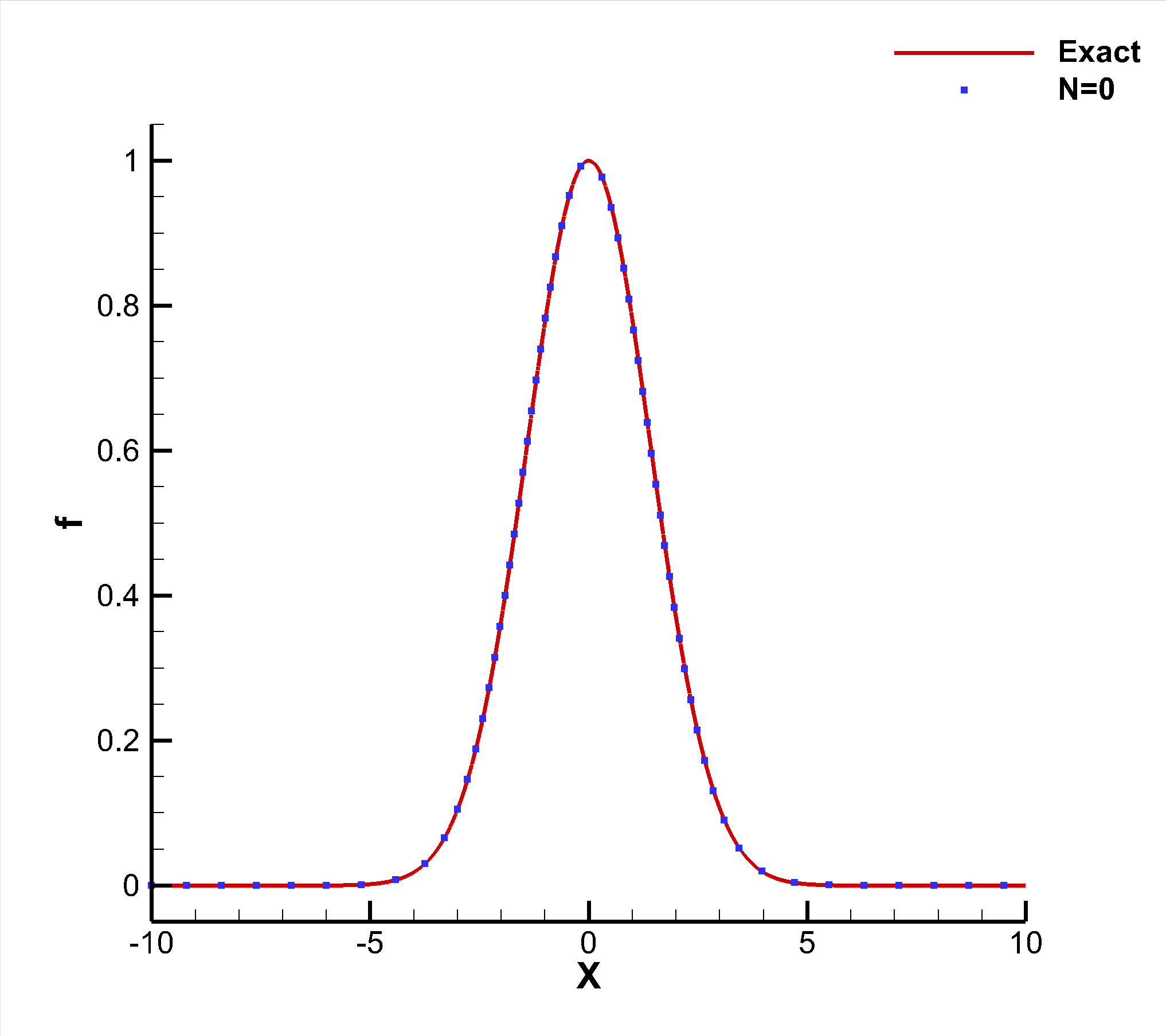}
    }
    \subfigure[$f = \exp(-\frac{1}{4}u^2)(1+0.5\sin(u))$]{
        \includegraphics[width=0.47\textwidth]{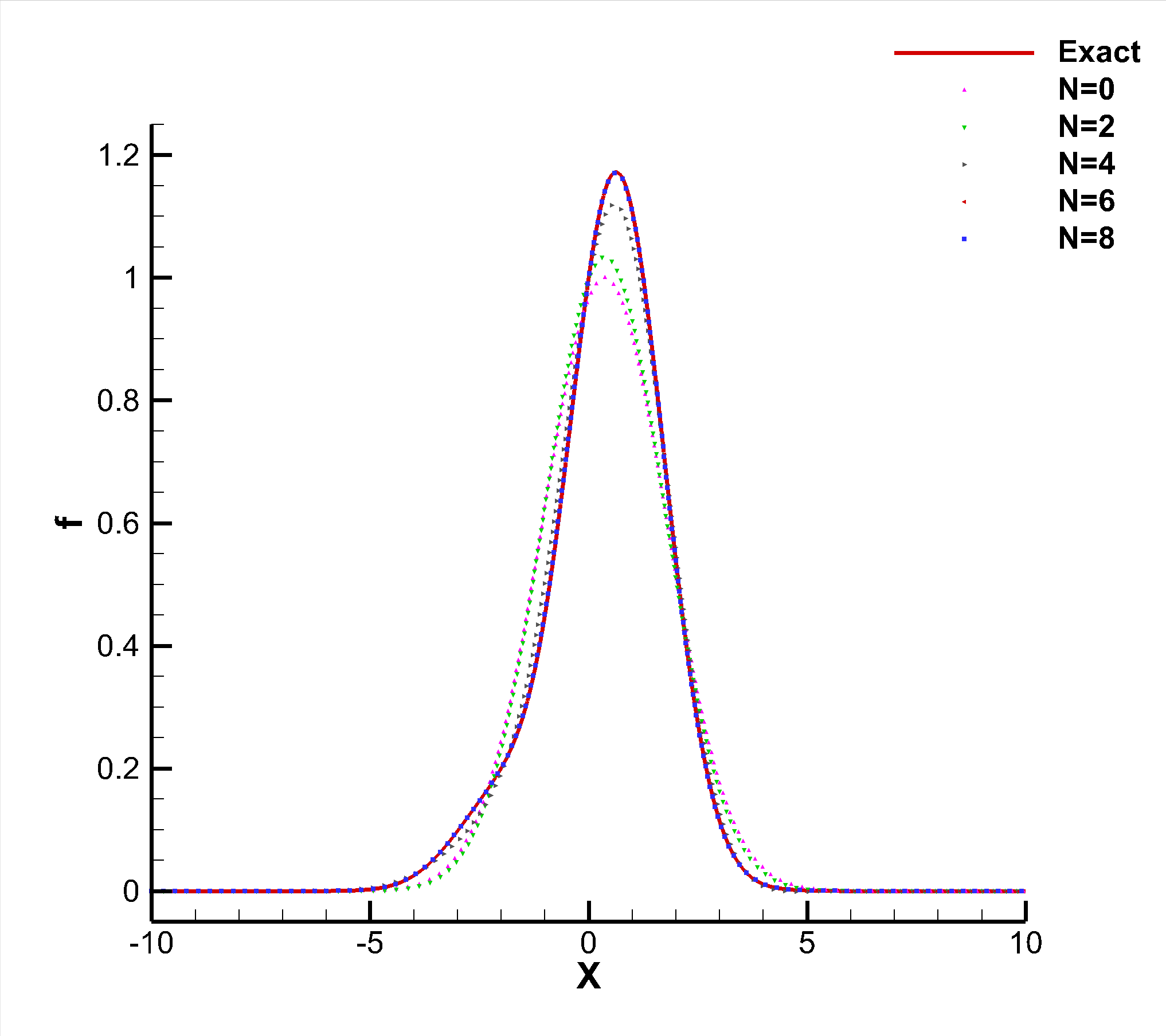}
    }
    \subfigure[$f=\exp(-\frac{1}{4}(u-2)^2)+\frac{1}{2}(-\frac{1}{4}(u+2)^2)$]{
        \includegraphics[width=0.47\textwidth]{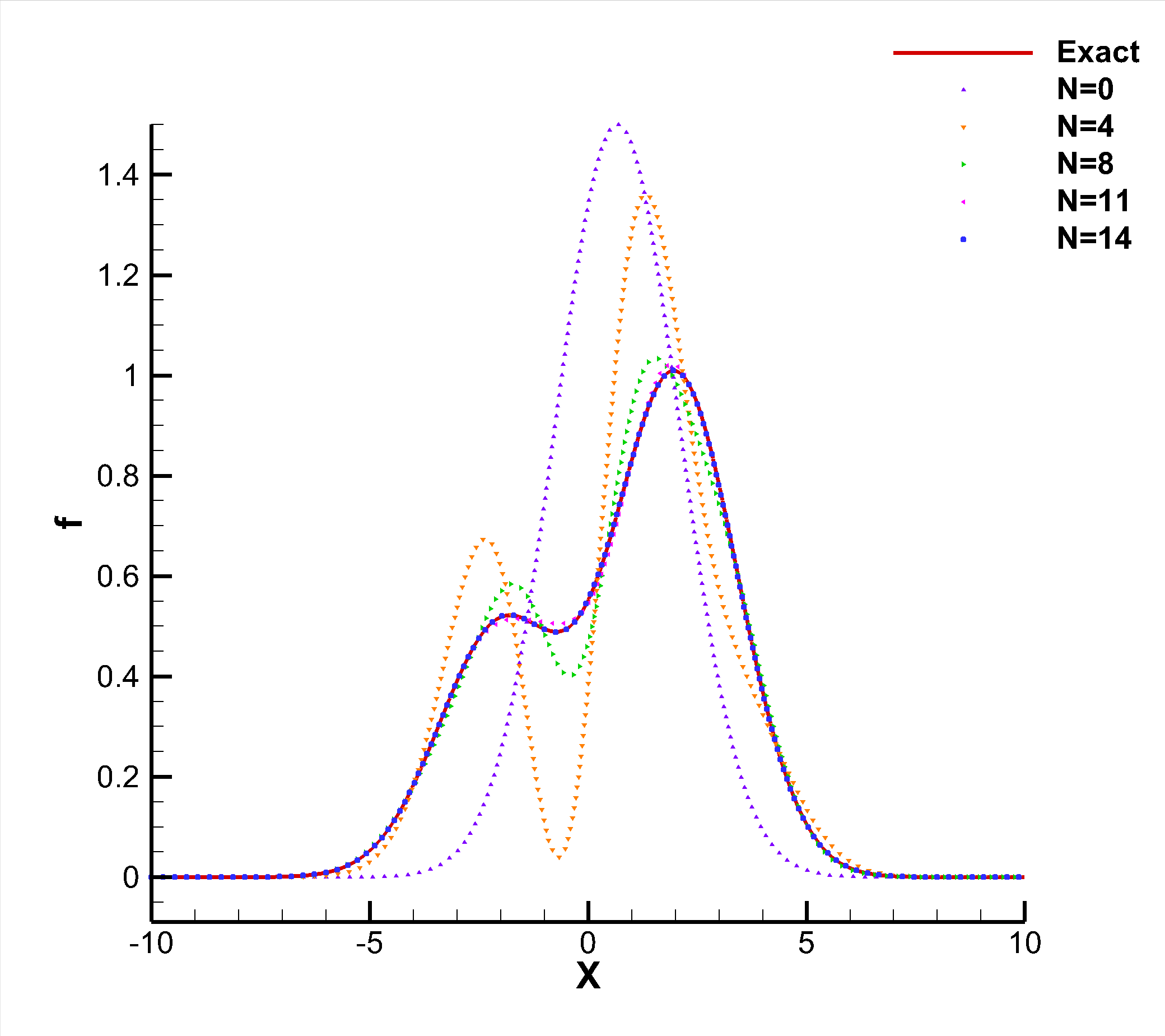}
    }
    \subfigure[$f = \exp(-\frac{1}{4}u^4)$]{
        \includegraphics[width=0.47\textwidth]{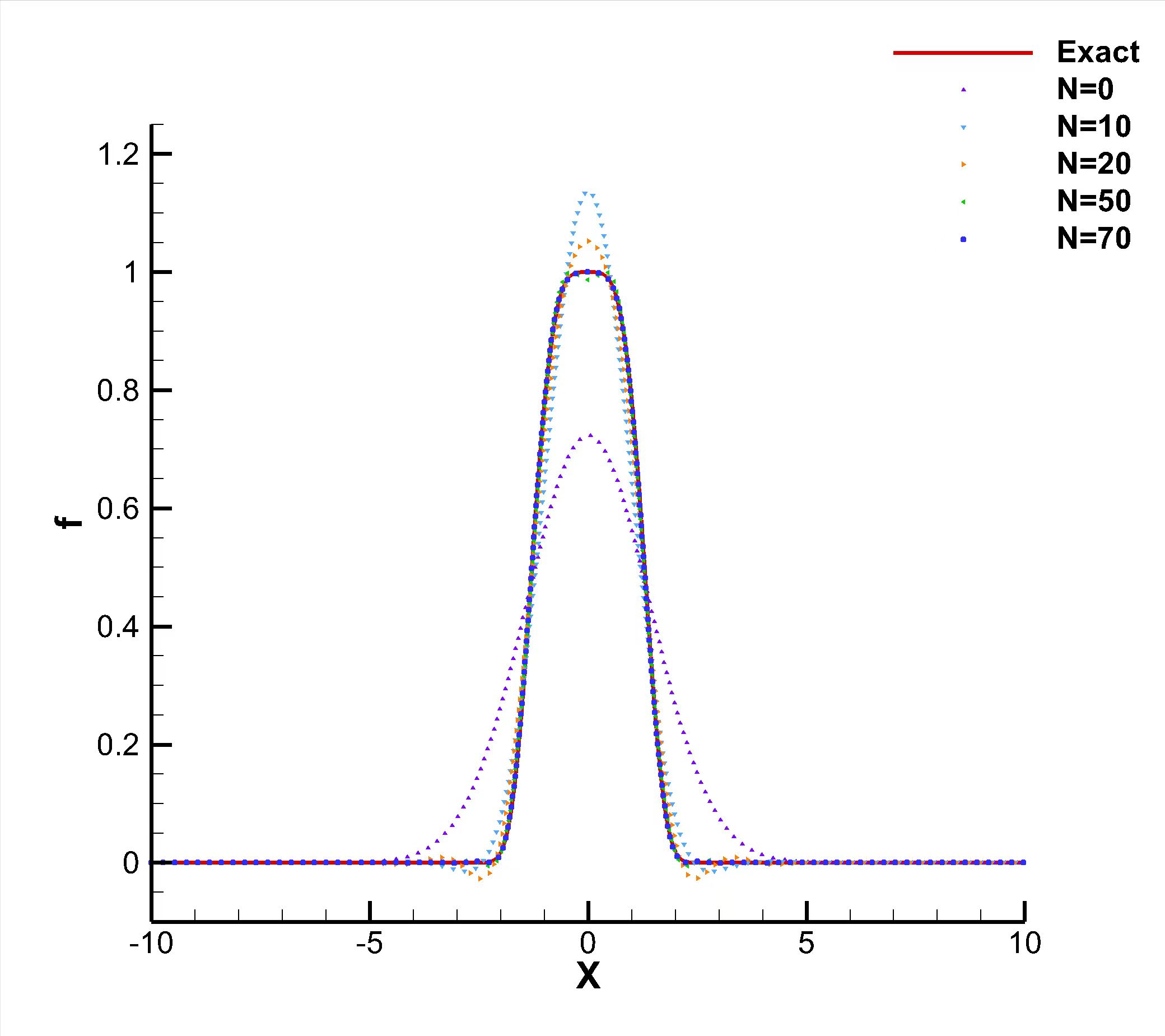}
    }
    \caption{Convergence tests of the Hermite approximation of near-equilibrium and non-equilibrium particle distribution functions. Here, $N$ represents the order of the Hermite expansion.}
    \label{fig:convergence}
\end{figure}

Figure \ref{fig:convergence} illustrates the approximation of four distribution functions using the Hermite expansion.
It can be seen that the numerical accuracy and rate of convergence are strongly correlated with the form of the distribution function.
For equilibrium and near-equilibrium distributions, since the first term of the Hermite expansion corresponds to the Maxwellian distribution, the distribution function can be accurately characterized with minimal expansion orders.
For non-equilibrium distributions, the Hermite spectral method also exhibits good convergence properties.
Merely $O(10)$ expansion terms are required to precisely resolve typical non-equilibrium distributions, e.g., the bimodal distribution shown in Figure \ref{fig:convergence}(c).
Compared to the finite difference discrete velocity method, this method offers substantially superior computational efficiency (which will be corroborated by our numerical experiments starting from Section 4.2).
When the distribution function is sharply located or discontinuous in the velocity space, e.g., in Figure \ref{fig:convergence}(d), Gibbs' phenomenon leads to the need for higher orders of expansion to converge to the accurate distribution.
Fortunately, such non-trivial distributions rarely occur in space and time in real-world flow problems.
Throughout the computational cycle, the Hermite approximation is capable of providing efficient and accurate characterization of distribution functions in most cases.

\subsection{One-dimensional hydrostatic solution}

We then verify the well-balanced property of the solution algorithm.
The one-dimensional, dimensionless hydrostatic solution is set as
\begin{equation}
    \rho_0=\exp(ax),\quad U_0=0,\quad p_0=\exp(ax).
\end{equation}
The pressure field is then subjected to the following perturbation,
\begin{equation}
    p= p_{0}(x)+0.01e^{-100(x-0.5)^2}.
\end{equation}
The initial particle distribution function is set as Maxwellian everywhere in correspondence with the macroscopic variables.
The detailed computational setup is provided in Table \ref{tab:hydrostatic}.

\begin{figure}[htb!]
    \centering
    \subfigure[Kn=0.0001]{
        \includegraphics[width=0.47\textwidth]{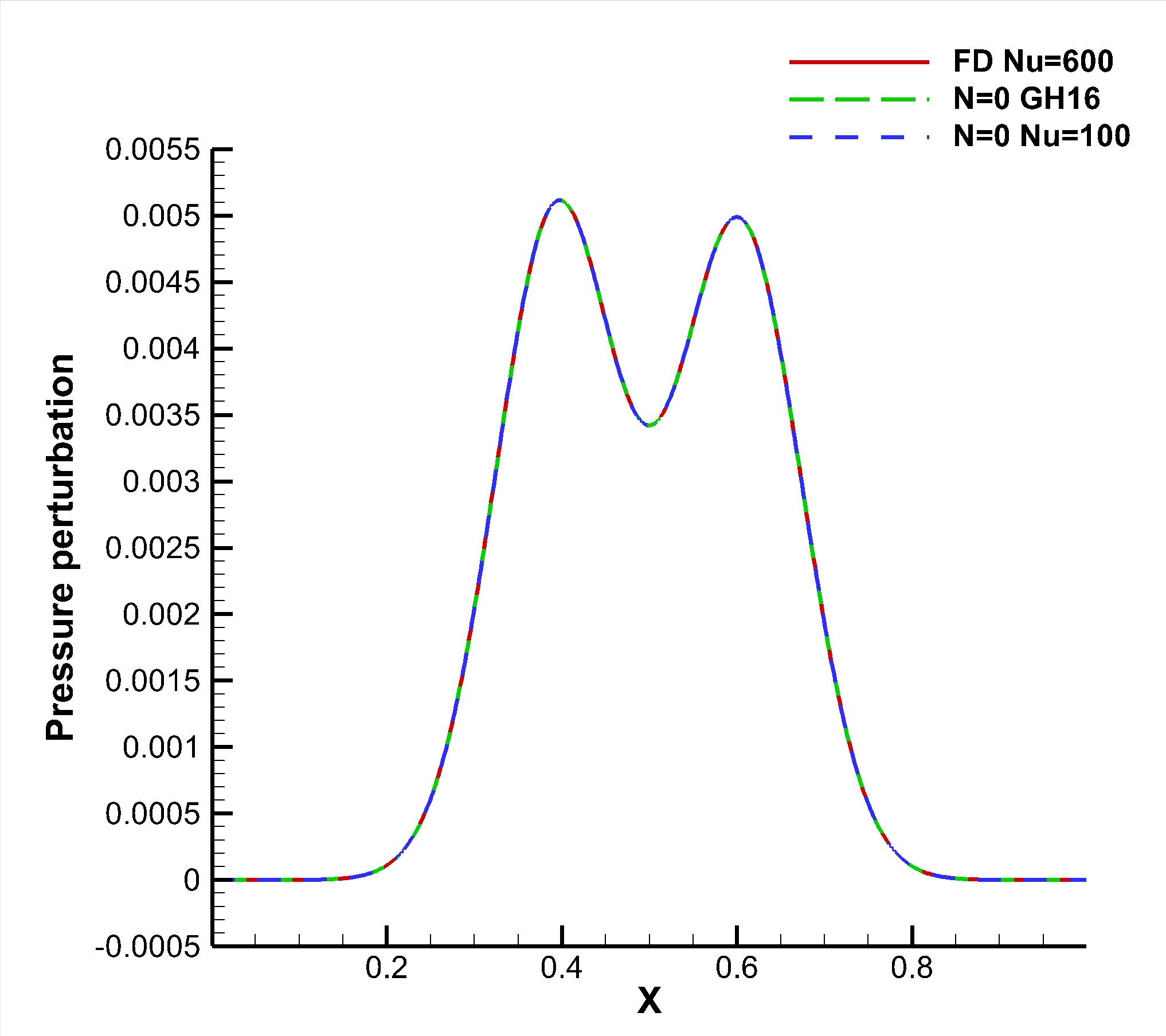}
    }
    \subfigure[Kn=0.01]{
        \includegraphics[width=0.47\textwidth]{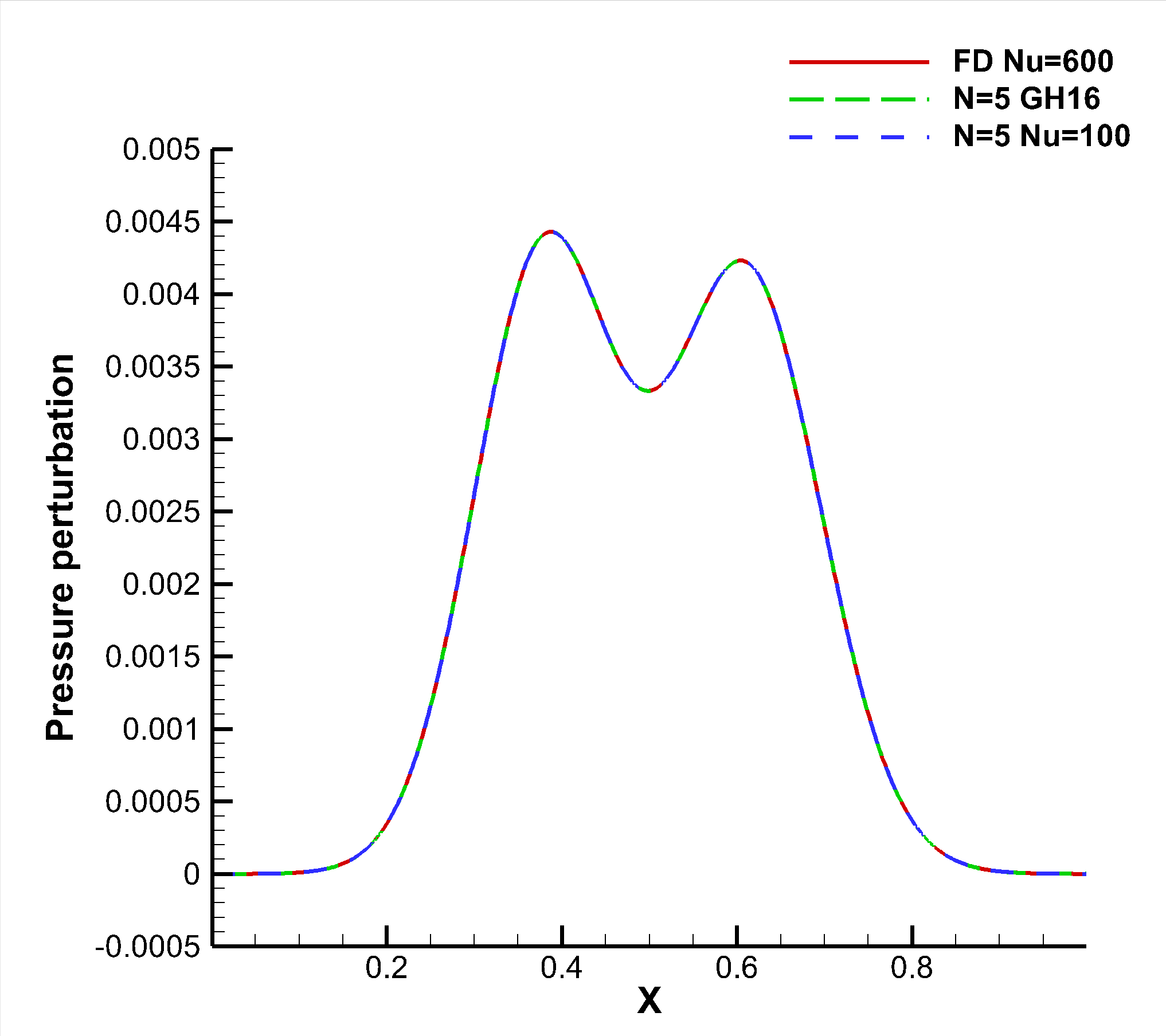}
    }
    \subfigure[Kn=0.1]{
        \includegraphics[width=0.47\textwidth]{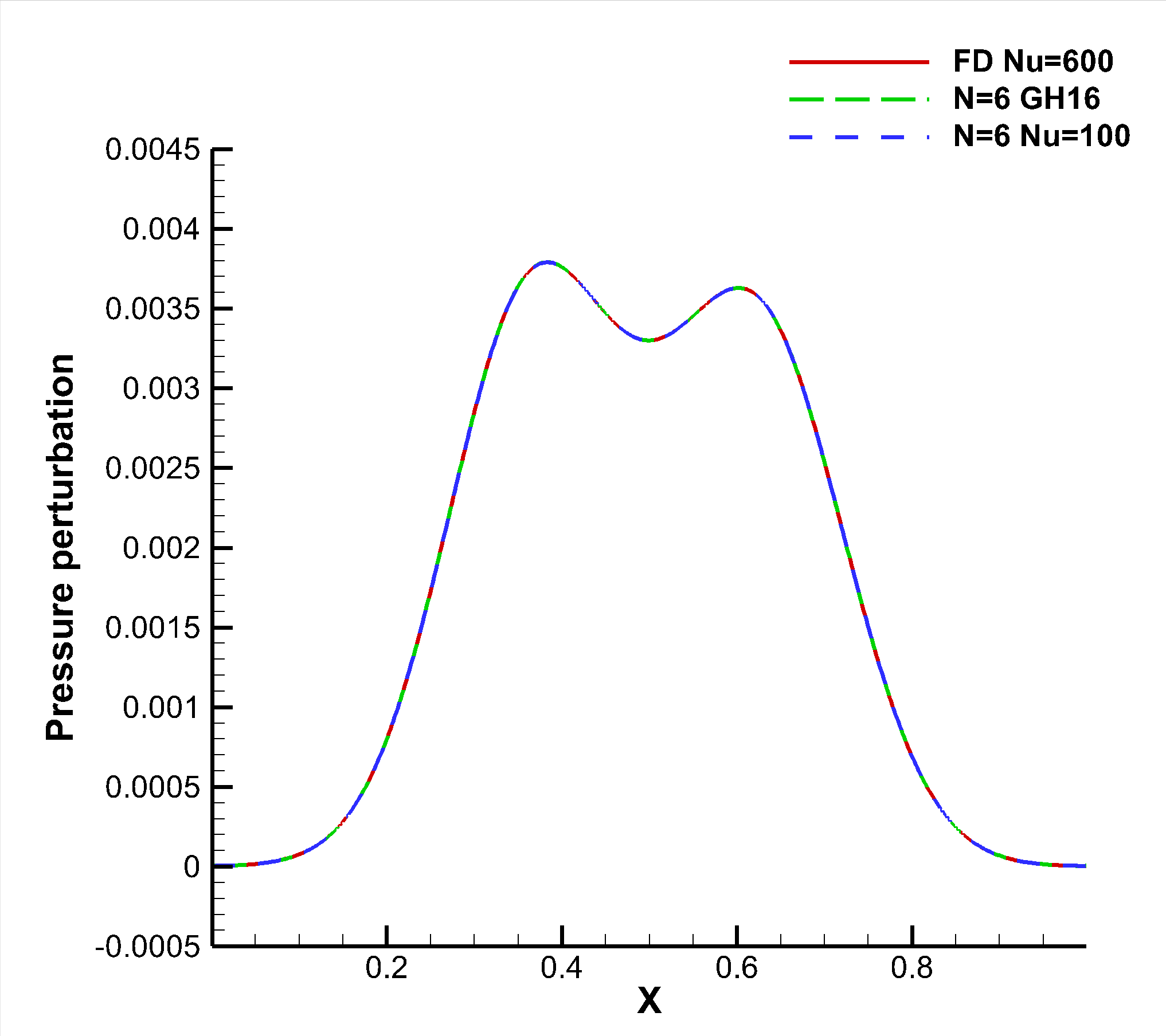}
    }
    \subfigure[Kn=1]{
        \includegraphics[width=0.47\textwidth]{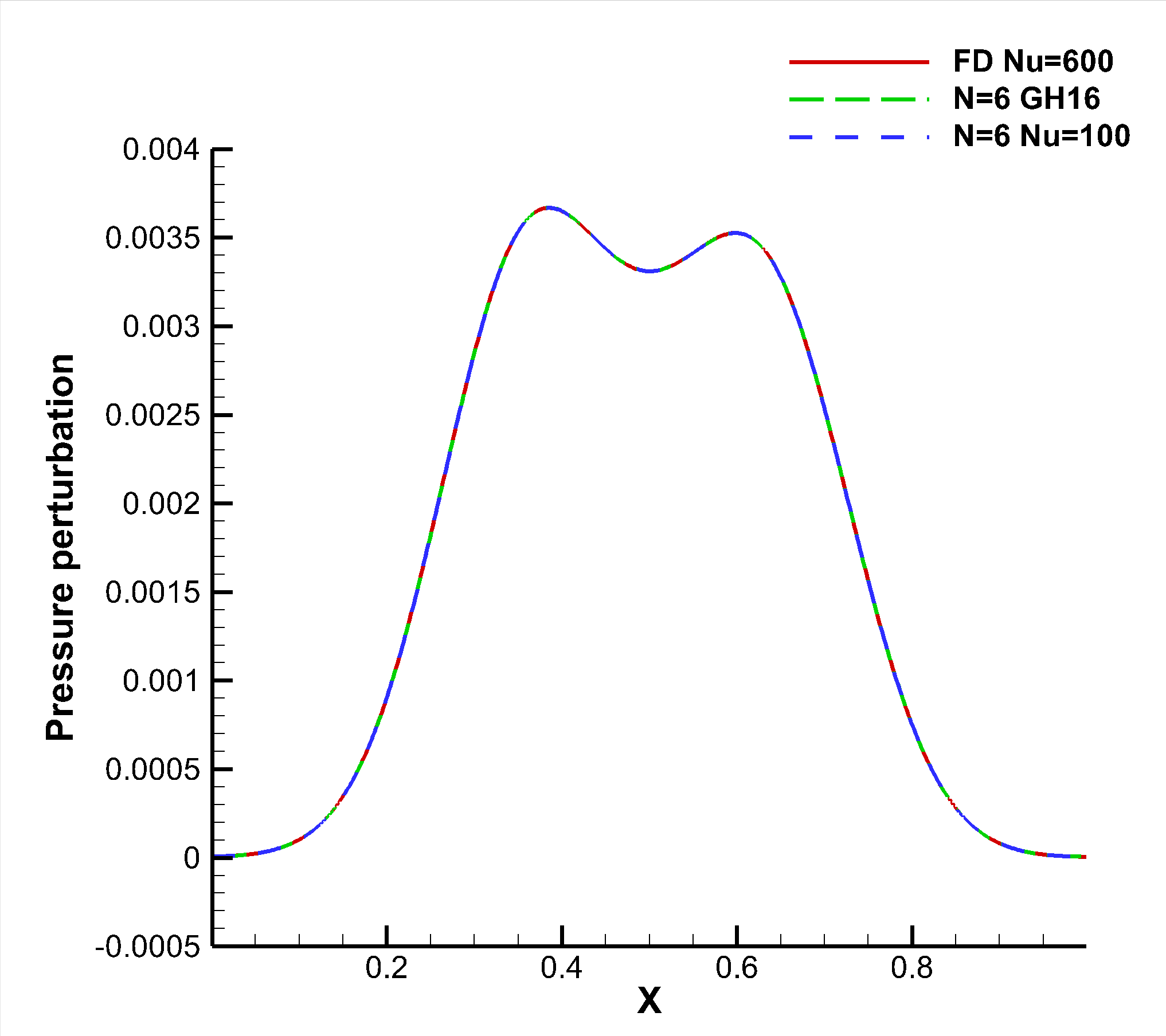}
    }
    \caption{Pressure perturbation distributions from hydrostatic solution at different reference Knudsen numbers. Here, $N$ represents the order of the Hermite expansion, GH denotes the Gauss-Hermite quadrature, and the rest that are not specifically labeled denote the Newton-Cotes quadrature.}
    \label{fig:hydrostatic}
\end{figure}

\begin{figure}[htb!]
    \centering
    \includegraphics[width=0.5\textwidth]{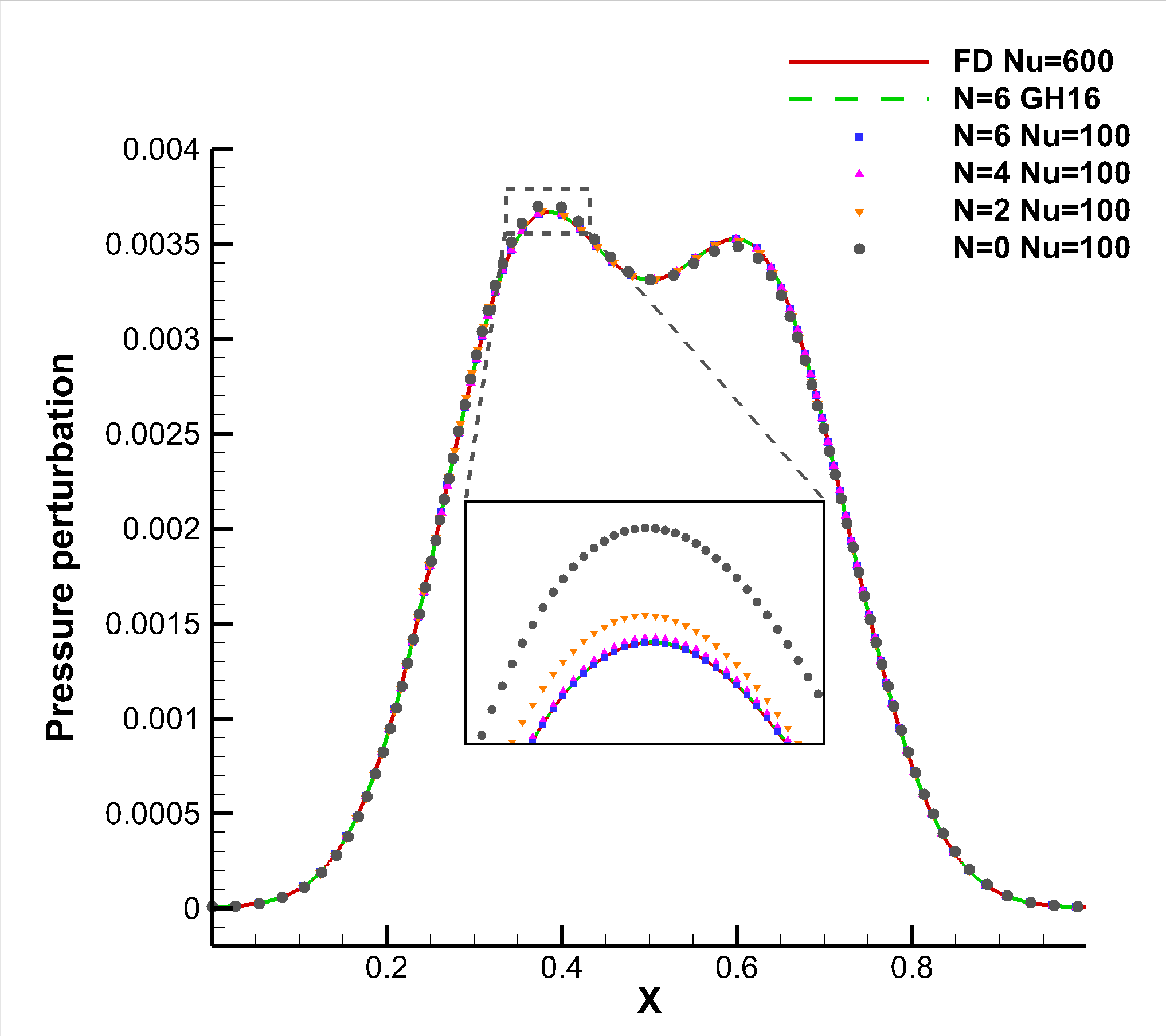}
    \caption{Convergence test and criterion for determining the order of Hermite polynomials in one-dimensional hydrostatic problem (with Kn=1).}
    \label{fig:hydro converge}
\end{figure}

Figure \ref{fig:hydrostatic} presents the spatial distribution of pressure perturbations at the output moment.
As discussed in \cite{leveque1999wave}, a naive treatment of the external force term can fail to capture such a small perturbation without destroying the global hydrostatic solution.
Thanks to the well-balanced property, the current solution algorithm accurately captures the propagation of pressure perturbations in different flow regimes.
Table \ref{tab:hydrostatic cost} lists the computational costs of different methods (i.e., the Hermite approximation in Section \ref{sec:hermite} and second-order central difference) in the same framework.
The results show that the Hermite spectral method, along with the Gauss-Hermite quadrature, significantly improves the computational efficiency of the numerical method at different reference Knudsen numbers (more than one order of magnitude improvement over the finite difference method).
The computational parameters in the table are determined by the grid convergence tests.
Figure \ref{fig:hydro converge} illustrates a typical convergence test case for determining the order of Hermite polynomials.

\begin{table}[htbp]
	\caption{Computational setup of one-dimensional hydrostatic problem (NC/GH refers to the Newton-Cotes/Gauss-Hermite quadrature).}
	\centering
	\begin{tabular}{lllllllll} 
	    \hline
	    Gas & $t$ & $x$ & $N_x$ & $u$ & $N_u$ & Order \\
	    Argon & $(0,0.1]$ & $[0,1]$ & 600 & $[-5,5]$ & $[16,600]$ & 2 \\
		%\hline
		\rule{0pt}{0ex}\\
		$N_m$ & $a$ & Kn & Quadrature & Integrator & Boundary & CFL \\
            $[0,6]$ & -1 & $[0.0001,1]$ & NC/GH & IMEX & Dirichlet & 0.5 \\
		\hline
	\end{tabular}
	\label{tab:hydrostatic}
\end{table}

\begin{table}[h]
    \centering
    \caption{Computational costs of different methods for solving the one-dimensional hydrostatic problem (NC/GH refers to the Newton-Cotes/Gauss-Hermite quadrature).}
    \begin{tabular}{llll}
        \toprule
        Knudsen number & Numerical method & Number of collocation points & Time (s) \\
        \midrule
        0.0001 & finite difference (NC) & 600 & 85.19 \\
         & Hermite (NC, $N_m=0$) & 100 & 20.17 \\
         & Hermite (GH, $N_m=0$) & 16 & 2.48 \\
        \midrule
        0.01 & finite difference (NC) & 600 & 85.14 \\
         & Hermite (NC, $N_m=5$) & 100 & 37.96 \\
         & Hermite (GH, $N_m=5$) & 22 & 4.20 \\
        \midrule
        0.1 & finite difference (NC) & 600 & 81.02 \\
         & Hermite (NC, $N_m=6$) & 100 & 41.69 \\
         & Hermite (GH, $N_m=6$) & 36 & 4.47 \\
        \midrule
        1 & finite difference (NC) & 600 & 81.12 \\
         & Hermite (NC, $N_m=6$) & 100 & 53.49 \\
         & Hermite (GH, $N_m=6$) & 36 & 4.59 \\
        \bottomrule
    \end{tabular}
    \label{tab:hydrostatic cost}
\end{table}

\subsection{Shock tube under external force}

We then test the capability of the solution algorithm for solving discontinuous solutions.
The Sod shock tube is employed as the numerical experiment.
The particle distribution function is initialized as Maxwellian in the whole flow domain, which corresponds to the following macroscopic variables,
\begin{equation}
    \left(\begin{array}{c}
    \rho \\
    U \\
    p \\
    \end{array}\right)_{{t=0,x<0.5}}=\left(\begin{array}{c}
    1 \\
    0 \\
    1 \\
    \end{array}\right), \quad 
    \left(\begin{array}{c}
    \rho \\
    U \\
    p \\
    \end{array}\right)_{{t=0,x\ge 0.5}}=\left(\begin{array}{c}
    0.125 \\
    0 \\
    0.1 \\
    \end{array}\right).
\end{equation}
The detailed computational setup is described in Table \ref{tab:sod}.

\begin{figure}[htb!]
    \centering
    \subfigure[Density]{
        \includegraphics[width=0.47\textwidth]{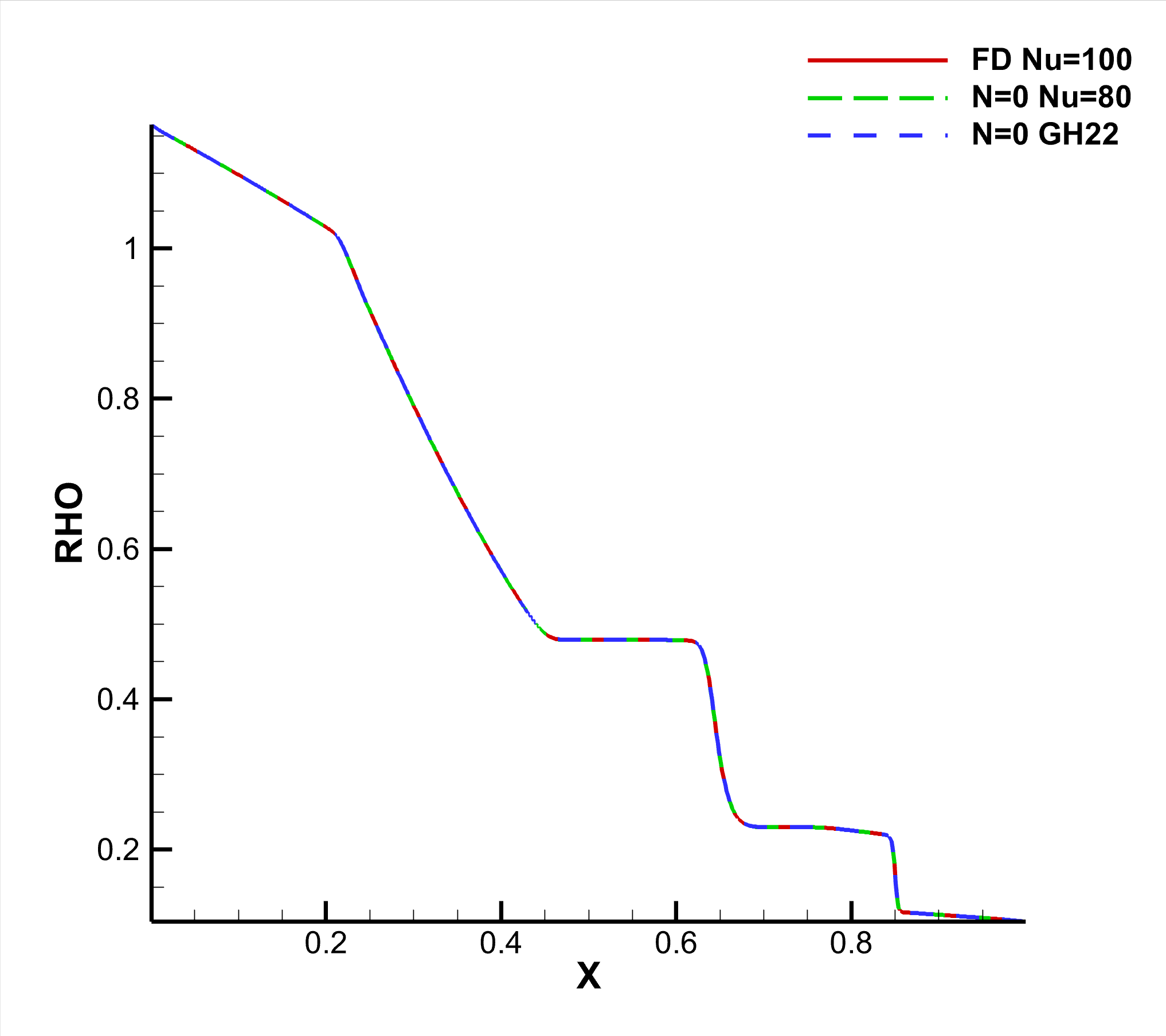}
    }
    \subfigure[Velocity]{
        \includegraphics[width=0.47\textwidth]{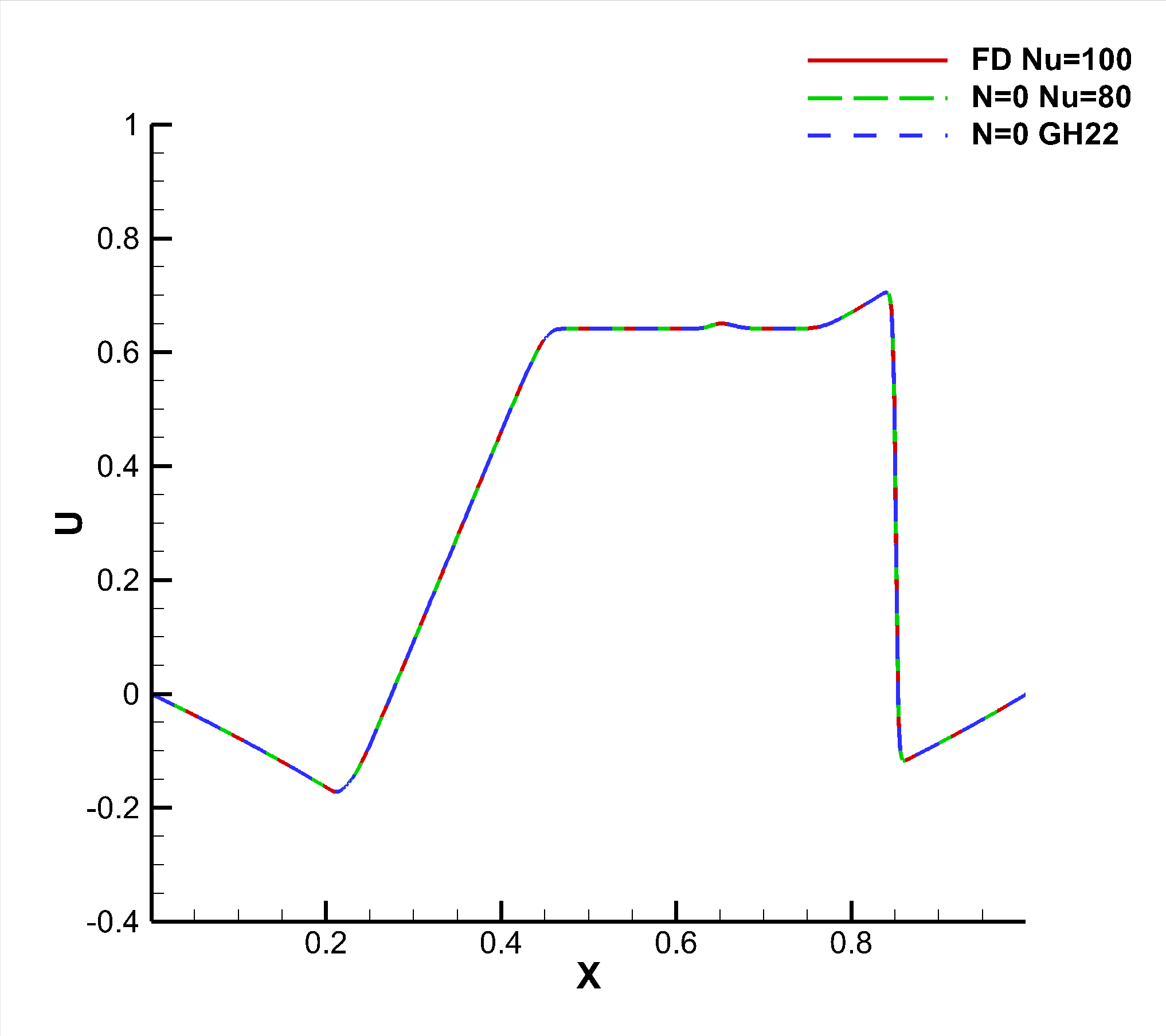}
    }
    \subfigure[Temperature]{
        \includegraphics[width=0.47\textwidth]{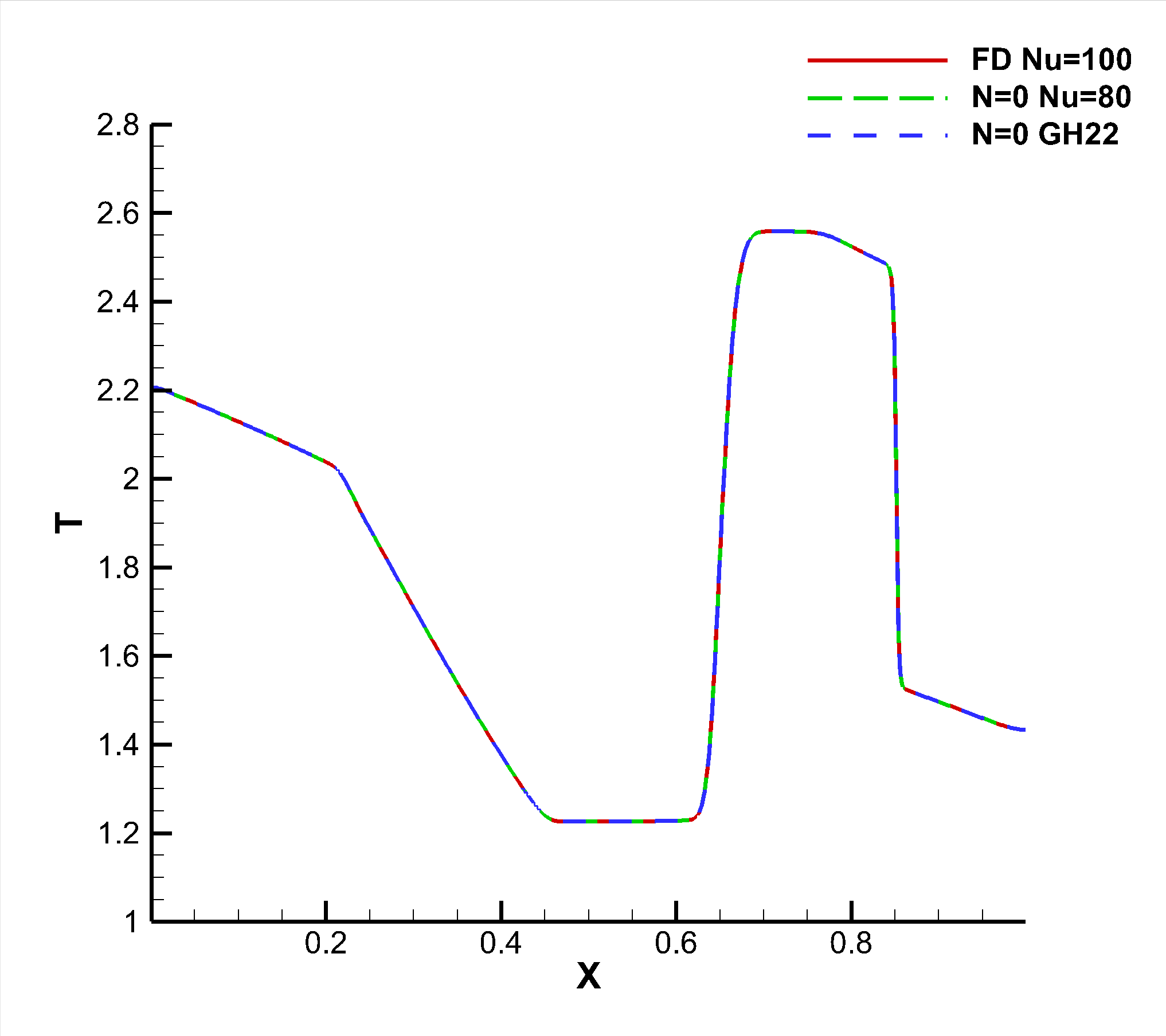}
    }
    \subfigure[Pressure]{
        \includegraphics[width=0.47\textwidth]{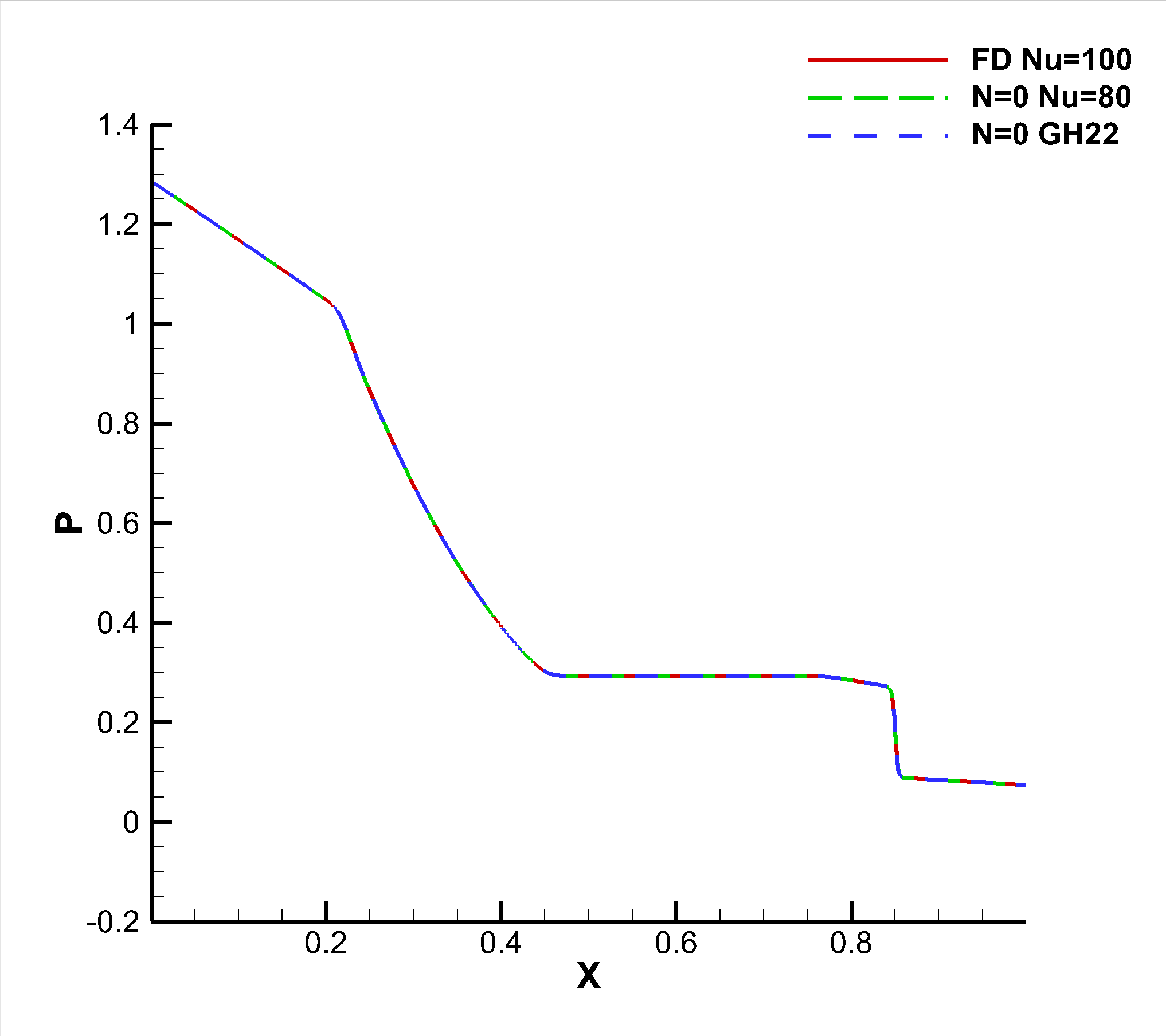}
    }
    \caption{Distributions of flow variables in the Sod shock tube under external force at Kn=0.0001. Here, $N$ represents the order of the Hermite expansion, GH denotes the Gauss-Hermite quadrature, and the rest that are not specifically labeled denote the Newton-Cotes quadrature.}
    \label{fig:sod1}
\end{figure}

\begin{figure}[htb!]
    \centering
    \subfigure[Density]{
        \includegraphics[width=0.47\textwidth]{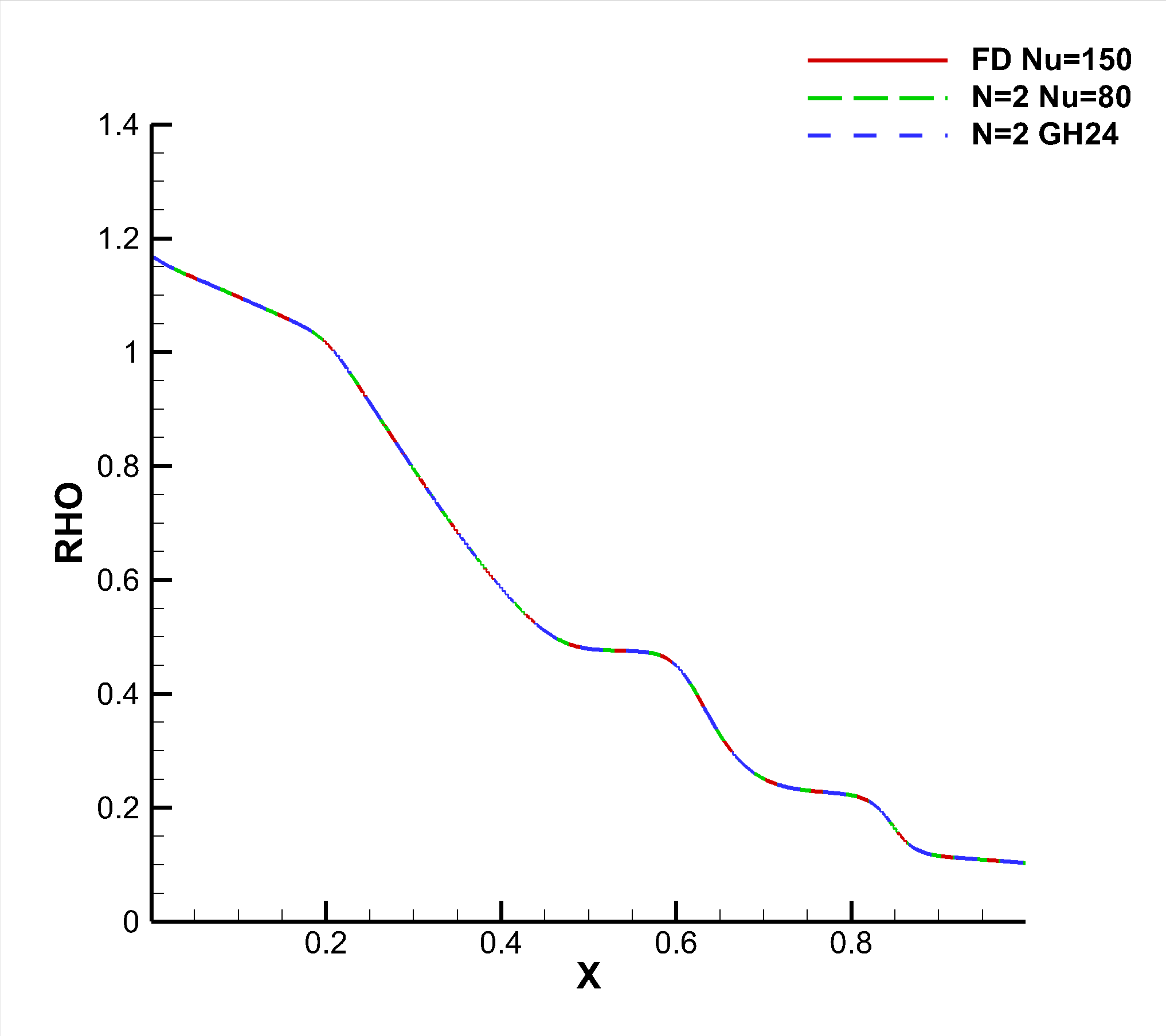}
    }
    \subfigure[Velocity]{
        \includegraphics[width=0.47\textwidth]{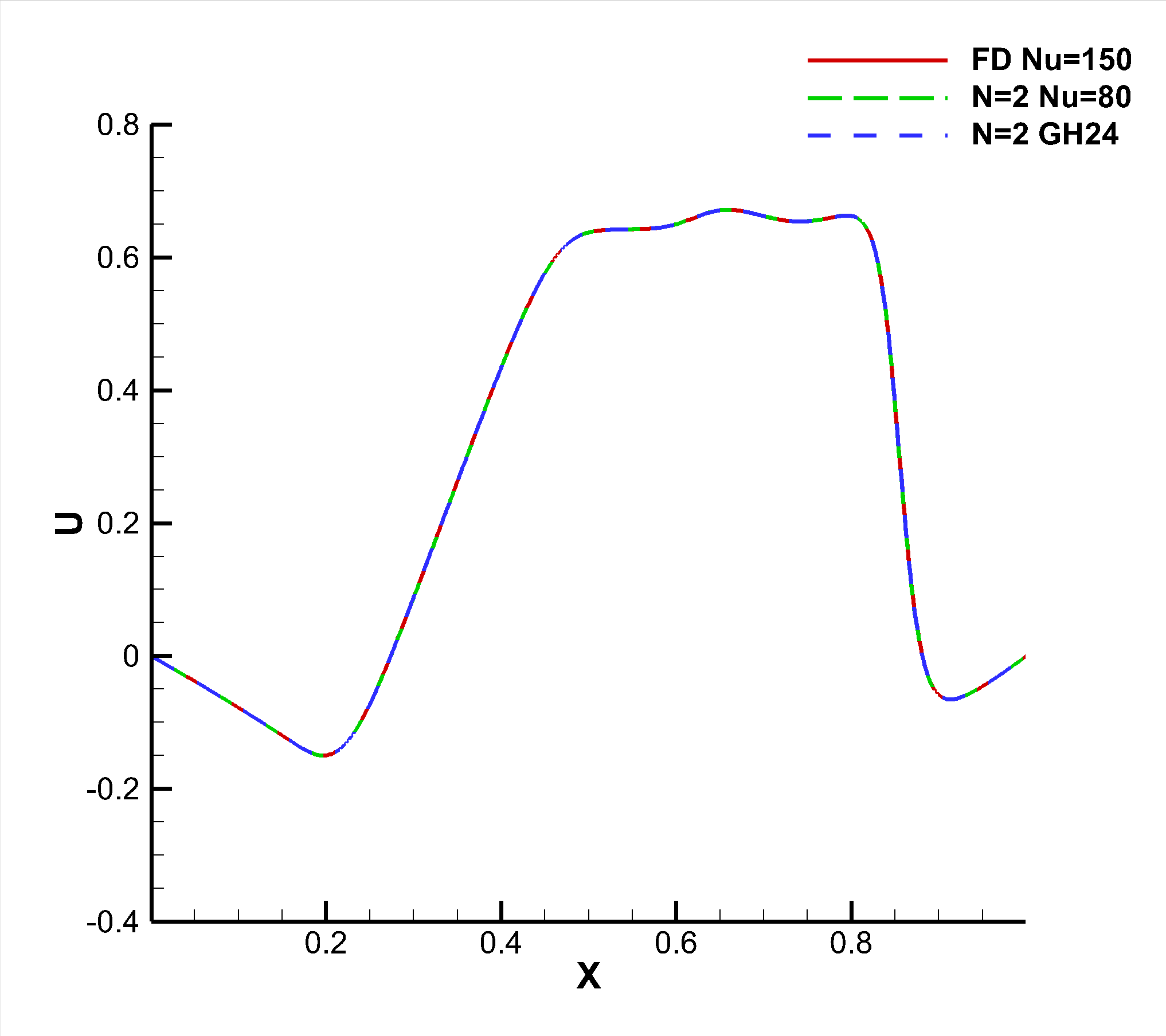}
    }
    \subfigure[Temperature]{
        \includegraphics[width=0.47\textwidth]{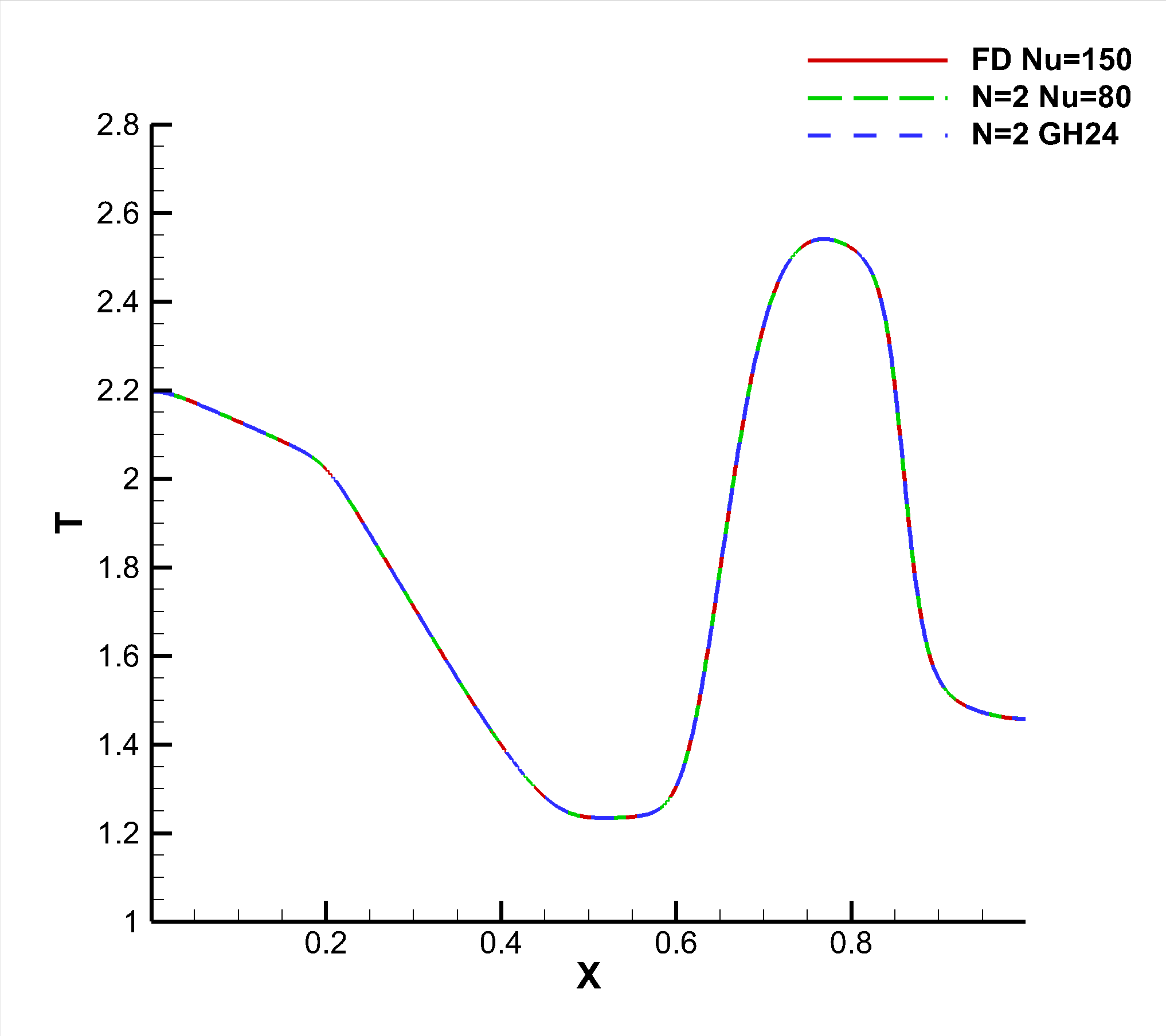}
    }
    \subfigure[Pressure]{
        \includegraphics[width=0.47\textwidth]{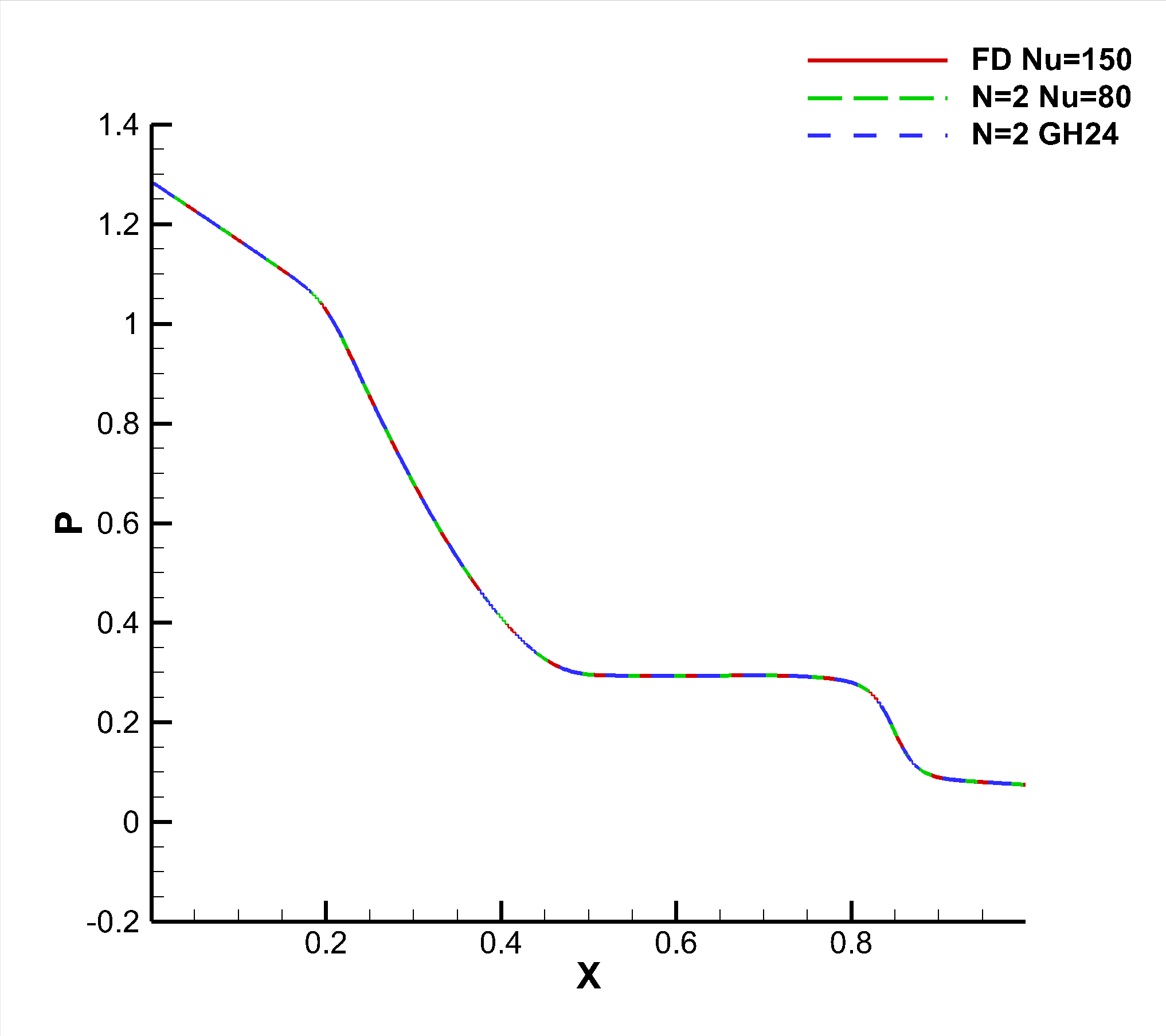}
    }
    \caption{Distributions of flow variables in the Sod shock tube under external force at Kn=0.001. Here, $N$ represents the order of the Hermite expansion, GH denotes the Gauss-Hermite quadrature, and the rest that are not specifically labeled denote the Newton-Cotes quadrature.}
    \label{fig:sod2}
\end{figure}

\begin{figure}[htb!]
    \centering
    \subfigure[Density]{
        \includegraphics[width=0.47\textwidth]{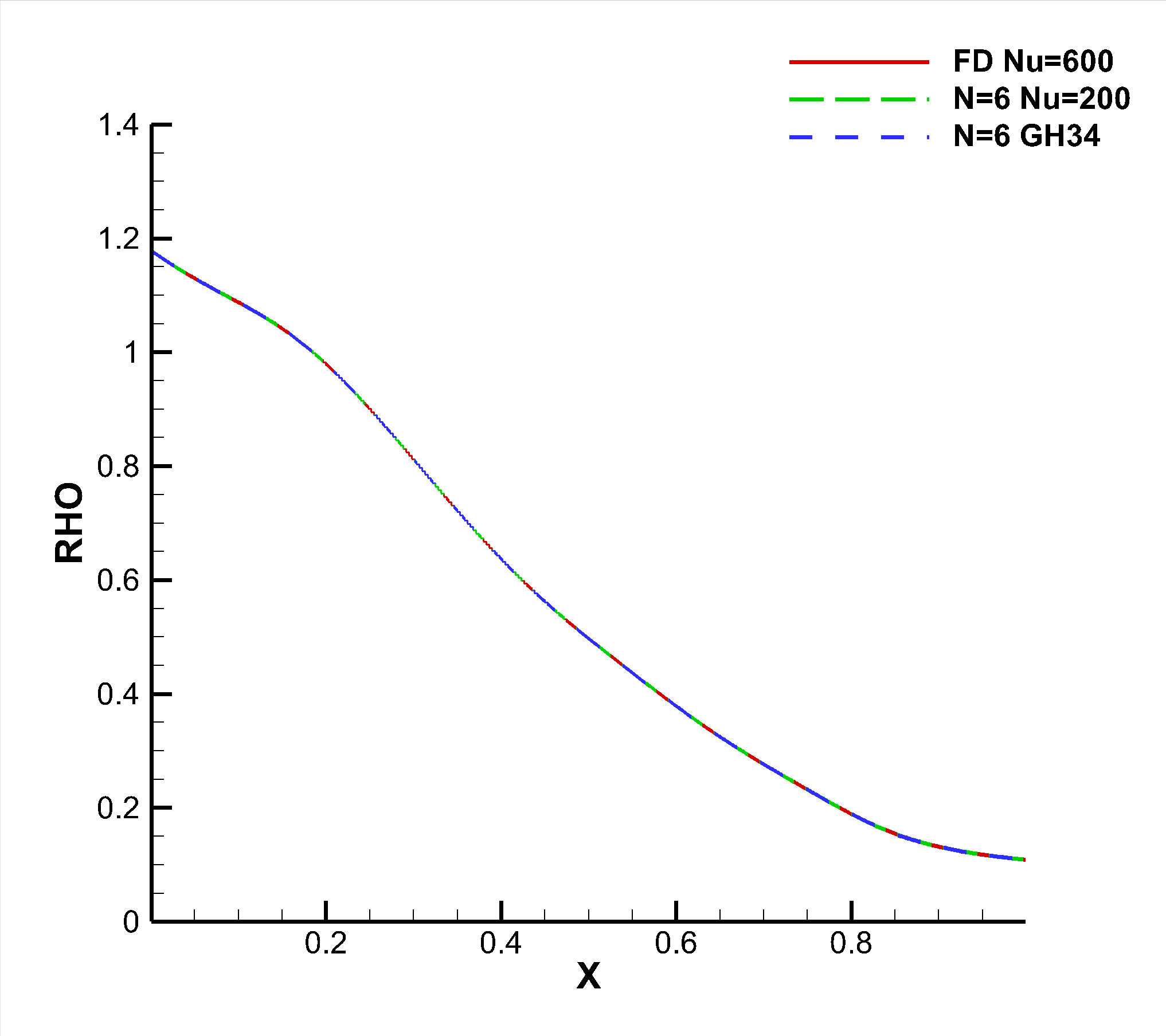}
    }
    \subfigure[Velocity]{
        \includegraphics[width=0.47\textwidth]{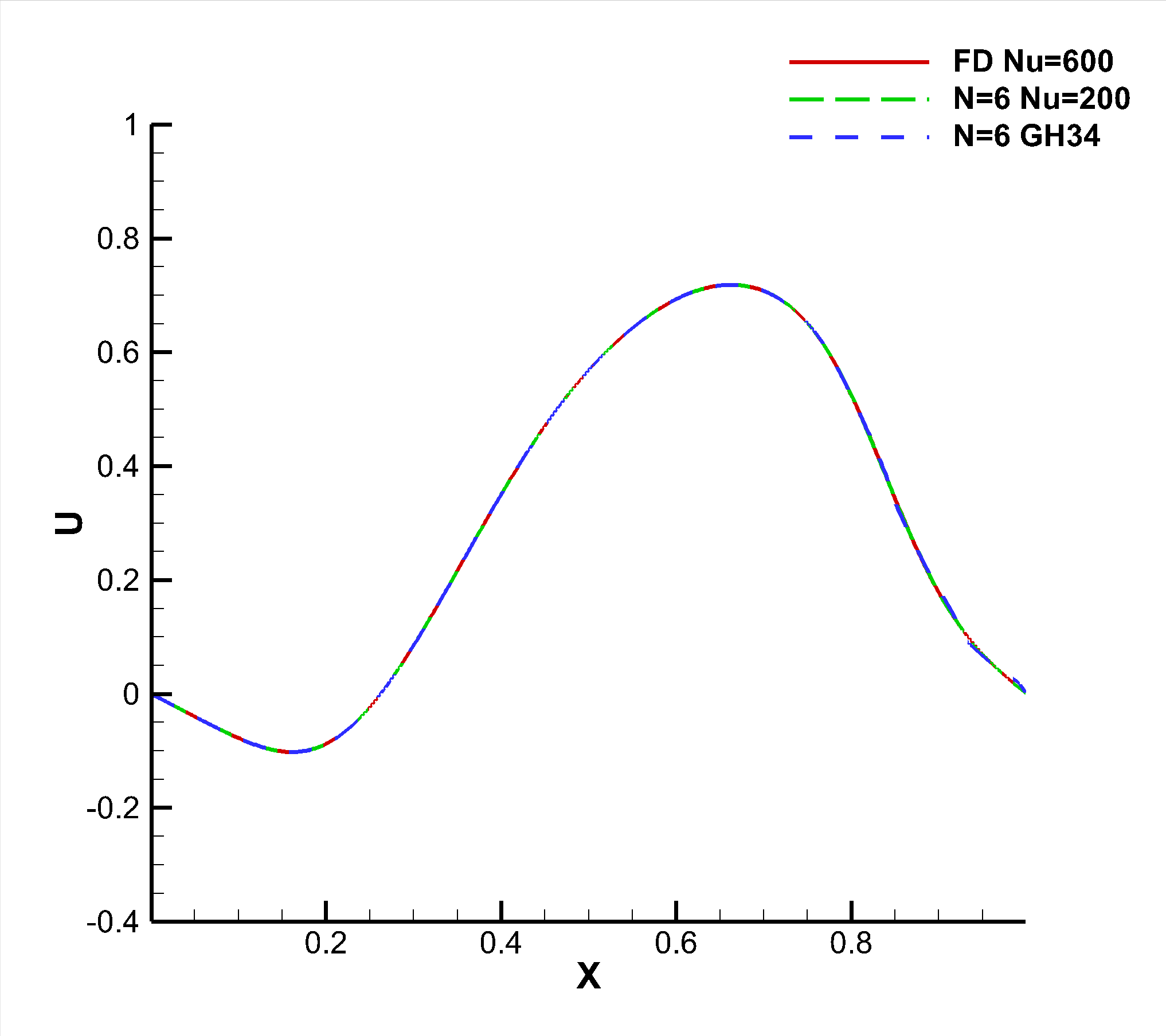}
    }
    \subfigure[Temperature]{
        \includegraphics[width=0.47\textwidth]{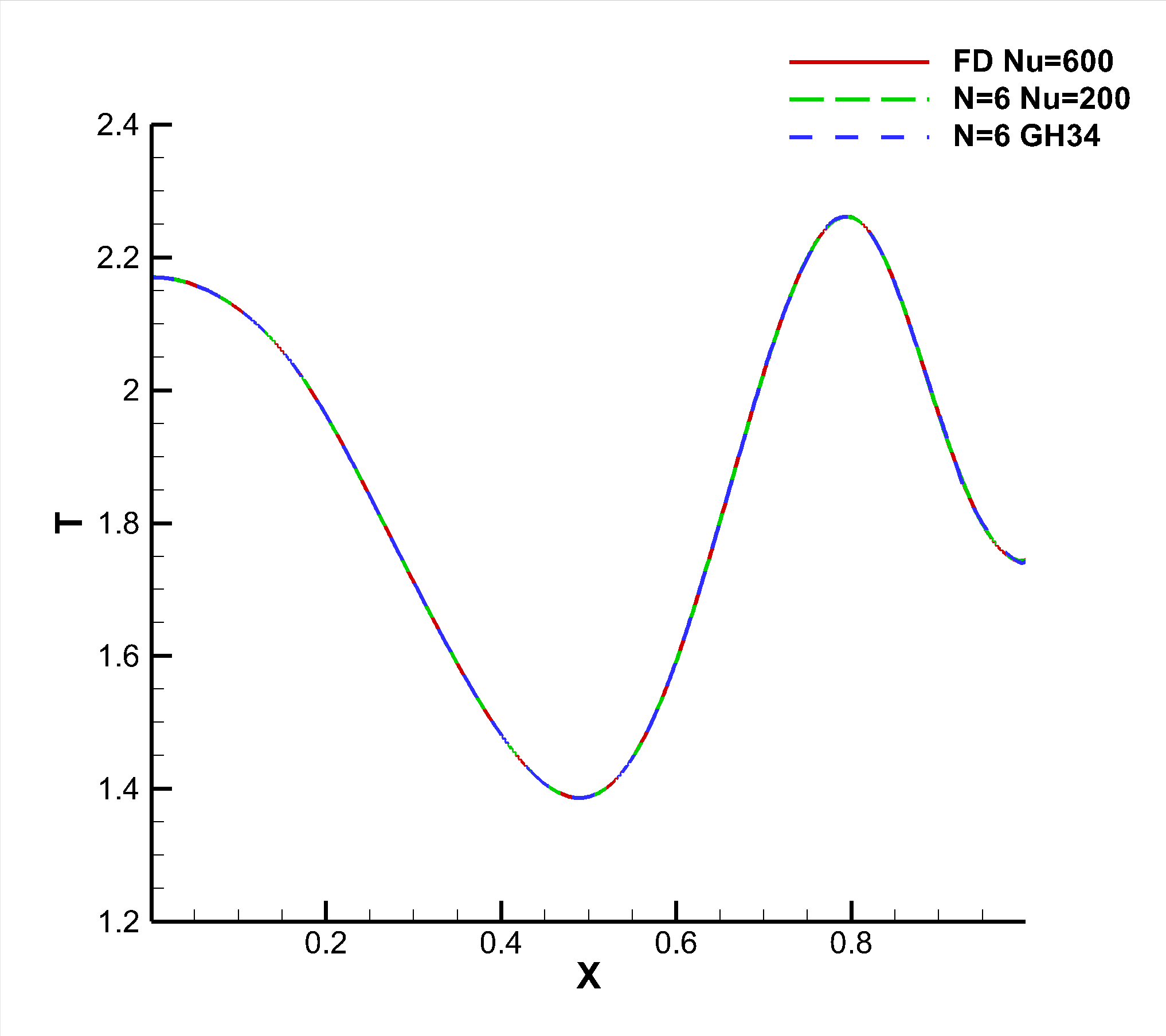}
    }
    \subfigure[Pressure]{
        \includegraphics[width=0.47\textwidth]{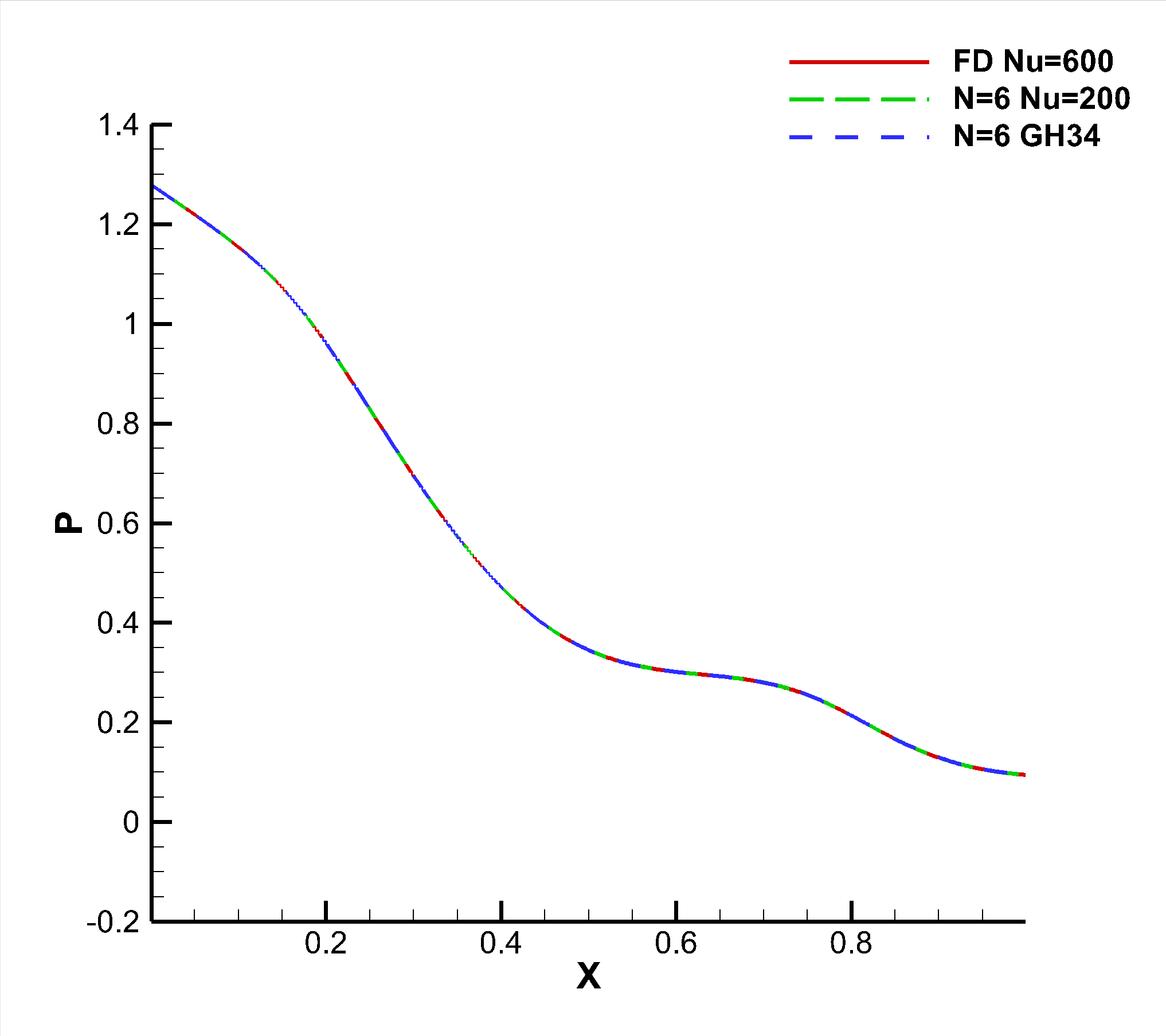}
    }
    \caption{Distributions of flow variables in the Sod shock tube under external force at Kn=0.01. Here, $N$ represents the order of the Hermite expansion, GH denotes the Gauss-Hermite quadrature, and the rest that are not specifically labeled denote the Newton-Cotes quadrature.}
    \label{fig:sod3}
\end{figure}

Figure \ref{fig:sod1}, \ref{fig:sod2}, and \ref{fig:sod3} present the spatial distribution of macroscopic variables at the output moment under different Knudsen numbers.
Under the external force field, particles in the region where the waves have not arrived are driven to the left.
This numerical experiment demonstrates the well-balanced and asymptotic-preserving properties of the current scheme.
The evolution of discontinuities is well captured under the force field in different flow regimes.
With the increasing Knudsen number, a continuous spectrum of flow physics is recovered from the Boltzmann solution to the Euler solution of the one-dimensional Riemann problem.

Table \ref{tab:sod cost} lists the computational costs of different methods (i.e., the Hermite approximation in Section \ref{sec:hermite} and second-order central difference) for the numerical solution in the same framework.
The results show that the Hermite spectral method, along with the Gauss-Hermite quadrature, significantly improves the computational efficiency of the numerical method at different reference Knudsen numbers.
The amount of computational degrees of freedom and overall time have been reduced by 80\%.

\begin{table}[htbp]
	\caption{Computational setup of Sod shock tube problem (NC/GH refers to the Newton-Cotes/Gauss-Hermite quadrature).}
	\centering
	\begin{tabular}{lllllllll} 
	    \hline
	    Gas & $t$ & $x$ & $N_x$ & $u$ & $N_u$ & Order \\
	    Argon & $(0,0.2]$ & $[0,1]$ & 600 & $[-5,5]$ & $[22,600]$ & 2 \\
		%\hline
		\rule{0pt}{0ex}\\
		$N_m$ & $a$ & Kn & Quadrature & Integrator & Boundary & CFL \\
            $[0,6]$ & -1 & $[0.0001,0.01]$ & NC/GH & IMEX & Dirichlet & 0.5 \\
		\hline
	\end{tabular}
	\label{tab:sod}
\end{table}

\begin{table}[h]
    \centering
    \caption{Computational costs of different methods for solving the Sod shock tube problem (NC/GH refers to the Newton-Cotes/Gauss-Hermite quadrature).}
    \begin{tabular}{llcl}
        \toprule
        Knudsen number & Numerical method & Number of collocation points & Time (s) \\
        \midrule
        0.0001 & finite difference (NC) & 100& 59.06\\
         & Hermite (NC, $N_m=0$) & 80& 41.19\\
         & Hermite (GH, $N_m=0$) & 22& 9.29\\
        \midrule
        0.001& finite difference (NC) & 150& 106.25\\
         & Hermite (NC, $N_m=2$)& 80& 68.82\\
         & Hermite (GH, $N_m=2$)& 24& 15.80\\
        \midrule
        0.01& finite difference (NC) & 600 & 227.54\\
         & Hermite (NC, $N_m=6$) & 200& 220.77\\
         & Hermite (GH, $N_m=6$) & 34& 44.82\\
         \bottomrule
    \end{tabular}
    \label{tab:sod cost}
\end{table}

\subsection{Fourier flow}

Subsequently, we examine the capability of the solution algorithm for solving energy transport.
The Fourier flow between two infinitely large parallel plates is employed as the numerical experiment, where the temperatures of the upper and lower plates are set as 1.1 and 1.0, respectively.
The initial flow field is set up with a fixed temperature, and the particle distribution function is initialized as Maxwellian in the whole computational domain.
The effect of external force fields of different magnitudes on the heat transfer at Kn=0.1 is investigated.
The detailed computational setup is provided in Table \ref{tab:fourier}.

\begin{figure}[htb!]
    \centering
    \subfigure[Density]{
        \includegraphics[width=0.47\textwidth]{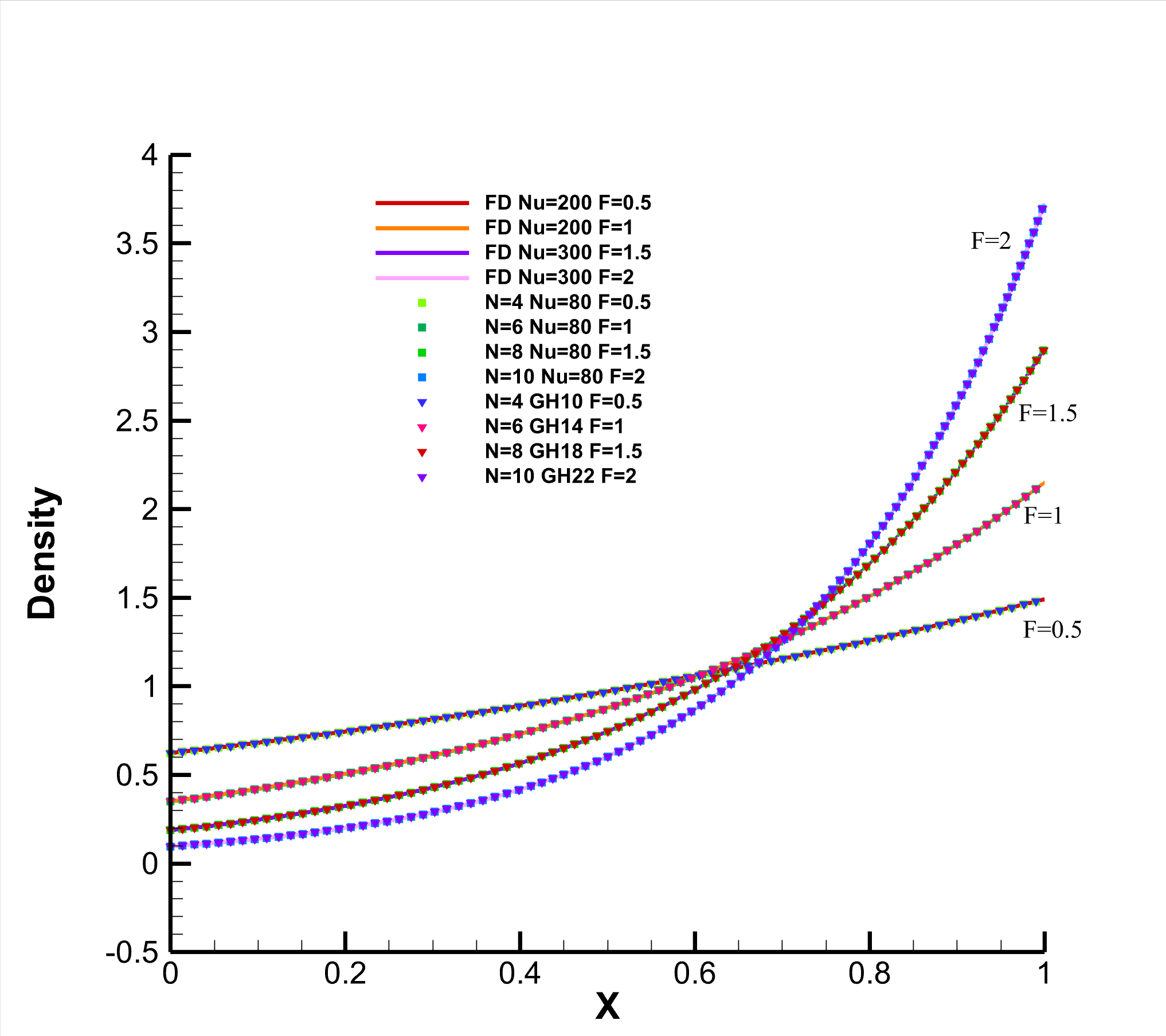}
    }
    \subfigure[Temperature]{
        \includegraphics[width=0.47\textwidth]{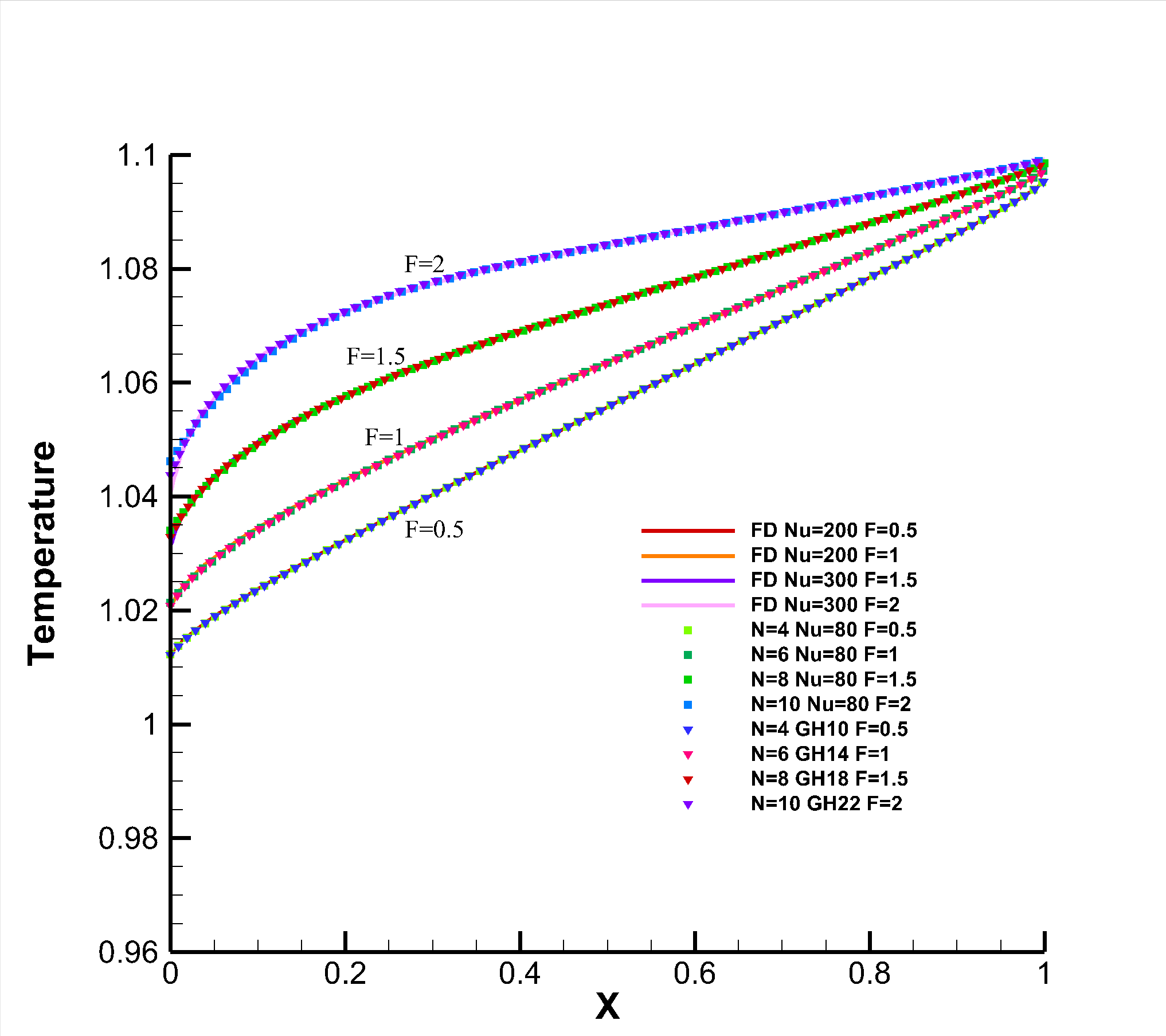}
    }
    \subfigure[Heat flux]{
        \includegraphics[width=0.47\textwidth]{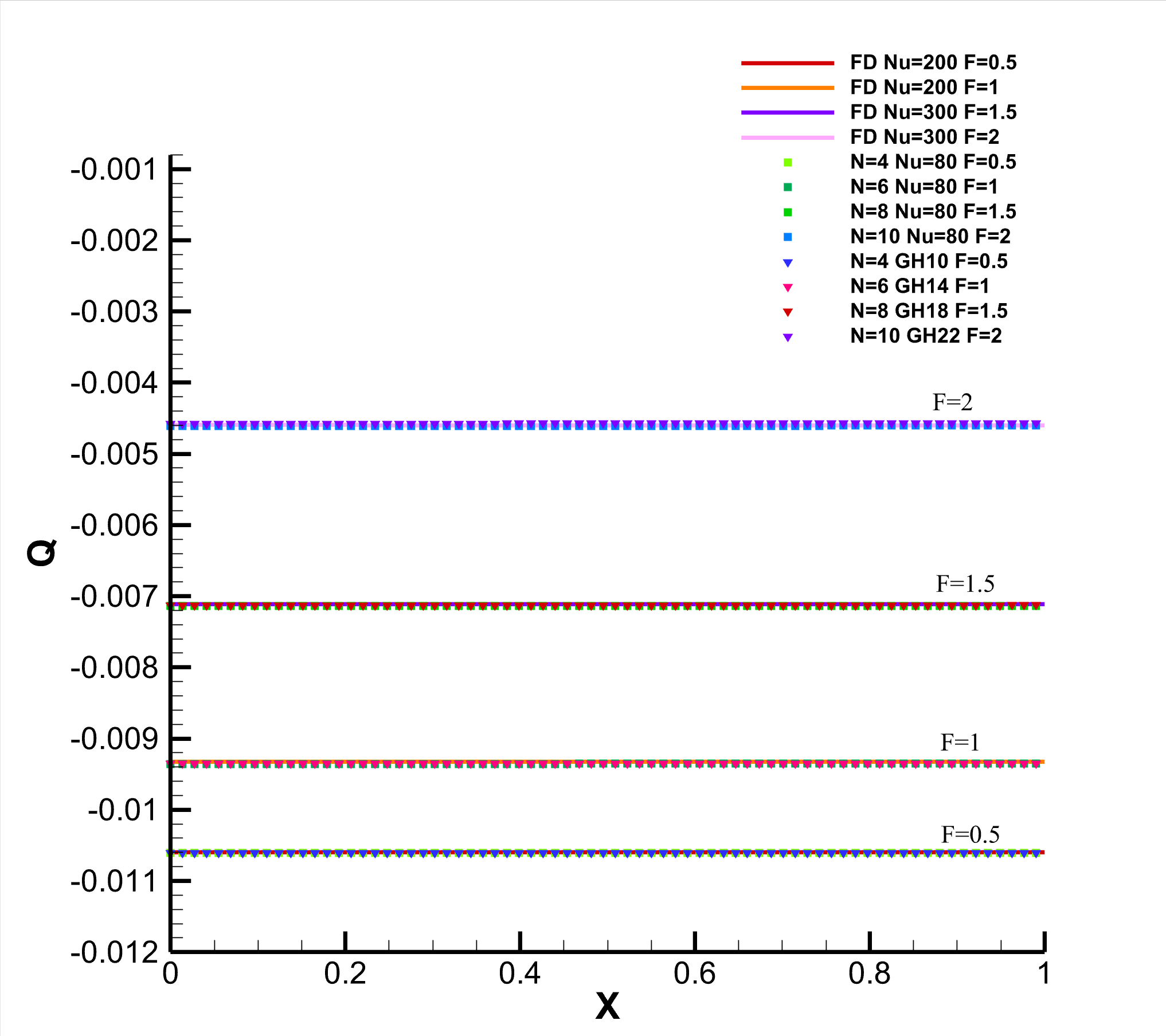}
    }
    \caption{Distributions of density, temperature, and heat flux in one-dimensional Fourier flow problem under different external forces at Kn=0.1. Here, $N$ represents the order of the Hermite expansion, GH denotes the Gauss-Hermite quadrature, and the rest that are not specifically labeled denote the Newton-Cotes quadrature.}
    \label{fig:fourier}
\end{figure}

Figure \ref{fig:fourier} presents the distributions of heat flux, density, and temperature under different external forces.
The modulation of density by external forces is precisely captured.
The non-equilibrium heat flux induced by the external force and its resulting changes in temperature profiles are quantitatively identified \cite{tij1994perturbation,xiao2018investigation}.

Table \ref{tab:fourier cost} lists the computational costs of different methods (i.e., the Hermite approximation in Section \ref{sec:hermite} and the second-order central difference).
The results show that the Hermite spectral method, along with the Gauss-Hermite quadrature, significantly improves the computational efficiency of the numerical method at different reference Knudsen numbers.
The amount of computational degrees of freedom and overall time have been reduced by 95\%.

\begin{table}[htbp]
	\caption{Computational setup of one-dimensional Fourier flow problem (NC/GH refers to the Newton-Cotes/Gauss-Hermite quadrature).}
	\centering
	\begin{tabular}{lllllllll} 
	    \hline
	    Gas & Force & $x$ & $N_x$ & $u$ & $N_u$ & Order \\
	    Argon & $[0.5,2]$ & $[0,1]$ & 800 & $[-5,5]$ & $[10,300]$ & 2 \\
		%\hline
		\rule{0pt}{0ex}\\
		$N_m$ & $a$ & Kn & Quadrature & Integrator & Boundary & CFL \\
            $[4,10]$ & $[0.5,2]$ & $[0.0001,0.1]$ & NC/GH & IMEX & Dirichlet & 0.5 \\
		\hline
	\end{tabular}
	\label{tab:fourier}
\end{table}

\begin{table}[h]
    \centering
    \caption{Computational costs of different methods for solving the one-dimensional Fourier flow problem (NC/GH refers to the Newton-Cotes/Gauss-Hermite quadrature).}
    \begin{tabular}{llcl}
        \toprule
        Force& Numerical method & Number of collocation points & Time (s) \\
        \midrule
        0.5& finite difference (NC) & 200& 5580.48\\
         & Hermite (NC, $N_m=4$)& 80& 1896.40\\
         & Hermite (GH, $N_m=4$)& 10& 263.15\\
        \midrule
        1.0& finite difference (NC) & 200& 5999.65\\
         & Hermite (NC, $N_m=6$)& 80& 2694.69\\
         & Hermite (GH, $N_m=6$)& 14& 423.10\\
        \midrule
        1.5& finite difference (NC) & 300& 14209.96\\
         & Hermite (NC, $N_m=8$)& 80& 2832.79\\
         & Hermite (GH, $N_m=8$)& 18& 865.73\\
         \midrule
        2.0& finite difference (NC) & 300& 12398.12\\
         & Hermite (NC, $N_m=10$)& 80& 4785.77\\
         & Hermite (GH, $N_m=10$)& 22& 1585.94\\
         \bottomrule
    \end{tabular}
    \label{tab:fourier cost}
\end{table}

\subsection{Poiseuille flow}

The force-driven Poiseuille flow of rarefied gas between two infinitely parallel isothermal plates is considered.
Initially stationary homogeneous gas is subjected to a constant body force acting in the positive $x$-direction parallel to the plates.
The particle distribution function is initialized as Maxwellian.
The detailed computational setup can be found in Table \ref{tab:poiseuille}.

\begin{figure}[htb!]
    \centering
    \subfigure[Density]{
        \includegraphics[width=0.47\textwidth]{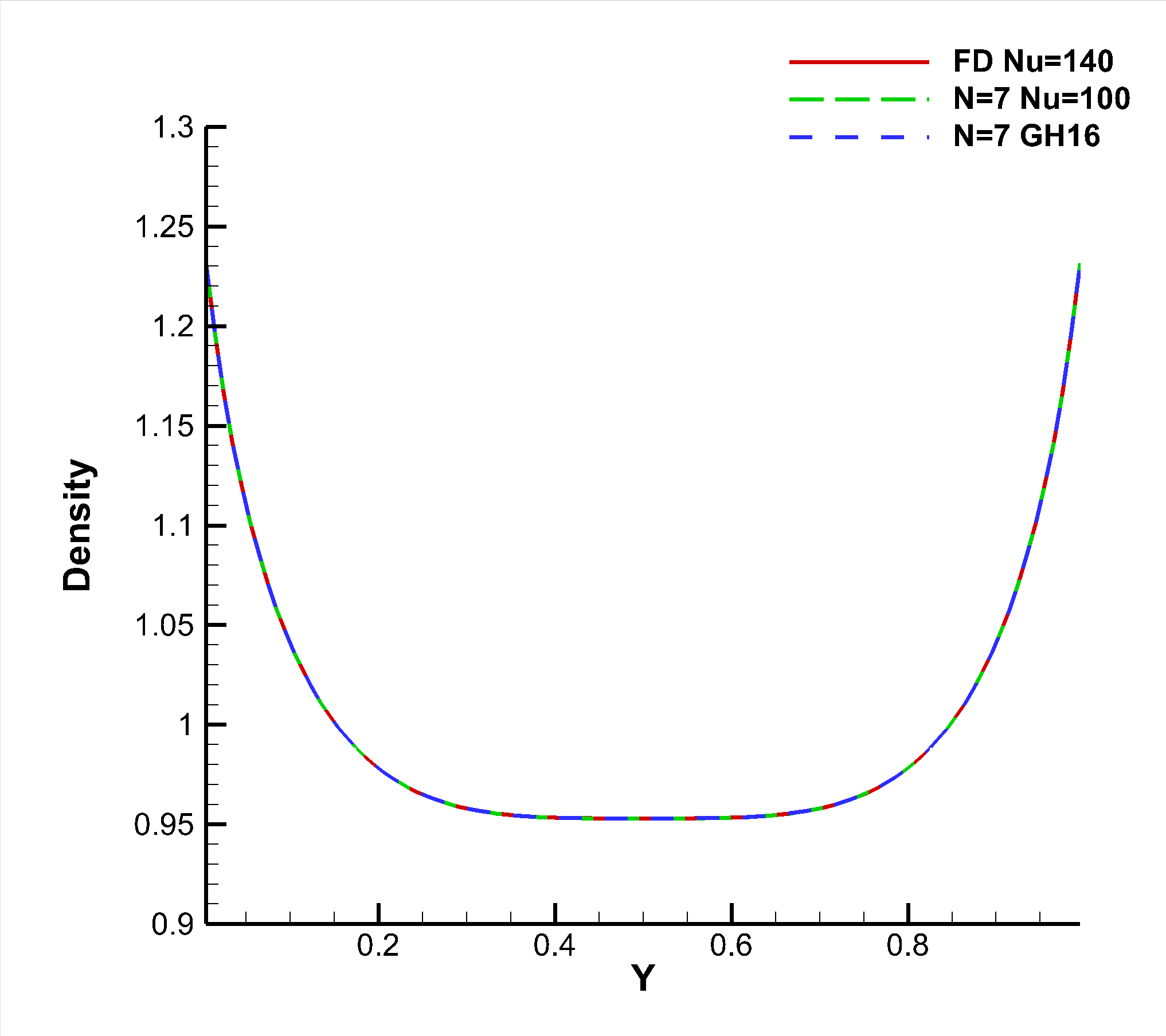}
    }
    \subfigure[Velocity]{
        \includegraphics[width=0.47\textwidth]{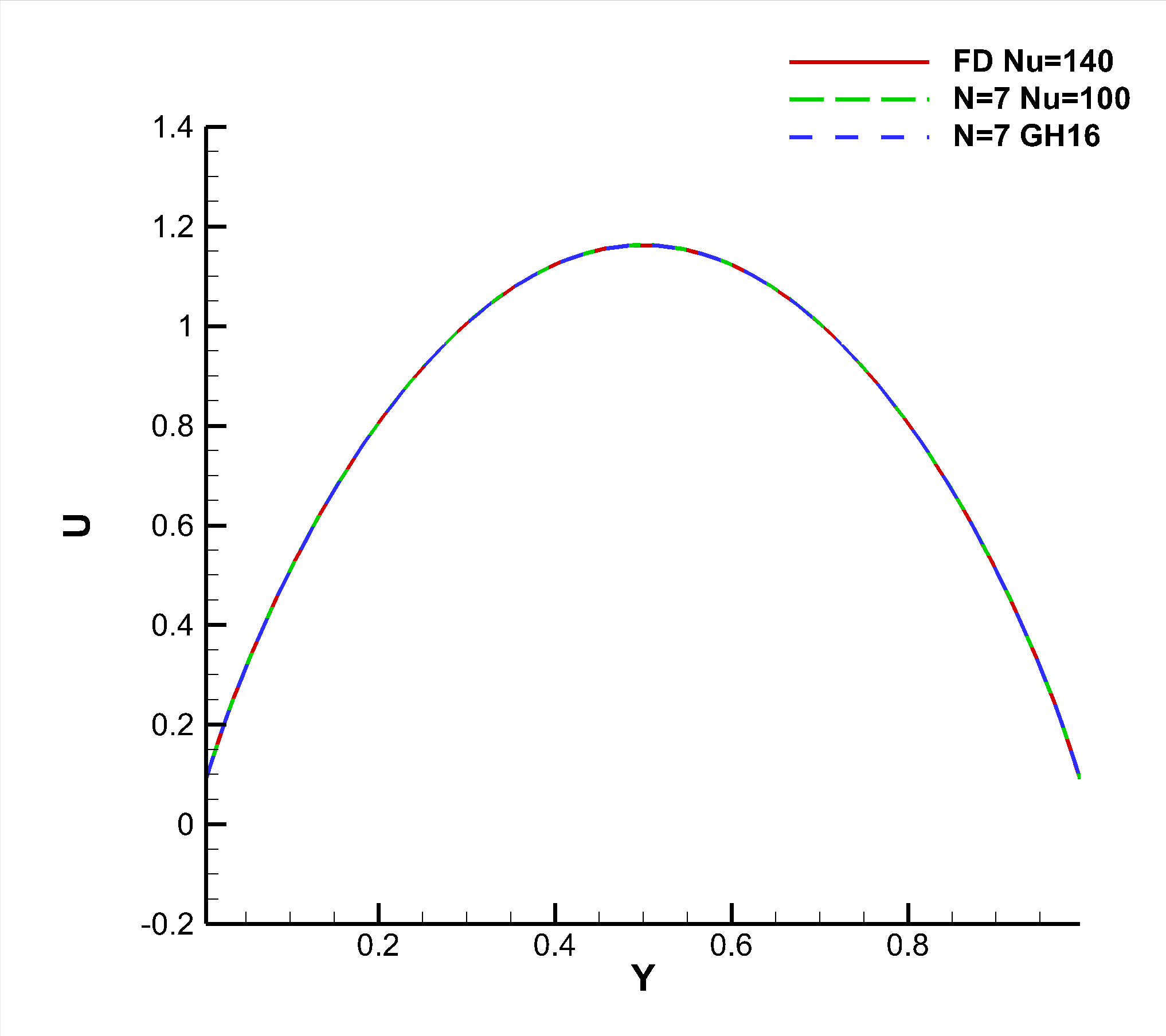}
    }
    \subfigure[Temperature]{
        \includegraphics[width=0.47\textwidth]{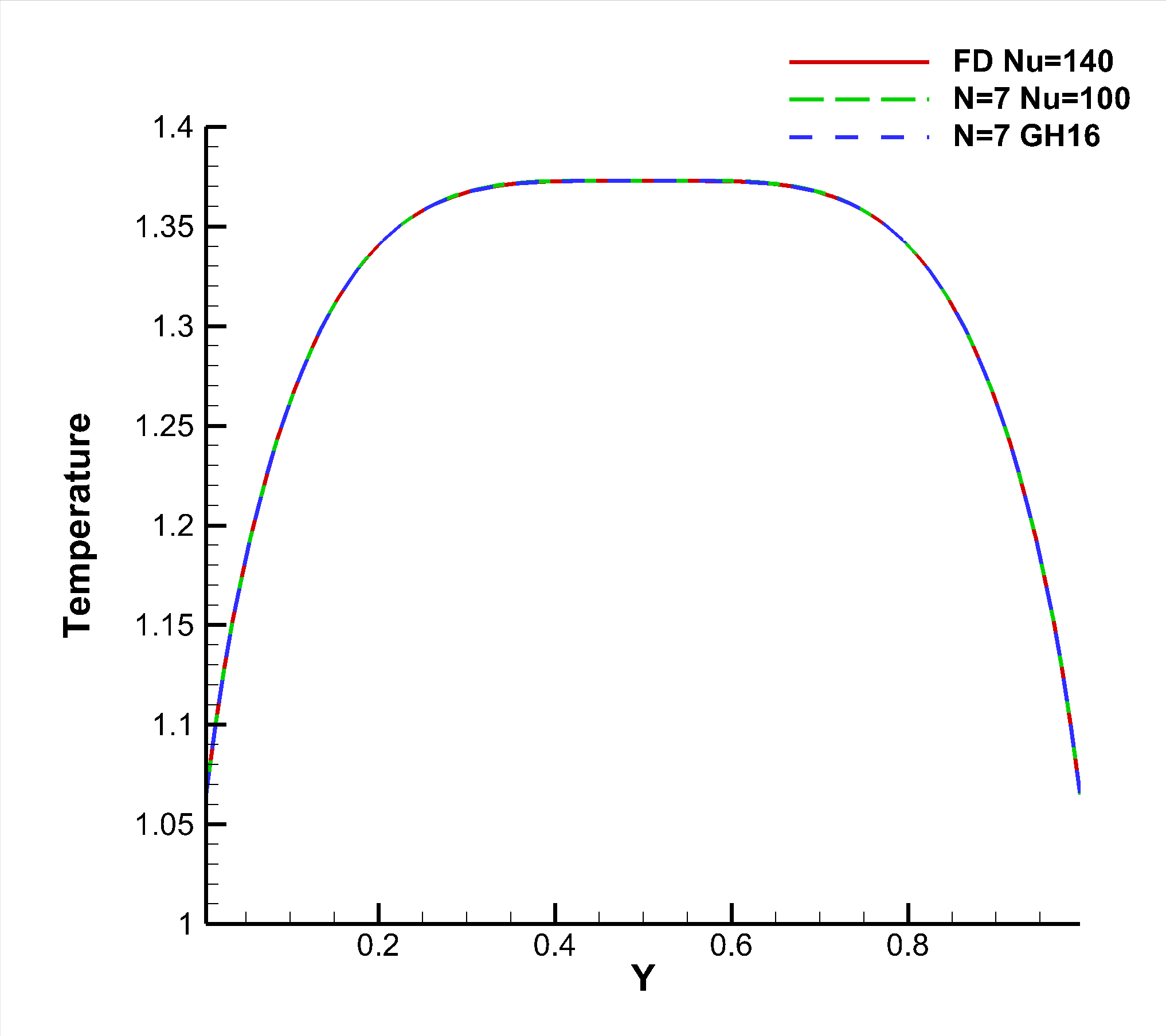}
    }
    \caption{Distributions of density, velocity, and temperature in the Poiseuille flow problem at Kn=0.02. Here, $N$ represents the order of the Hermite expansion, GH denotes the Gauss-Hermite quadrature, and the rest that are not specifically labeled denote the Newton-Cotes quadrature.}
    \label{fig:poiseuille1}
\end{figure}

\begin{figure}[htb!]
    \centering
    \subfigure[Density]{
        \includegraphics[width=0.47\textwidth]{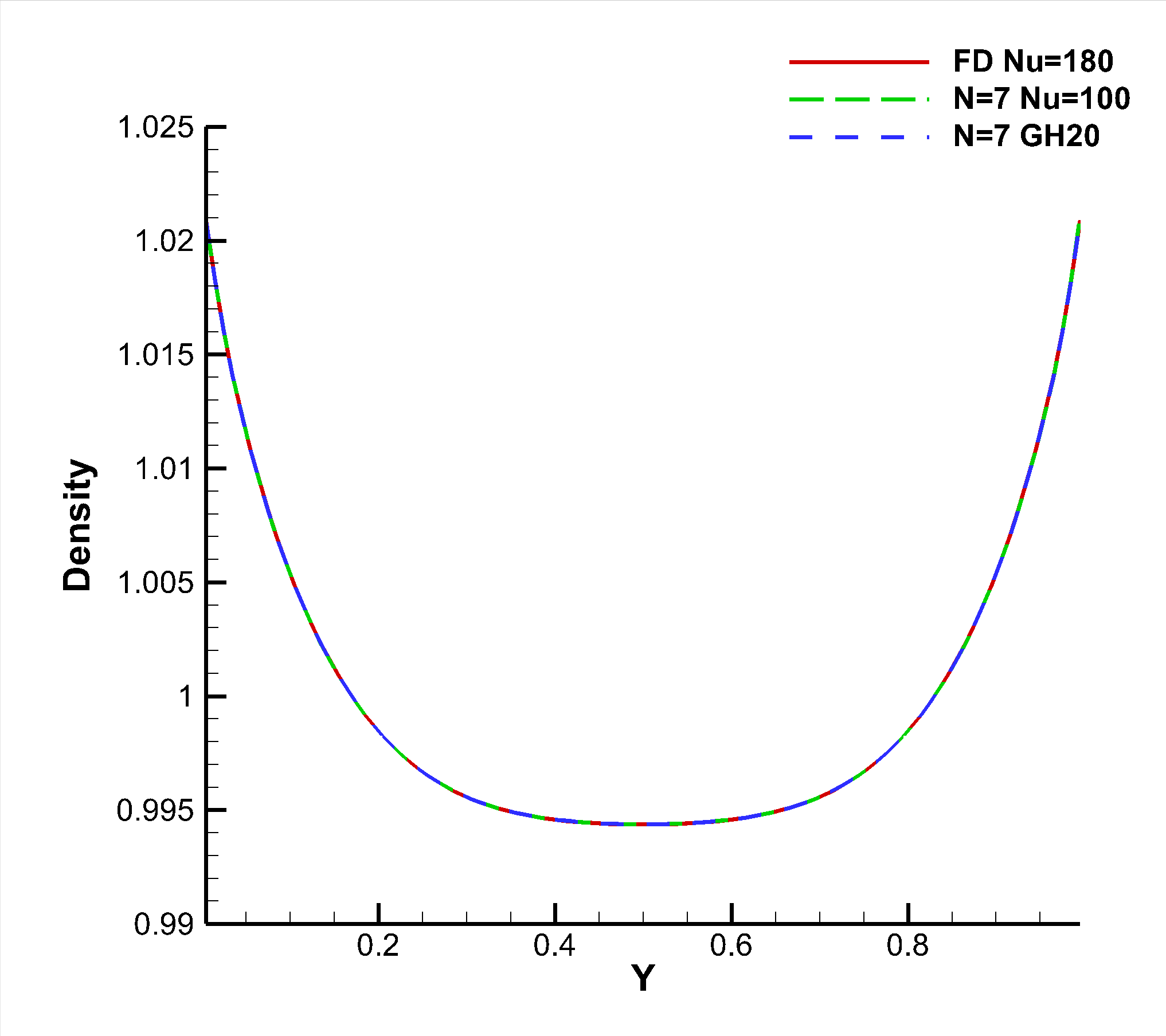}
    }
    \subfigure[Velocity]{
        \includegraphics[width=0.47\textwidth]{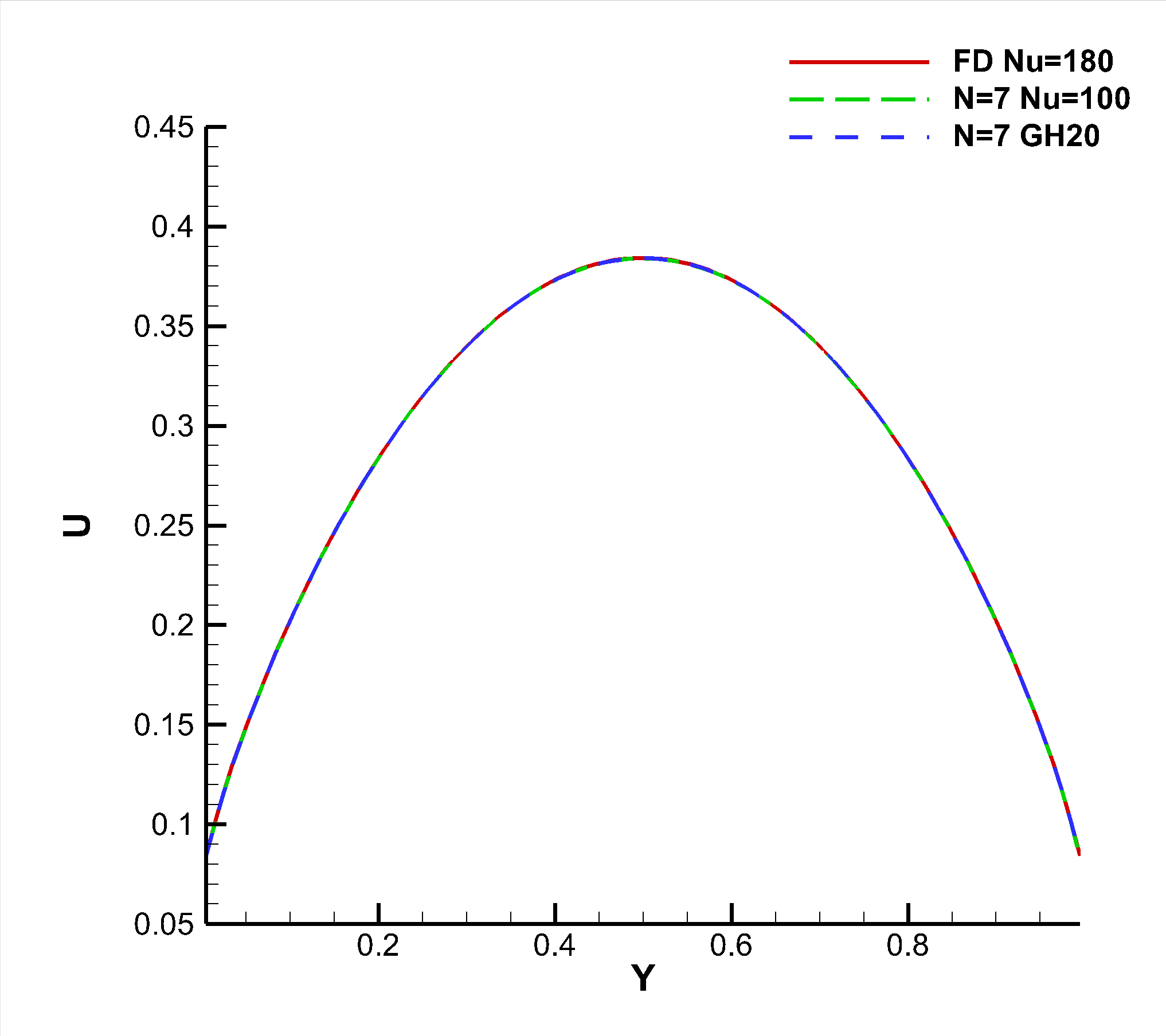}
    }
    \subfigure[Temperature]{
        \includegraphics[width=0.47\textwidth]{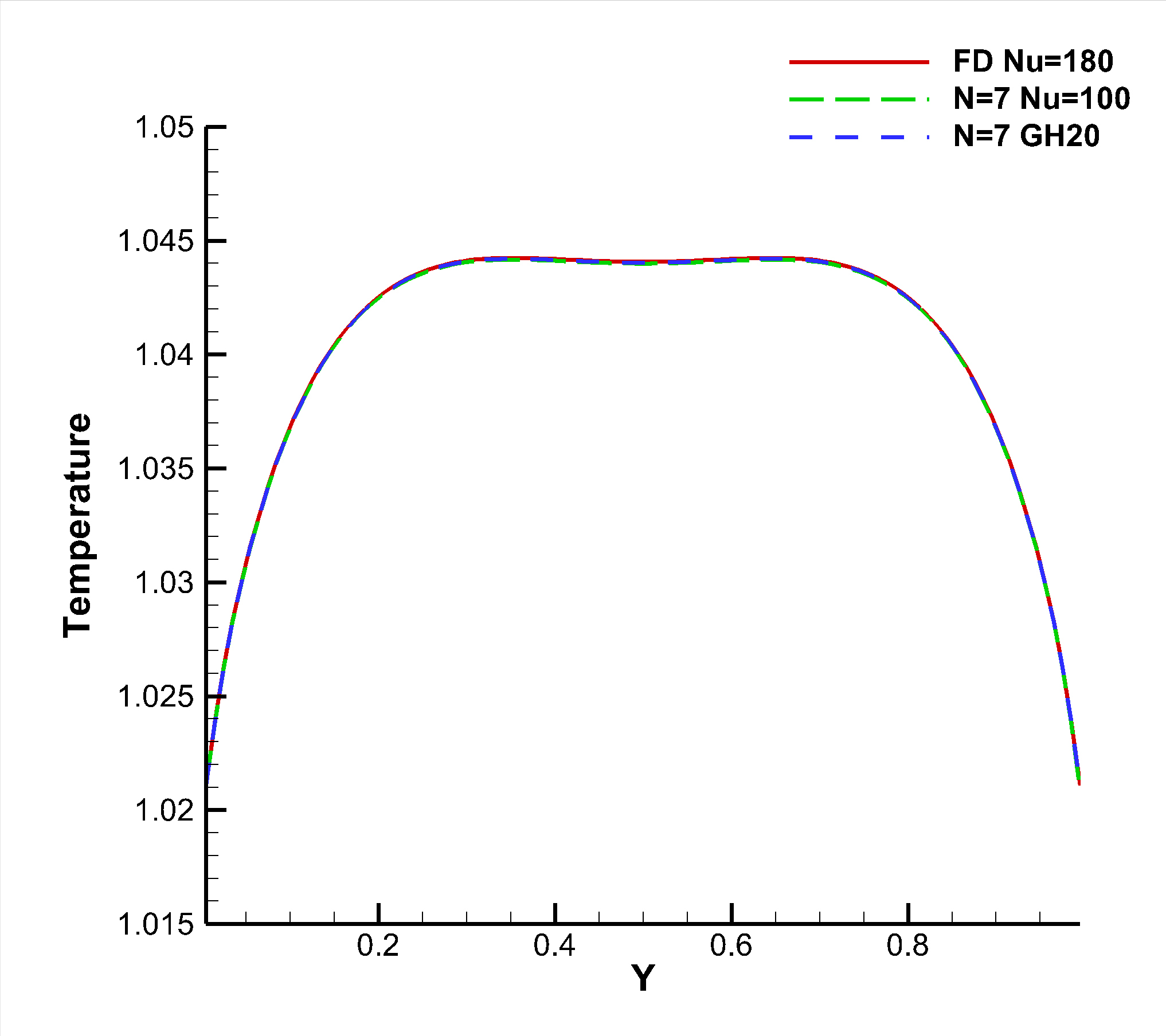}
    }
    \caption{Distributions of density, velocity, and temperature in the Poiseuille flow problem at Kn=0.1. Here, $N$ represents the order of the Hermite expansion, GH denotes the Gauss-Hermite quadrature, and the rest that are not specifically labeled denote the Newton-Cotes quadrature.}
    \label{fig:poiseuille2}
\end{figure}

Figures \ref{fig:poiseuille1} and \ref{fig:poiseuille2} present the distributions of density, velocity, and temperature at different Knudsen numbers.
Table \ref{tab:poiseuille cost} lists the computational costs of different methods (i.e., the Hermite approximation in Section \ref{sec:hermite}.
It can be observed that the Hermite spectral method, along with the Gauss-Hermite quadrature, achieves a 30-fold efficiency improvement while maintaining equivalent computational accuracy.
The non-equilibrium distributions of density, velocity, and temperature under wall effects are clearly identified.

\begin{table}[htbp]
	\caption{Computational setup of the Poiseuille flow problem (NC/GH refers to the Newton-Cotes/Gauss-Hermite quadrature).}
	\centering
	\begin{tabular}{lllllllll} 
	    \hline
	    Gas & $t$ & $x$ & $N_x$ & $u$ & $N_u$ & Order \\
	    Argon & $(0,0.2]$ & $[0,1]$ & 100 & $[-5,5]$ & $[16,180]$ & 2 \\
		%\hline
		\rule{0pt}{0ex}\\
		$N_m$ & $a$ & Kn & Quadrature & Integrator & Boundary & CFL \\
            $7$ & 0.1 & $[0.02,0.1]$ & NC/GH & IMEX & Dirichlet & 0.5 \\
		\hline
	\end{tabular}
	\label{tab:poiseuille}
\end{table}

\begin{table}[h]
    \centering
    \caption{Computational costs of different methods for solving the Poiseuille flow problem (NC/GH refers to the Newton-Cotes/Gauss-Hermite quadrature).}
    \begin{tabular}{llcl}
        \toprule
        Knudsen number& Numerical method & Number of collocation points & Time (s) \\
        \midrule
        0.02& finite difference (NC) & 140& 126103.57\\
         & Hermite (NC, $N_m=7$)& 80& 41981.64\\
         & Hermite (GH, $N_m=7$)& 16& 8465.92\\
        \midrule
        0.1& finite difference (NC) & 180& 53638.11\\
         & Hermite (NC, $N_m=7$)& 100& 17172.96\\
         & Hermite (GH, $N_m=7$)& 20& 1682.34\\
        \bottomrule
    \end{tabular}
    \label{tab:poiseuille cost}
\end{table}

\subsection{Rayleigh–Taylor instability}

The last numerical experiment is the Rayleigh–Taylor problem.
The distribution of initially static gas in a two-dimensional polar coordinates $(r,\theta)$ is set as
\begin{equation}
\rho_{0}(r)= e^{-\alpha(r+r_0)}, \quad p_{0}(r)=\frac{1.5}{\alpha} e^{-\alpha(r+r_0)}, \quad \mathbf V_{0}=0,
\end{equation}
with
\begin{equation}
\begin{aligned}
\alpha =&2.68,r_0=0.258,\quad r\le r_1, 
\\
\alpha =&5.53,r_0=-0.308,\quad r>  r_1, 
\end{aligned}
\end{equation}
where
\begin{equation}
\begin{aligned}
r_1=&0.6(1+0.02 \cos(20\theta)),\quad \text{for density},\\
r_1=&0.62324965,\quad \text{ for pressure}.  
\end{aligned}
\end{equation}
The particle distribution function is initialized as Maxwellian.
The detailed computational setup can be found in Table \ref{tab:RT}.

\begin{figure}[htb!]
    \centering
    \subfigure[t=0]{
        \includegraphics[width=0.22\textwidth]{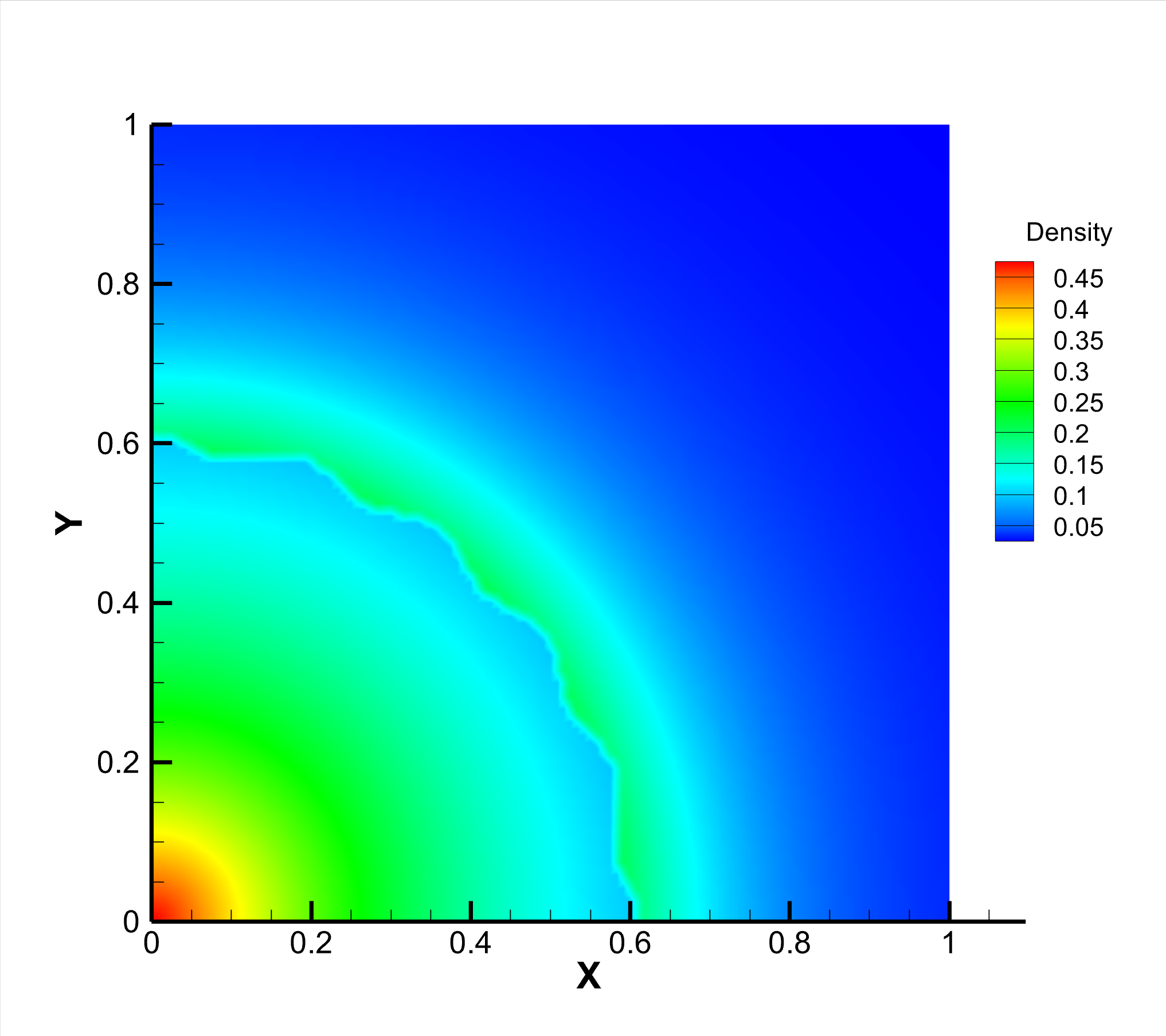}
    }
    \subfigure[t=0.8]{
        \includegraphics[width=0.22\textwidth]{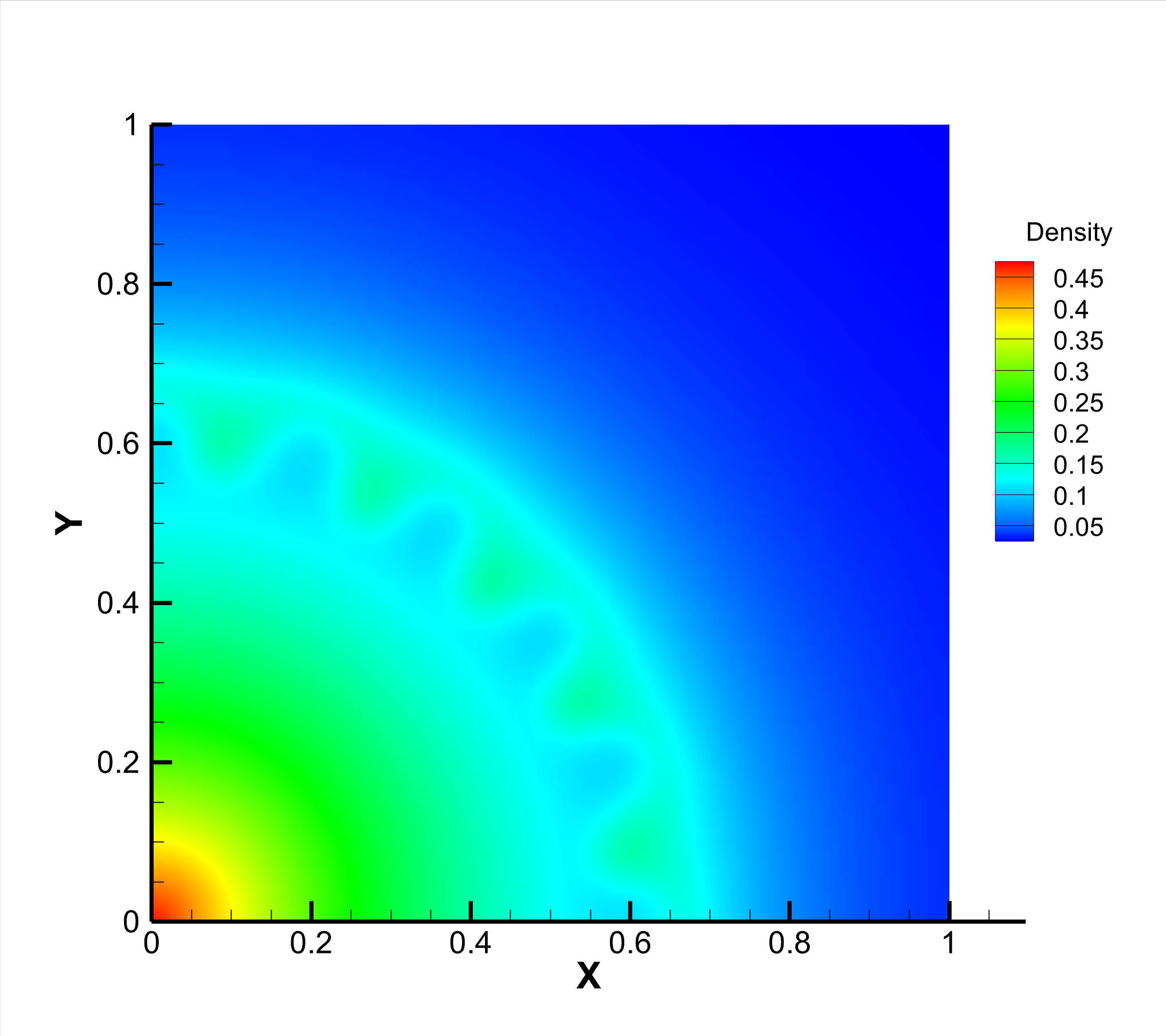}
    }
    \subfigure[t=1.2]{
        \includegraphics[width=0.22\textwidth]{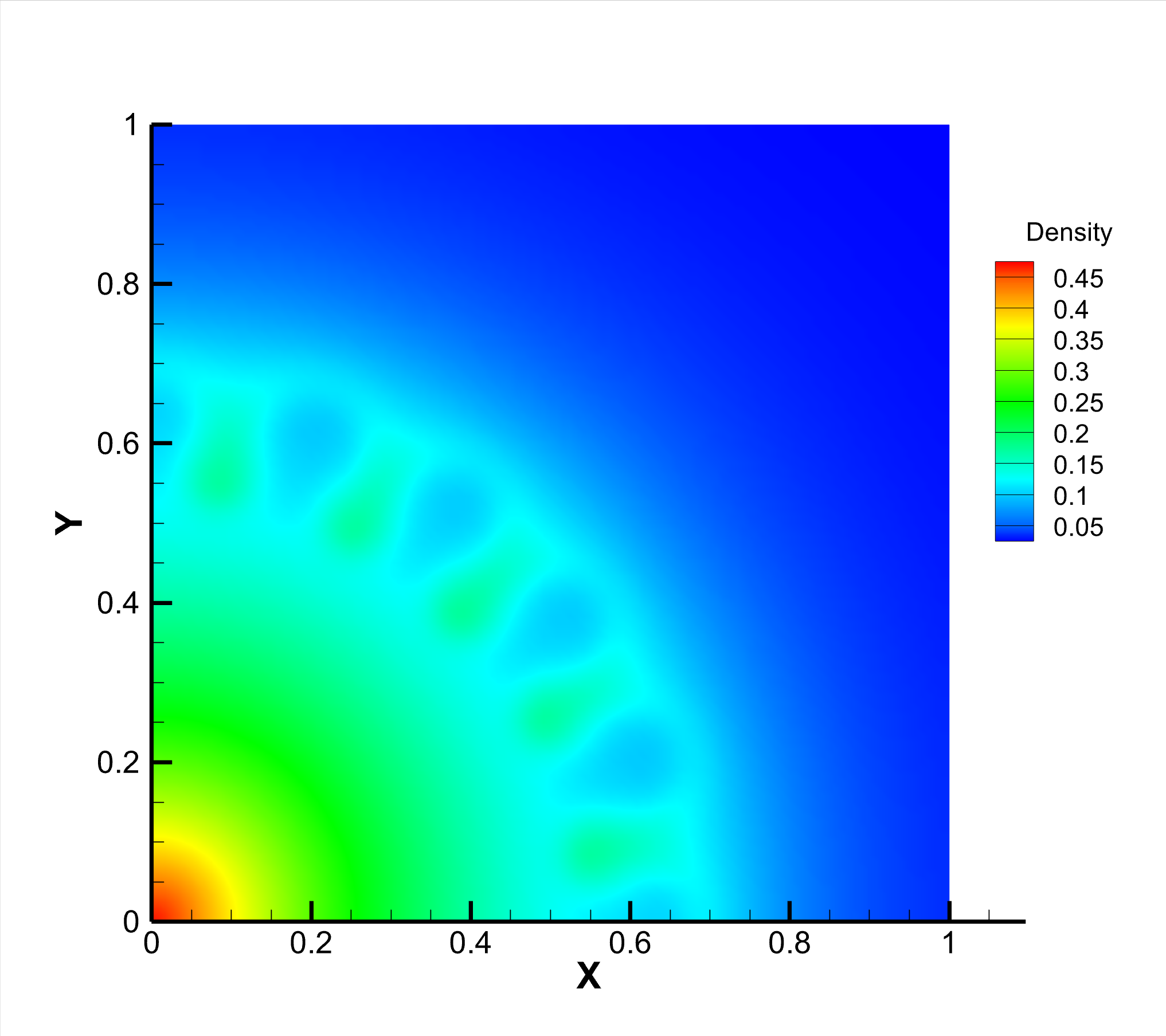}
    }
    \subfigure[t=2.0]{
        \includegraphics[width=0.22\textwidth]{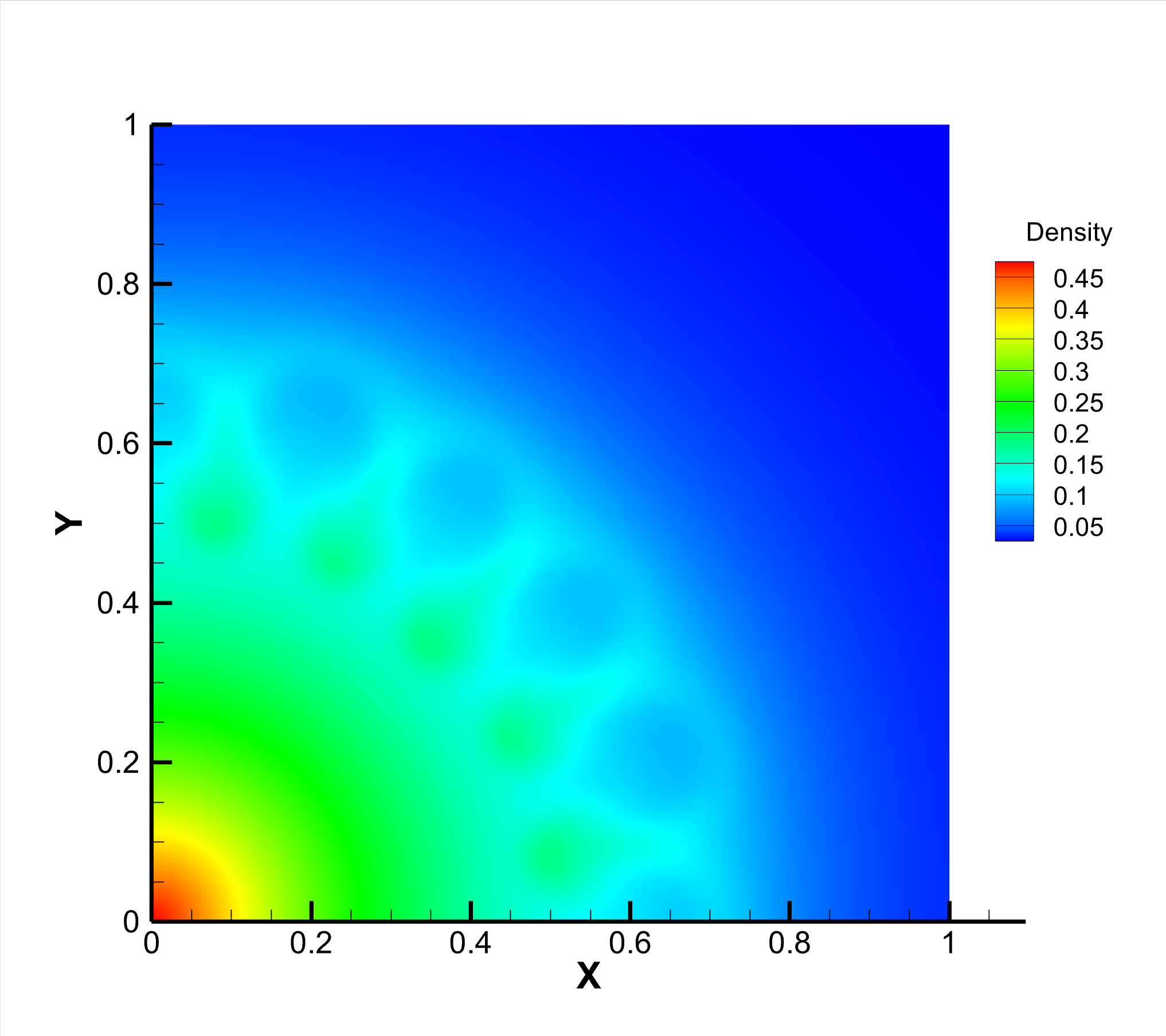}
    }
    \subfigure[t=0]{
        \includegraphics[width=0.22\textwidth]{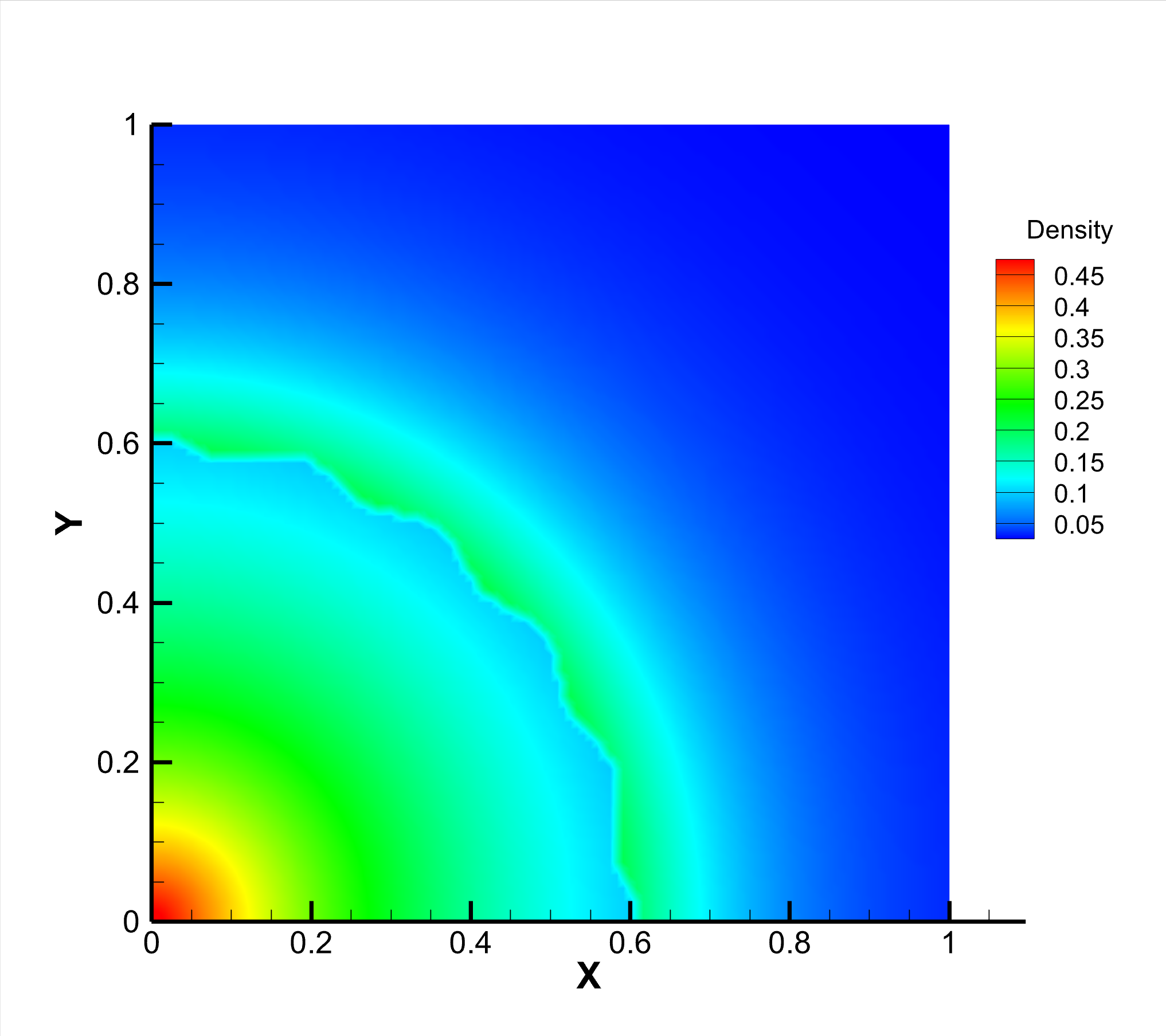}
    }
    \subfigure[t=0.04]{
        \includegraphics[width=0.22\textwidth]{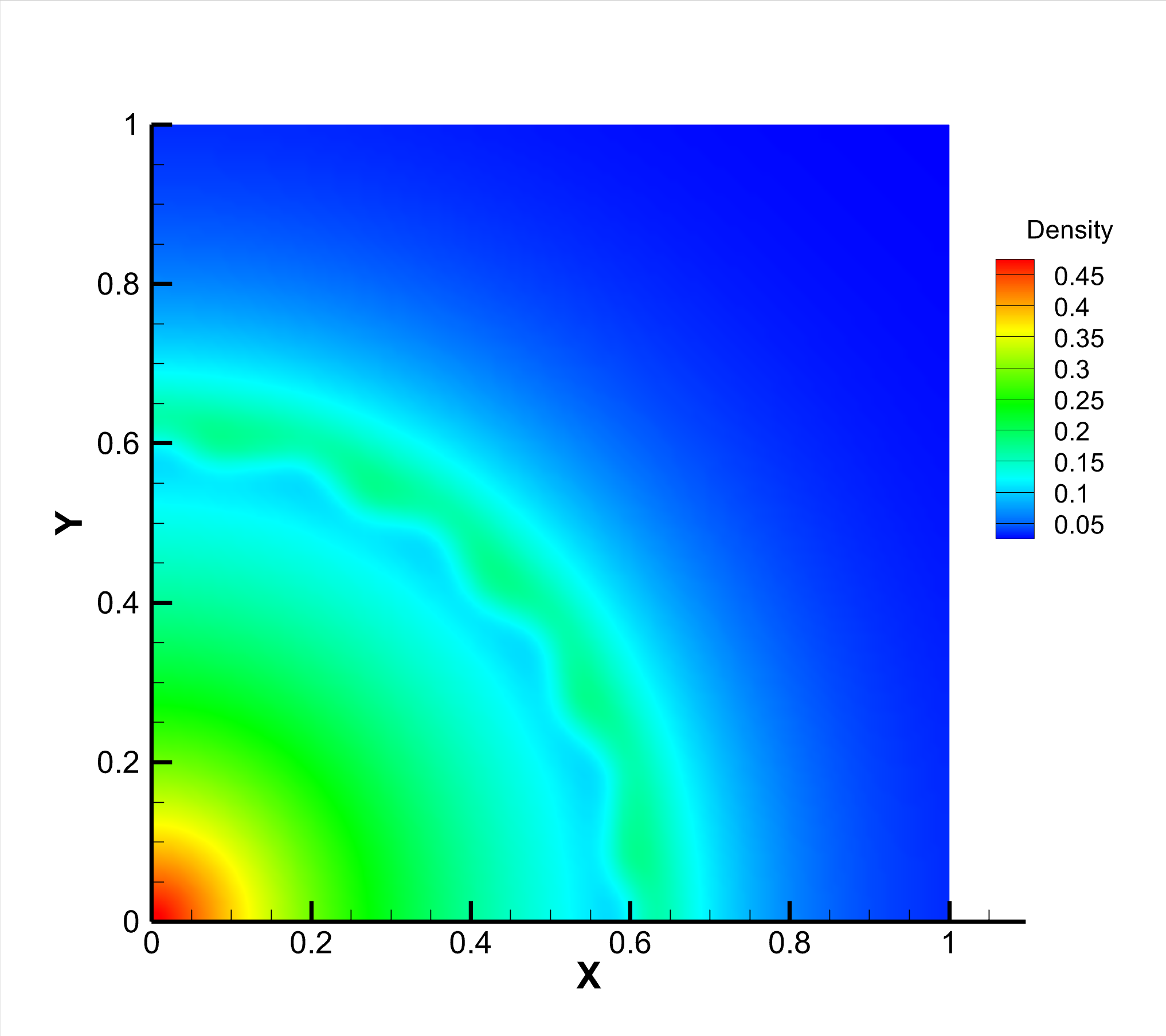}
    }
    \subfigure[t=0.08]{
        \includegraphics[width=0.22\textwidth]{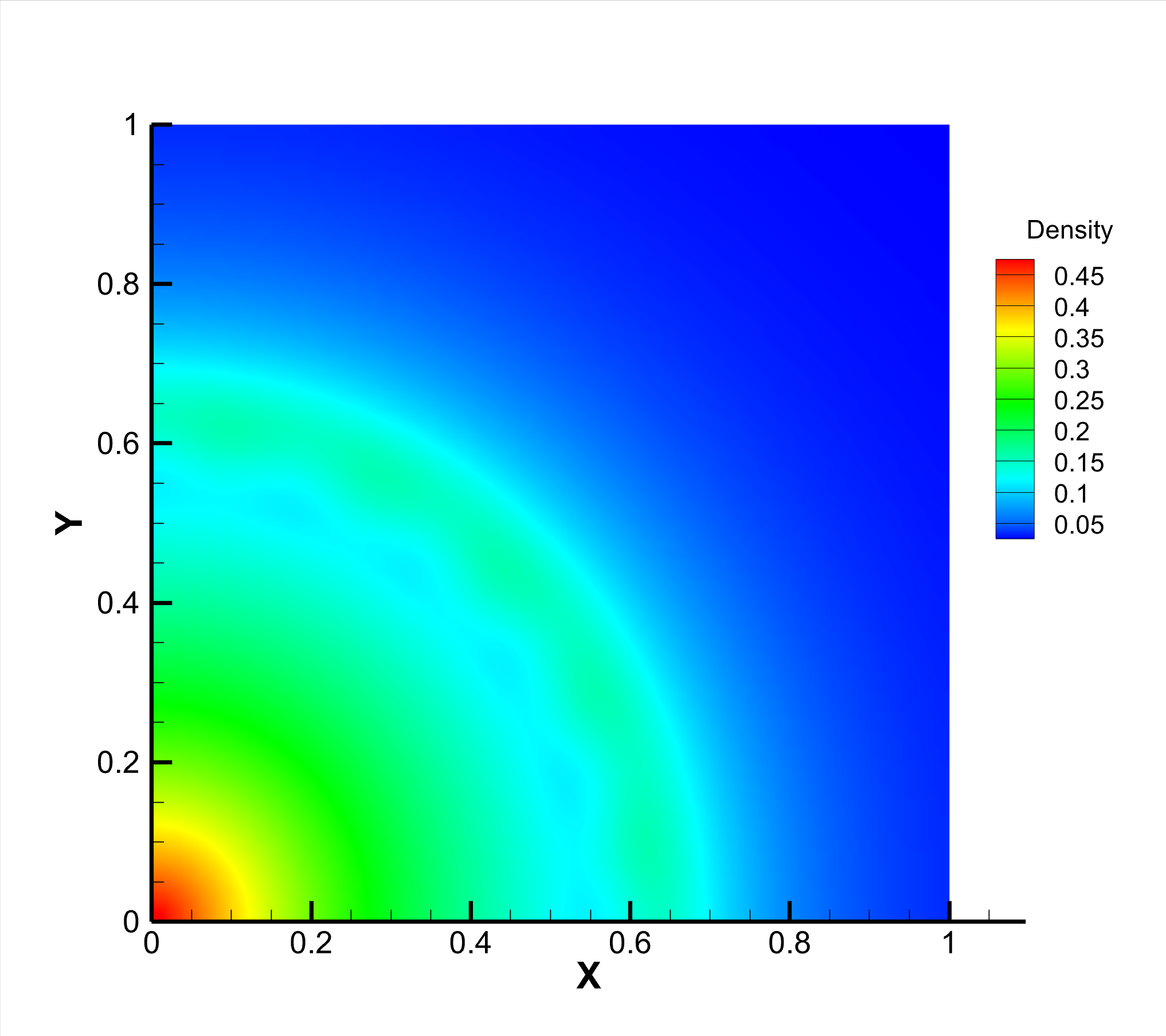}
    }
    \subfigure[t=0.14]{
        \includegraphics[width=0.22\textwidth]{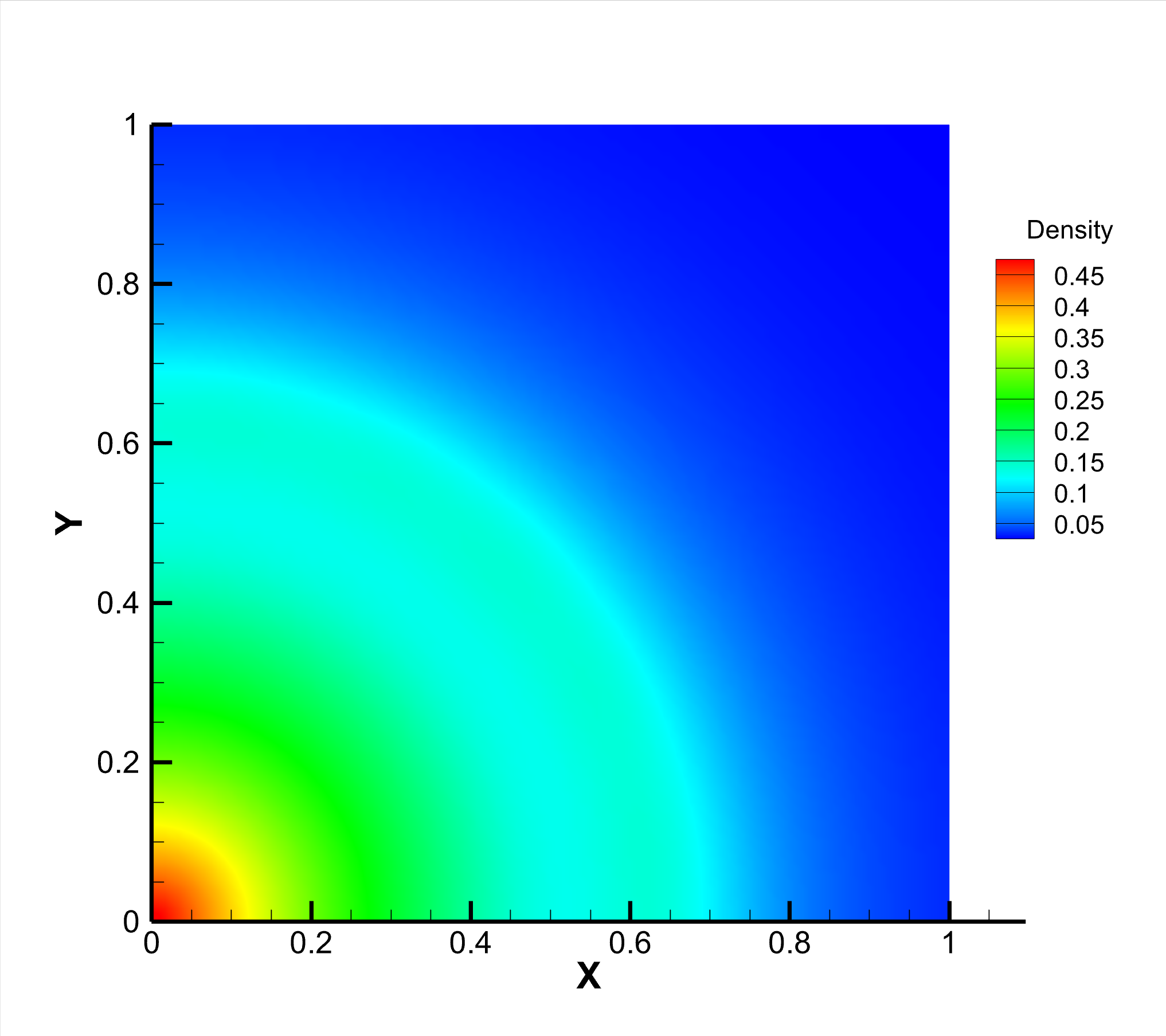}
    }
    \subfigure[t=0]{
        \includegraphics[width=0.22\textwidth]{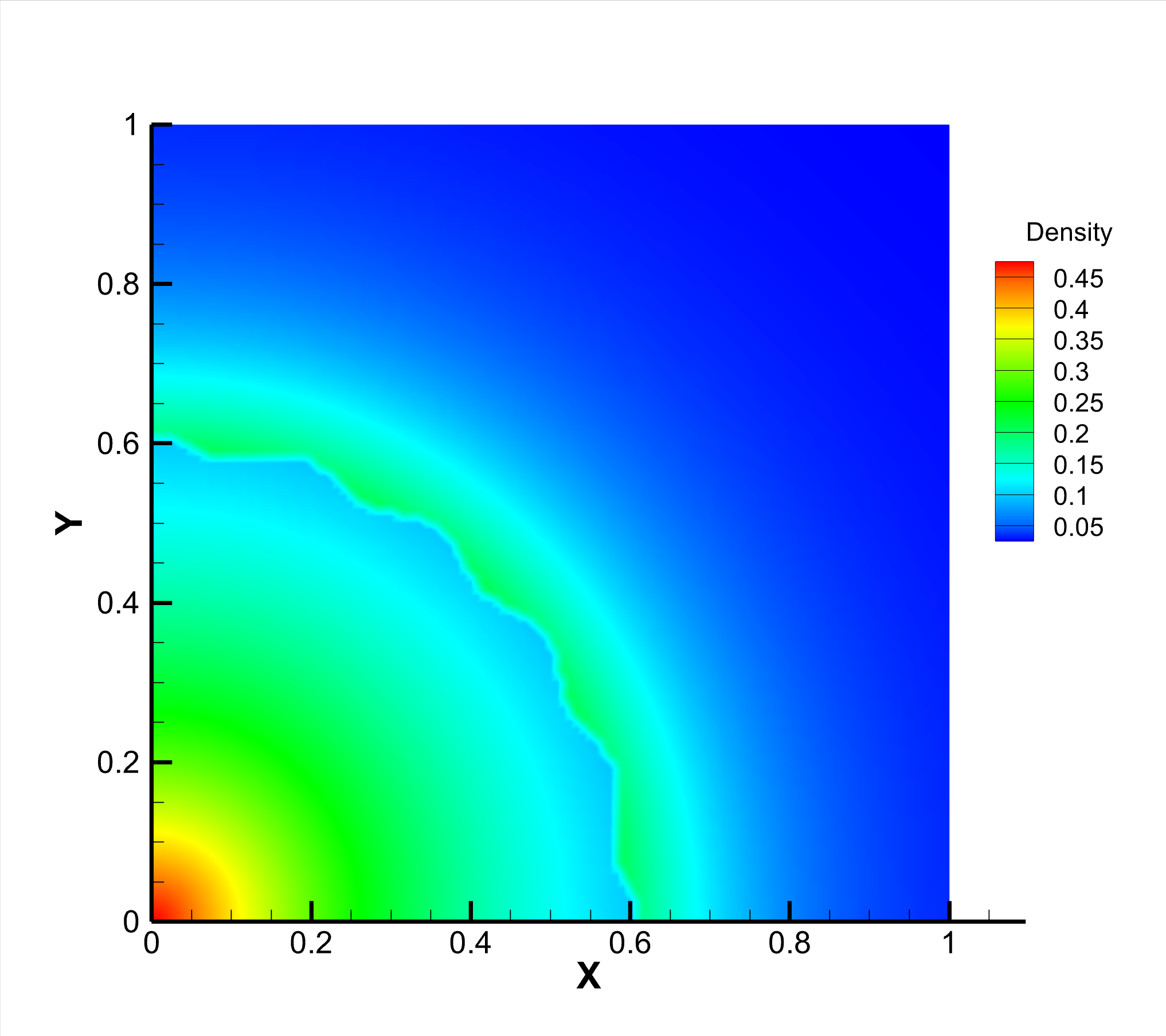}
    }
    \subfigure[t=0.04]{
        \includegraphics[width=0.22\textwidth]{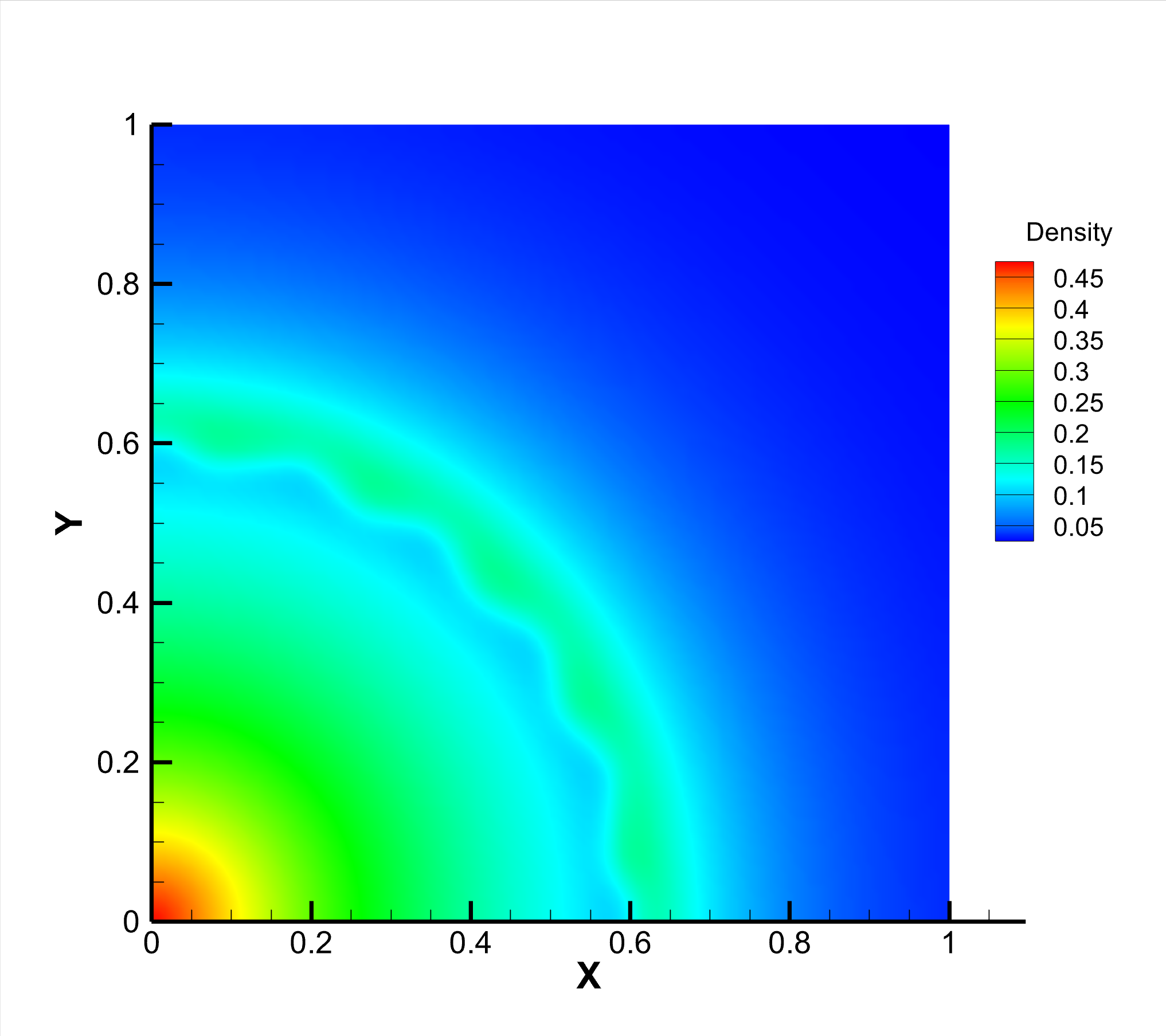}
    }
    \subfigure[t=0.08]{
        \includegraphics[width=0.22\textwidth]{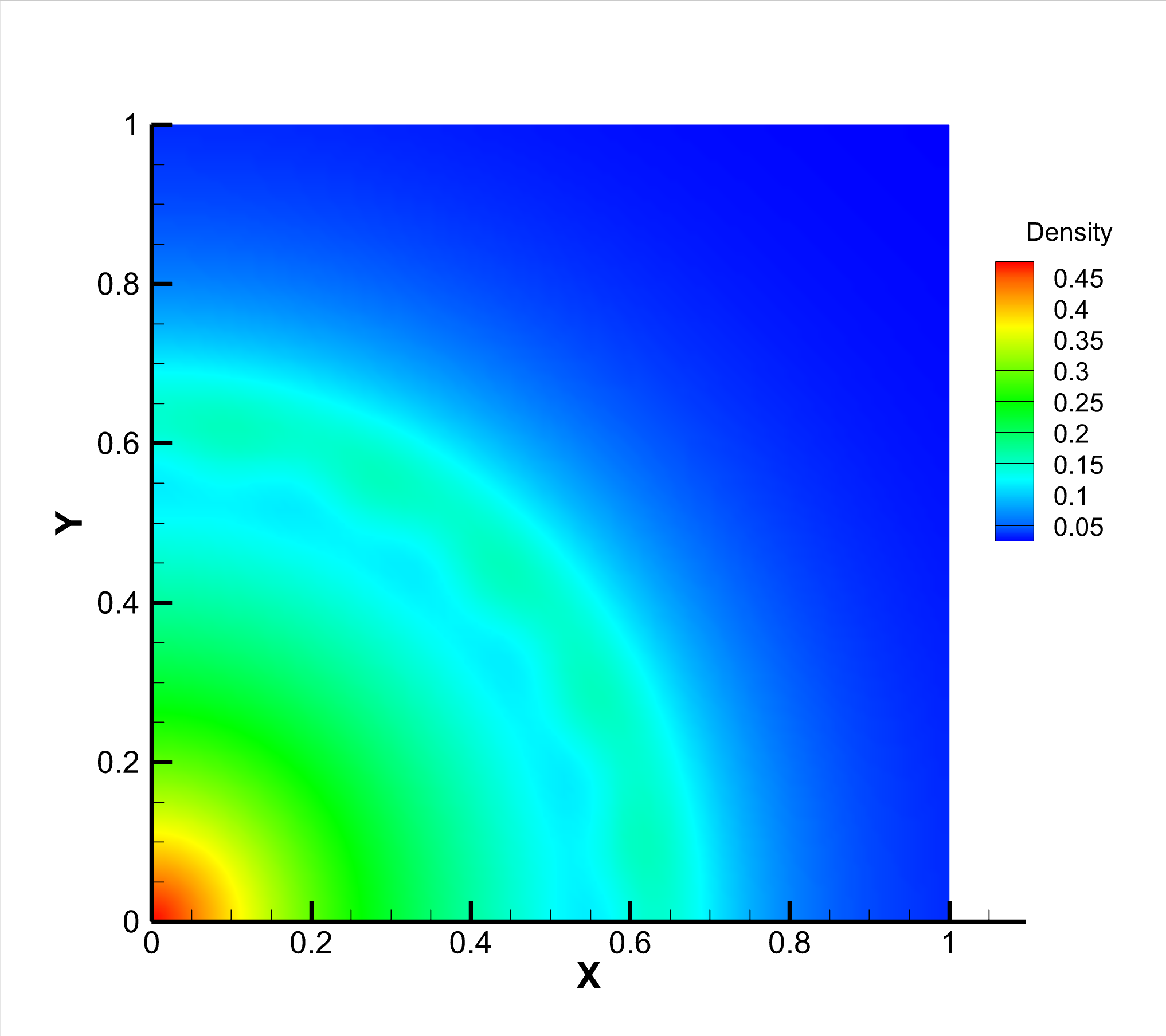}
    }
    \subfigure[t=0.14]{
        \includegraphics[width=0.22\textwidth]{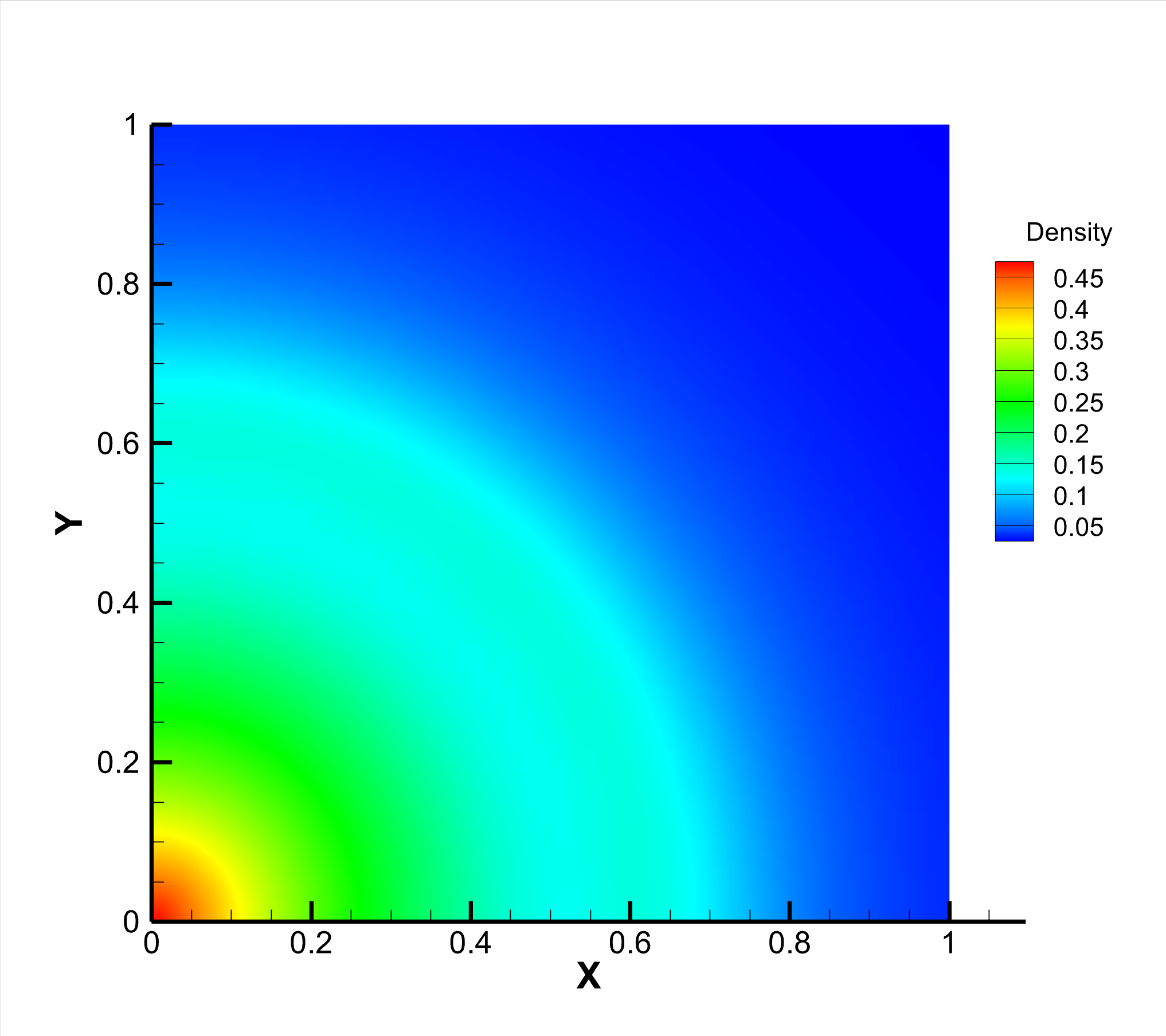}
    }
    \caption{Density evolution in the Rayleigh–Taylor instability problem at reference Knudsen numbers 0.0001(1st row), 0.01(2nd row), 1(3rd row).}
    \label{fig:RT}
\end{figure}

\begin{figure}[htb!]
    \centering
    \subfigure[t=0]{
        \includegraphics[width=0.22\textwidth]{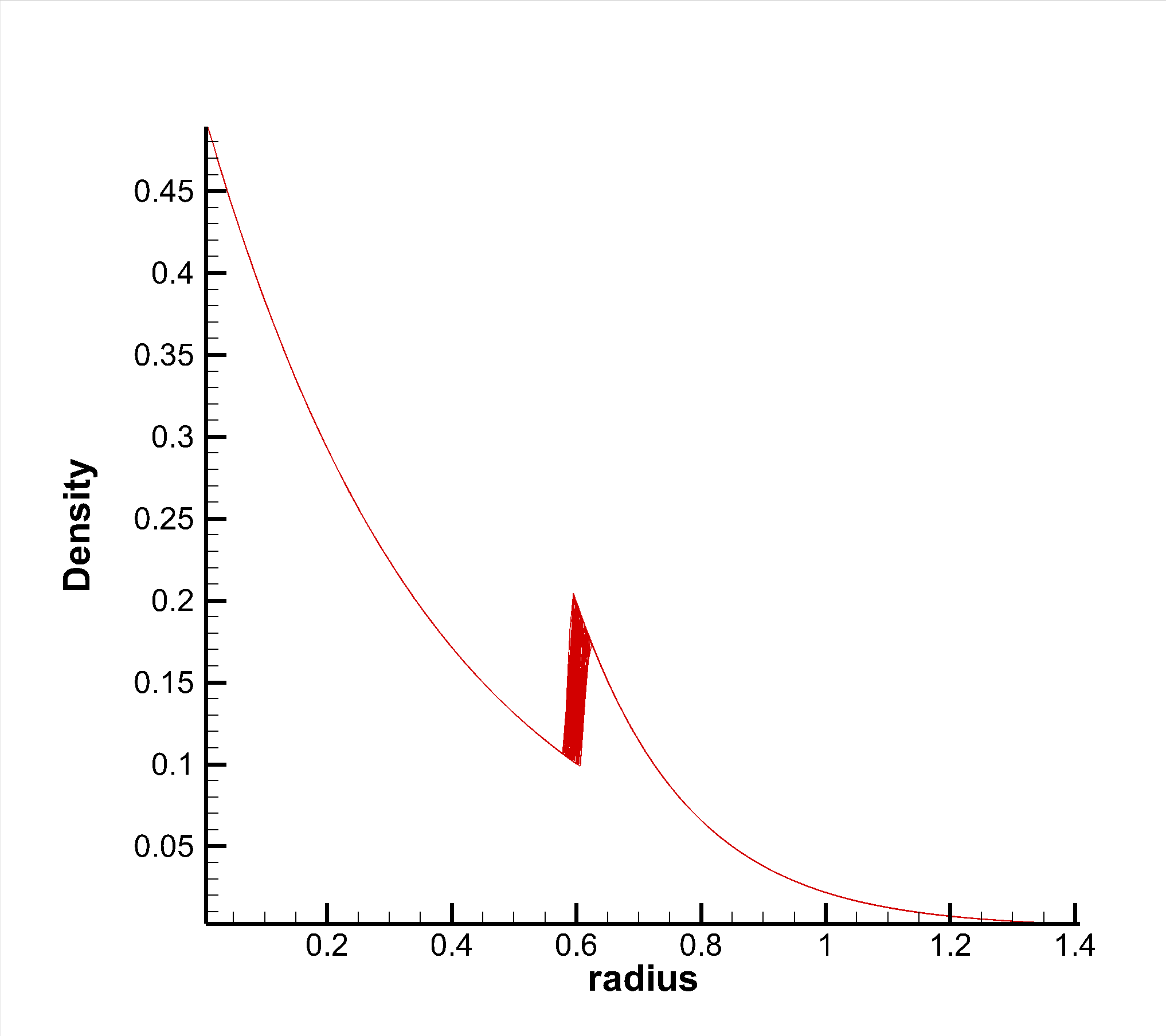}
    }
    \subfigure[t=0.8]{
        \includegraphics[width=0.22\textwidth]{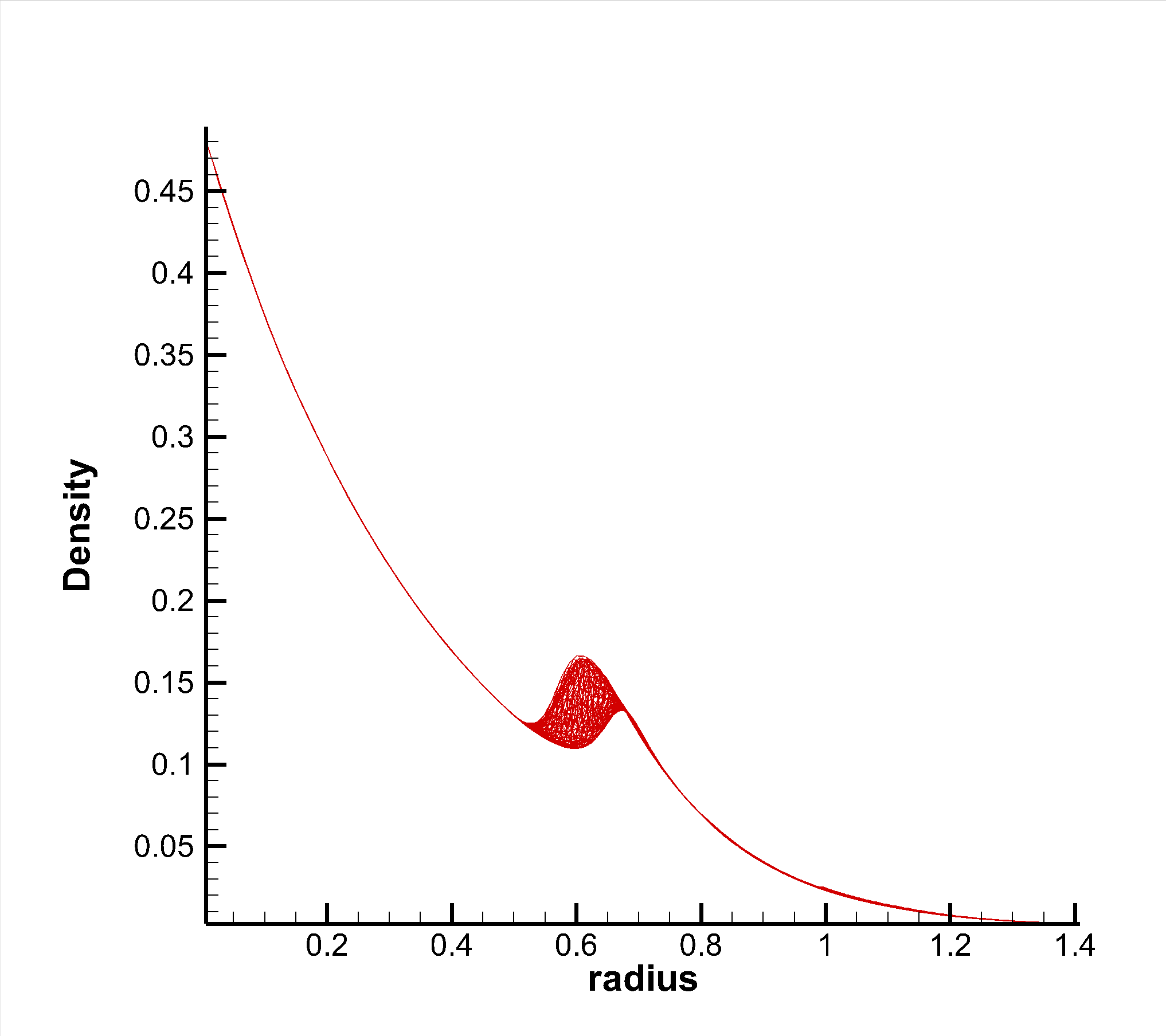}
    }
    \subfigure[t=1.2]{
        \includegraphics[width=0.22\textwidth]{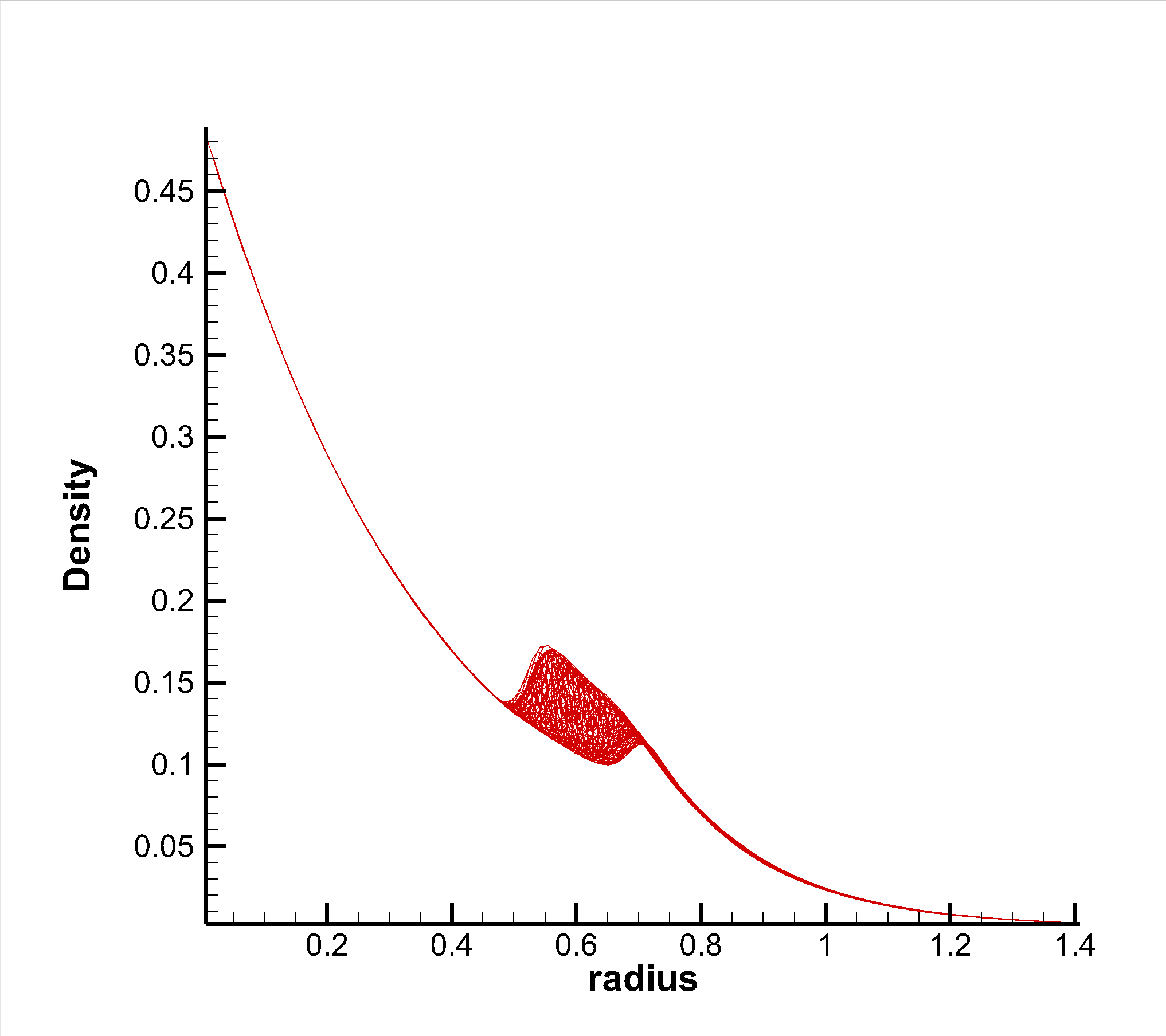}
    }
    \subfigure[t=2.0]{
        \includegraphics[width=0.22\textwidth]{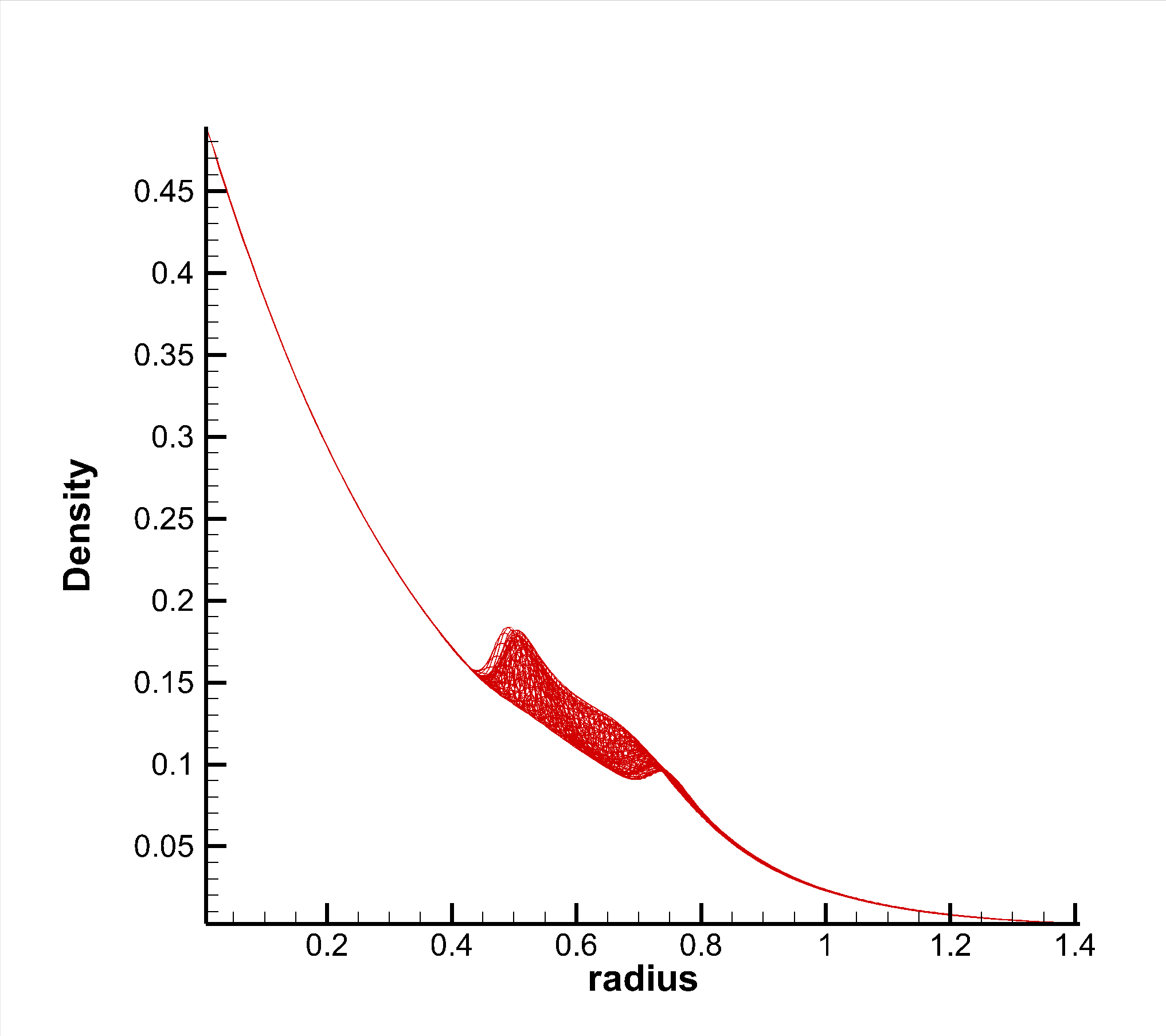}
    }
    \subfigure[t=0]{
        \includegraphics[width=0.22\textwidth]{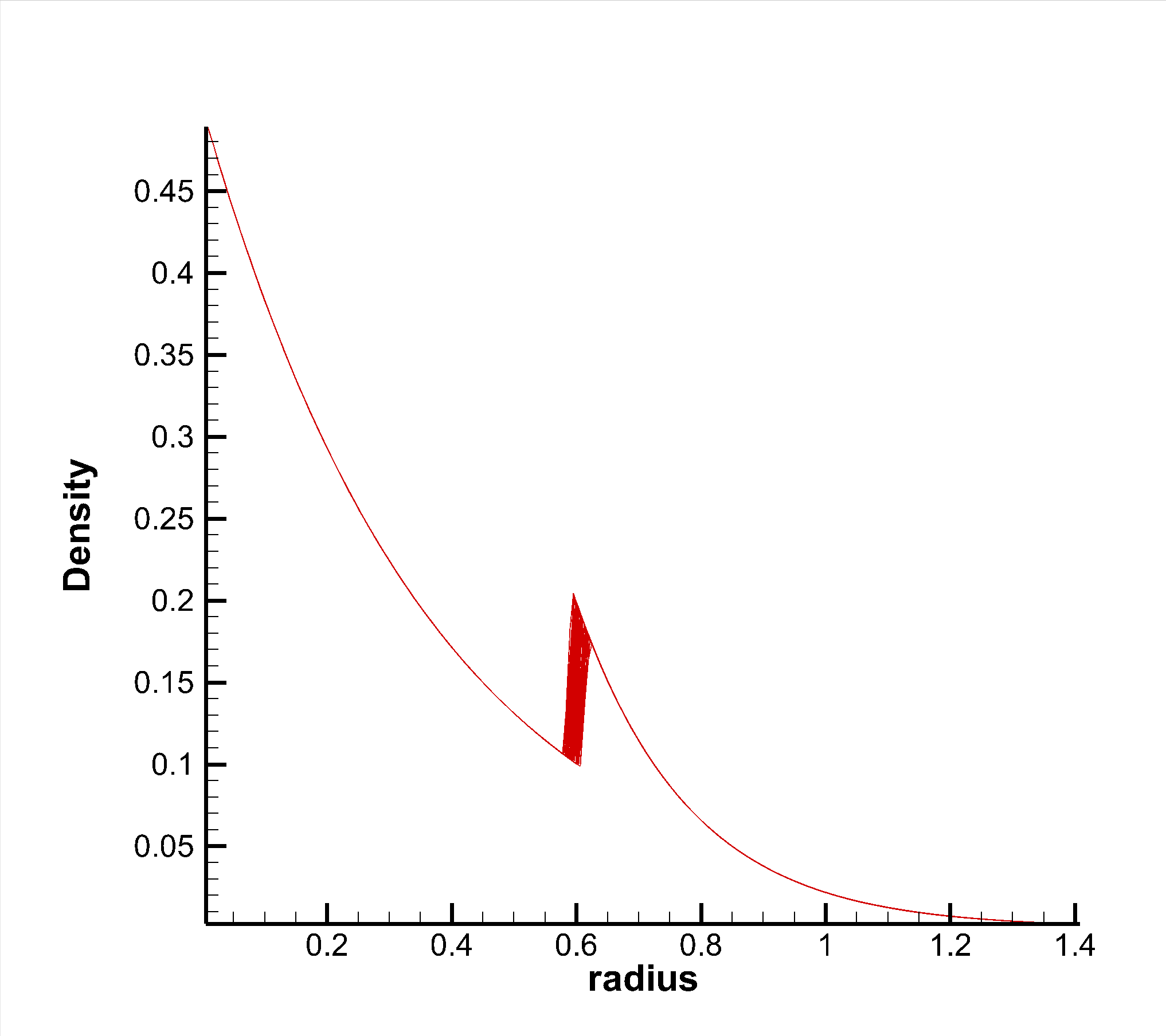}
    }
    \subfigure[t=0.04]{
        \includegraphics[width=0.22\textwidth]{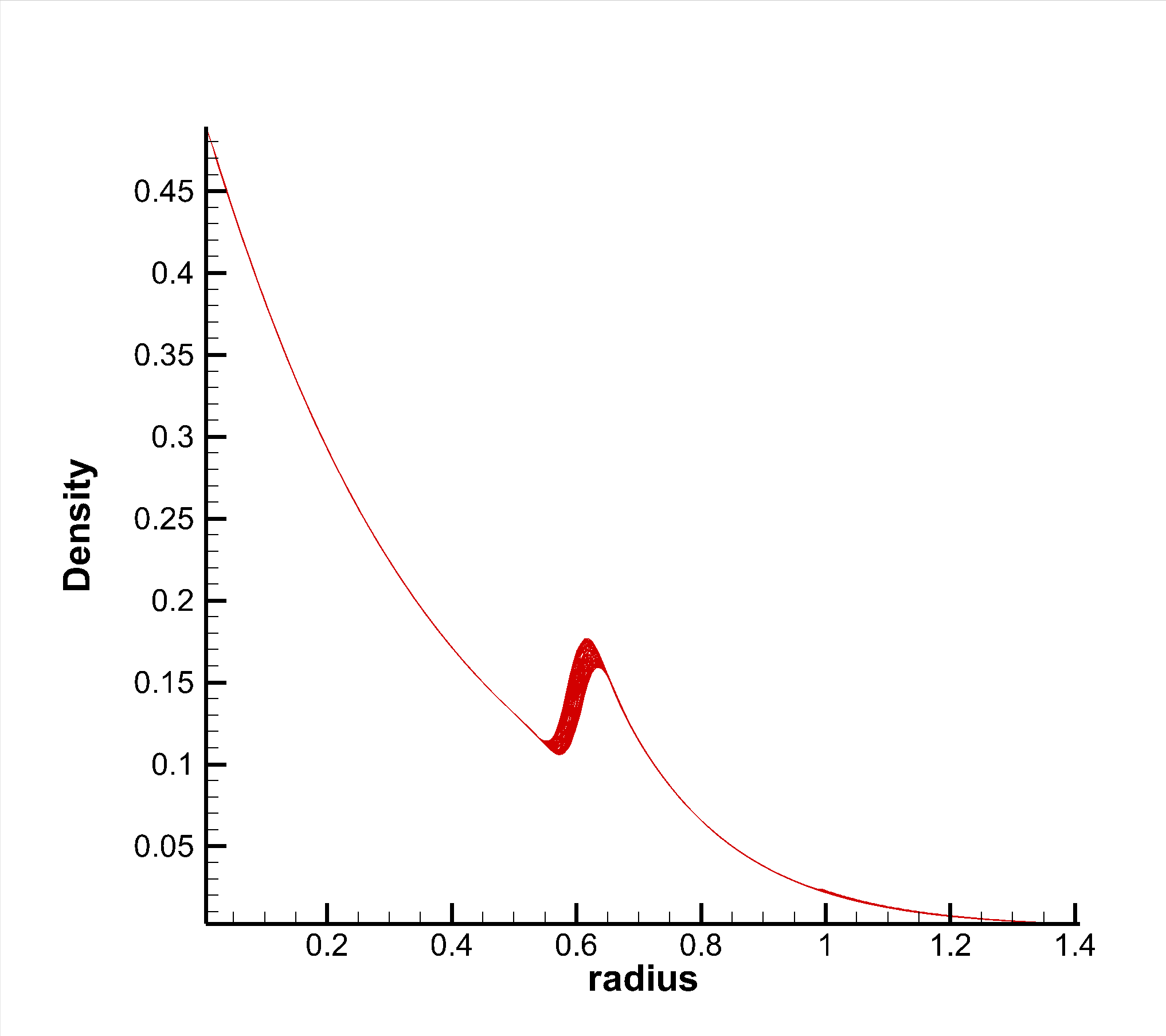}
    }
    \subfigure[t=0.08]{
        \includegraphics[width=0.22\textwidth]{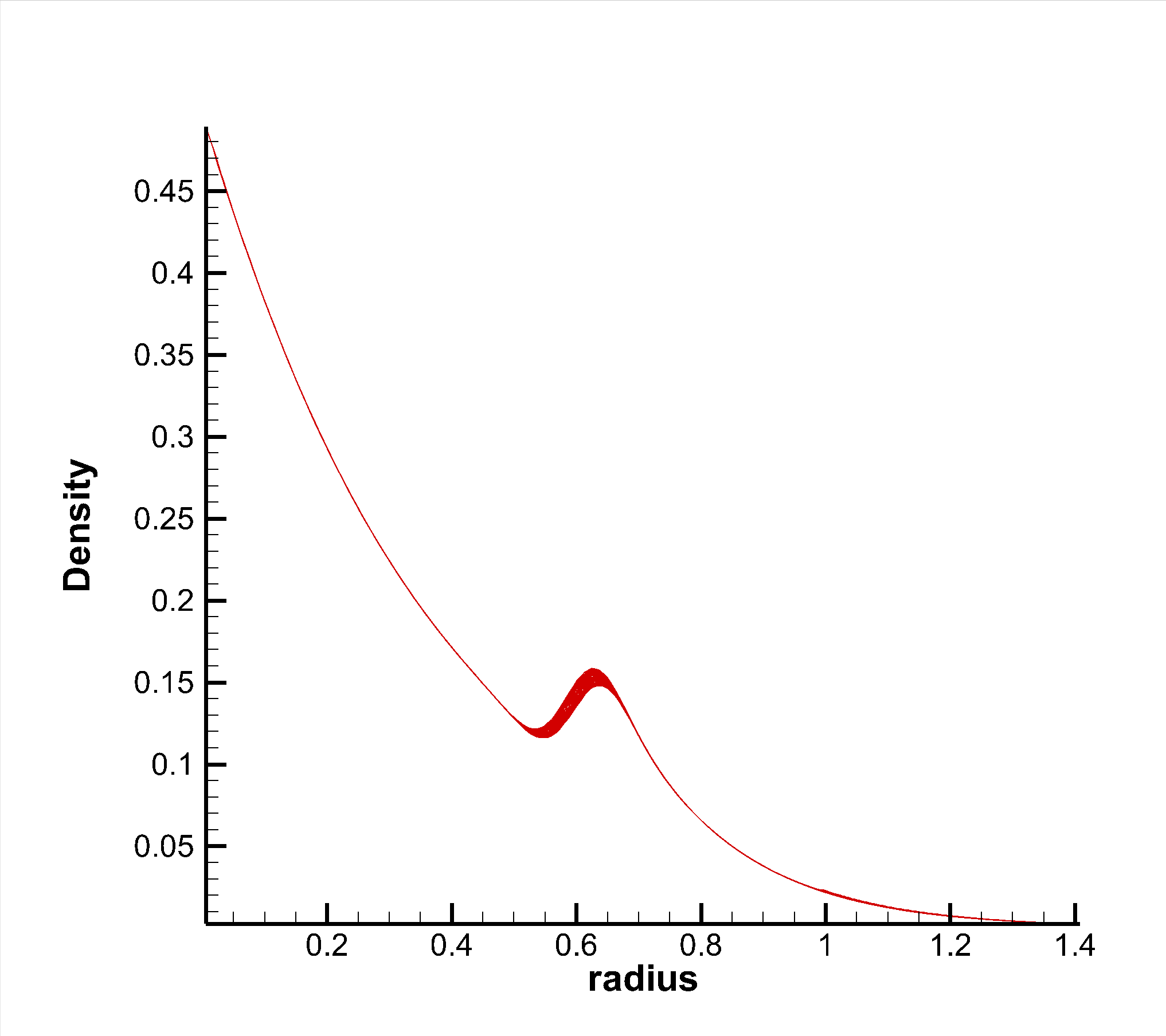}
    }
    \subfigure[t=0.14]{
        \includegraphics[width=0.22\textwidth]{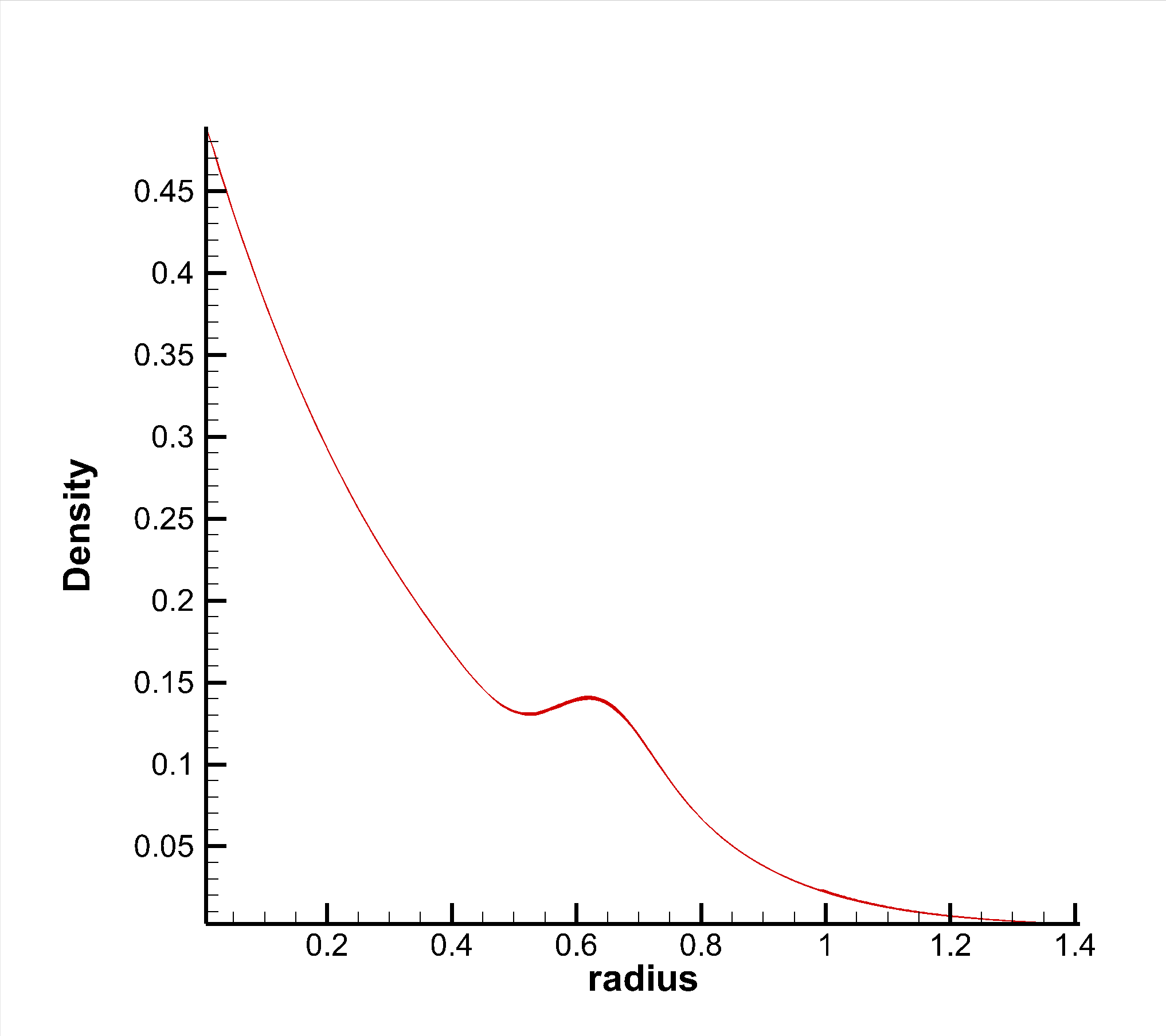}
    }
    \subfigure[t=0]{
        \includegraphics[width=0.22\textwidth]{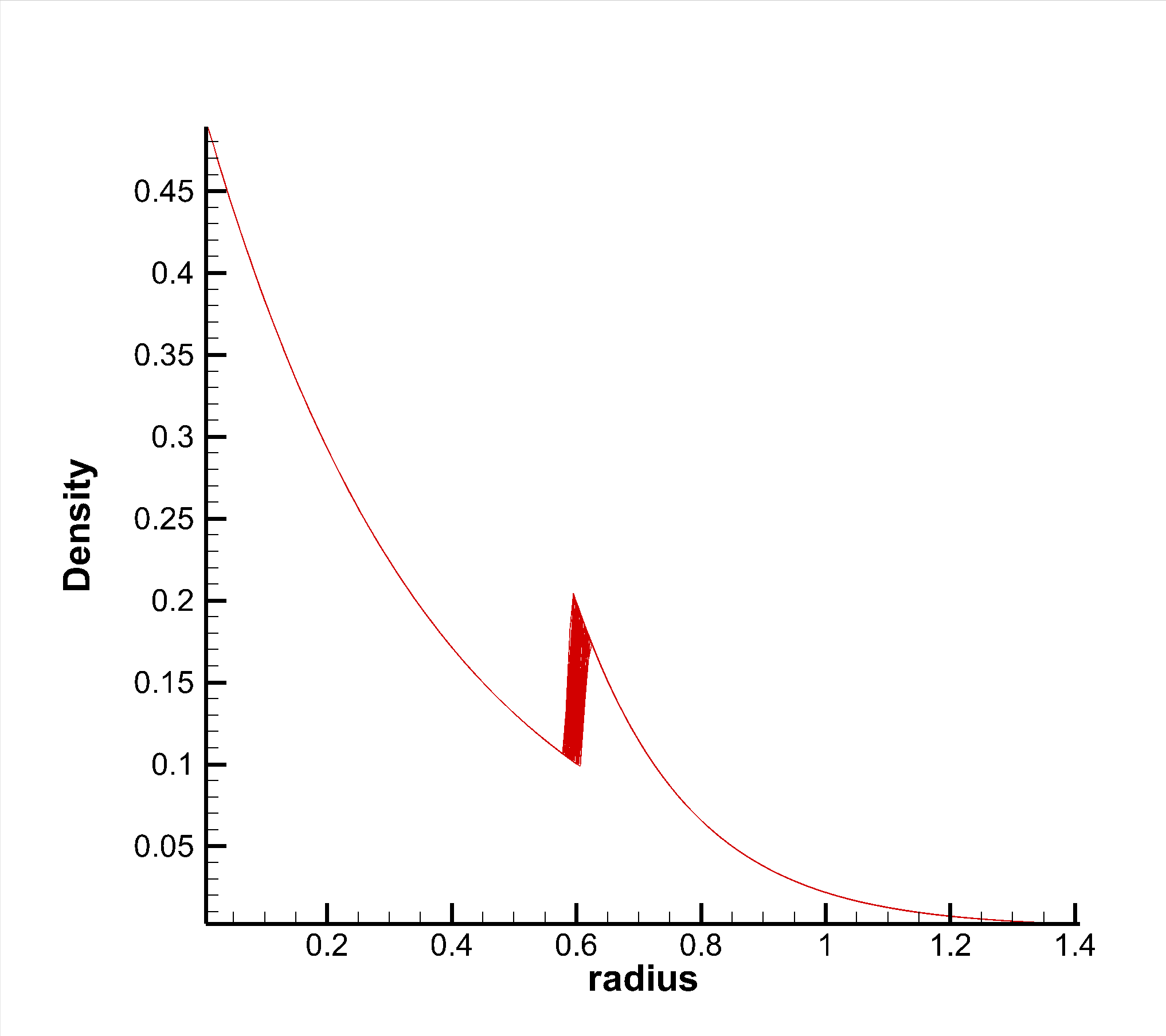}
    }
    \subfigure[t=0.04]{
        \includegraphics[width=0.22\textwidth]{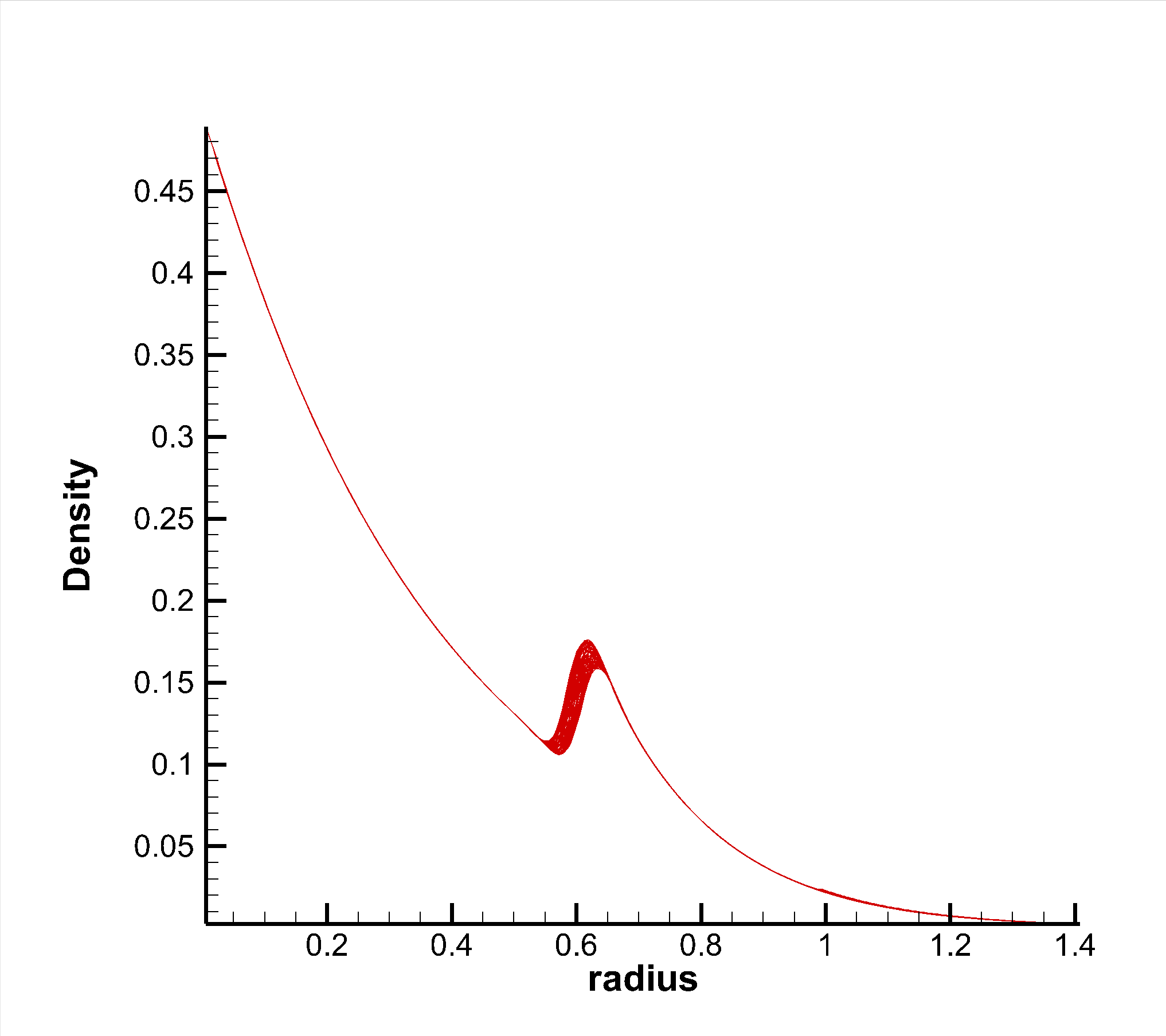}
    }
    \subfigure[t=0.08]{
        \includegraphics[width=0.22\textwidth]{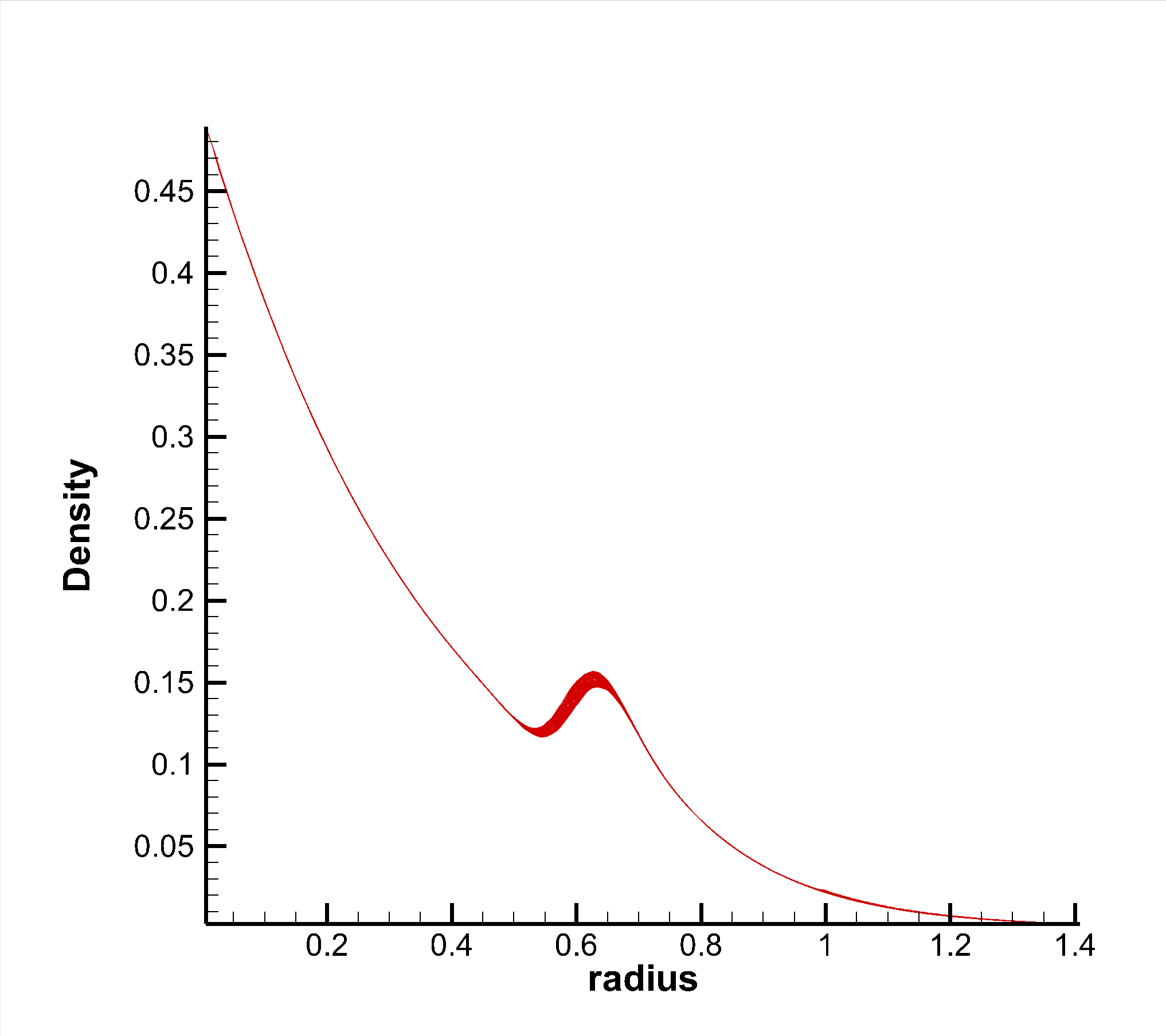}
    }
    \subfigure[t=0.14]{
        \includegraphics[width=0.22\textwidth]{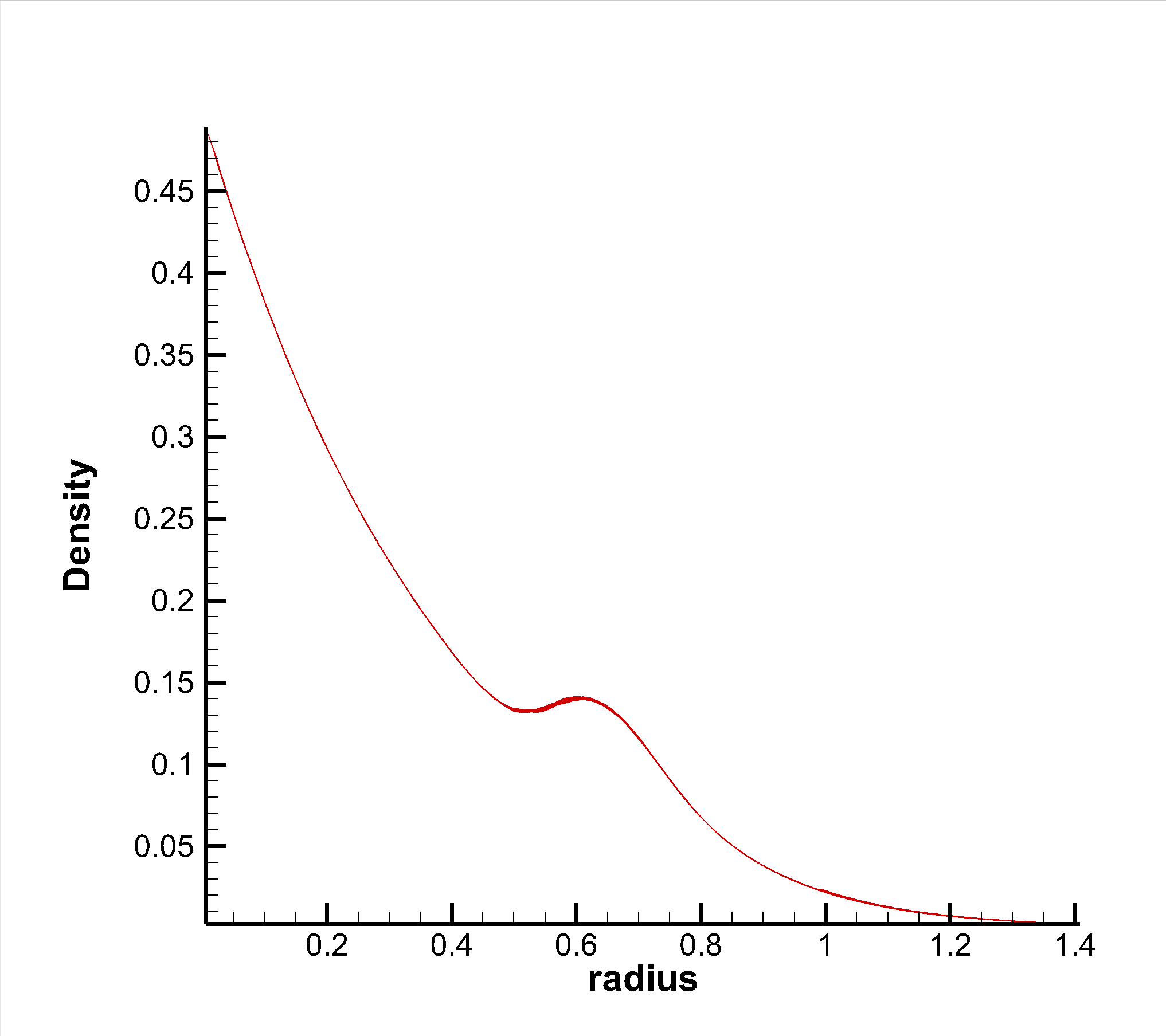}
    }
    \caption{Density distribution along the radial direction in the Rayleigh–Taylor instability problem at reference Knudsen numbers 0.0001(1st row), 0.01(2nd row), 1(3rd row).}
    \label{fig:RT line}
\end{figure}

\begin{figure}[htb!]
    \centering
    \subfigure[Kn=0.0001]{
        \includegraphics[width=0.31\textwidth]{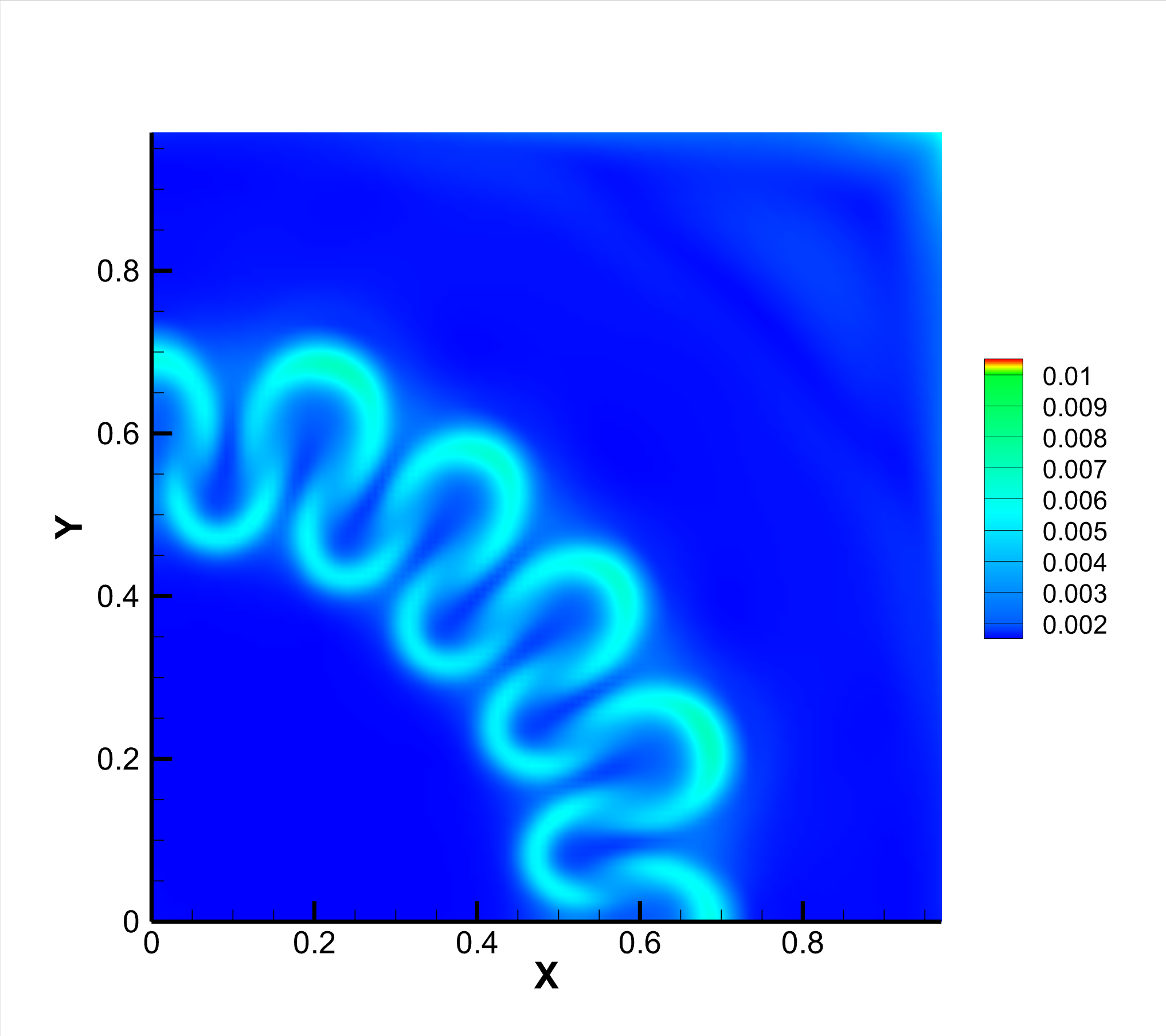}
    }
    \subfigure[Kn=0.01]{
        \includegraphics[width=0.31\textwidth]{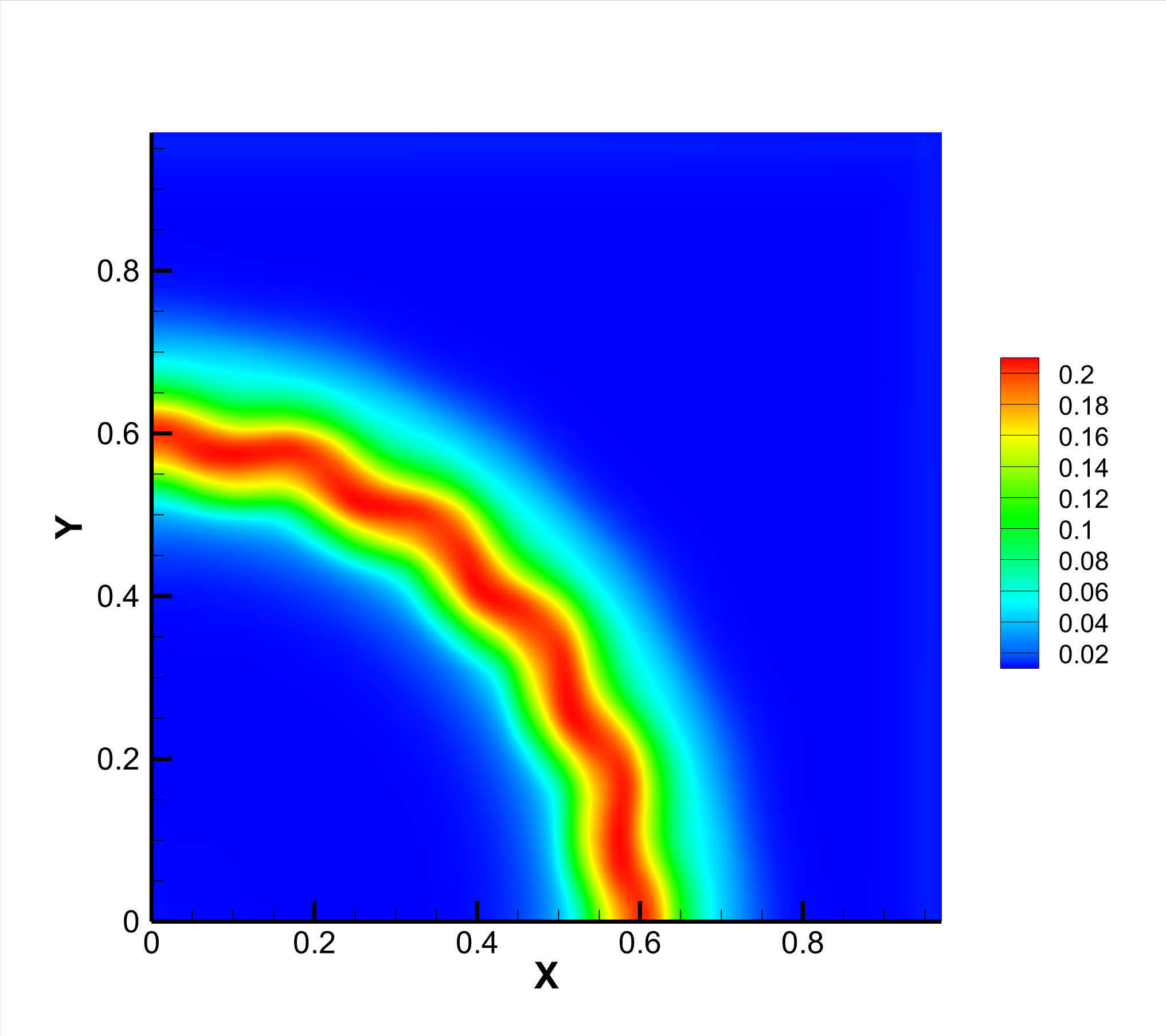}
    }
    \subfigure[Kn=1]{
        \includegraphics[width=0.31\textwidth]{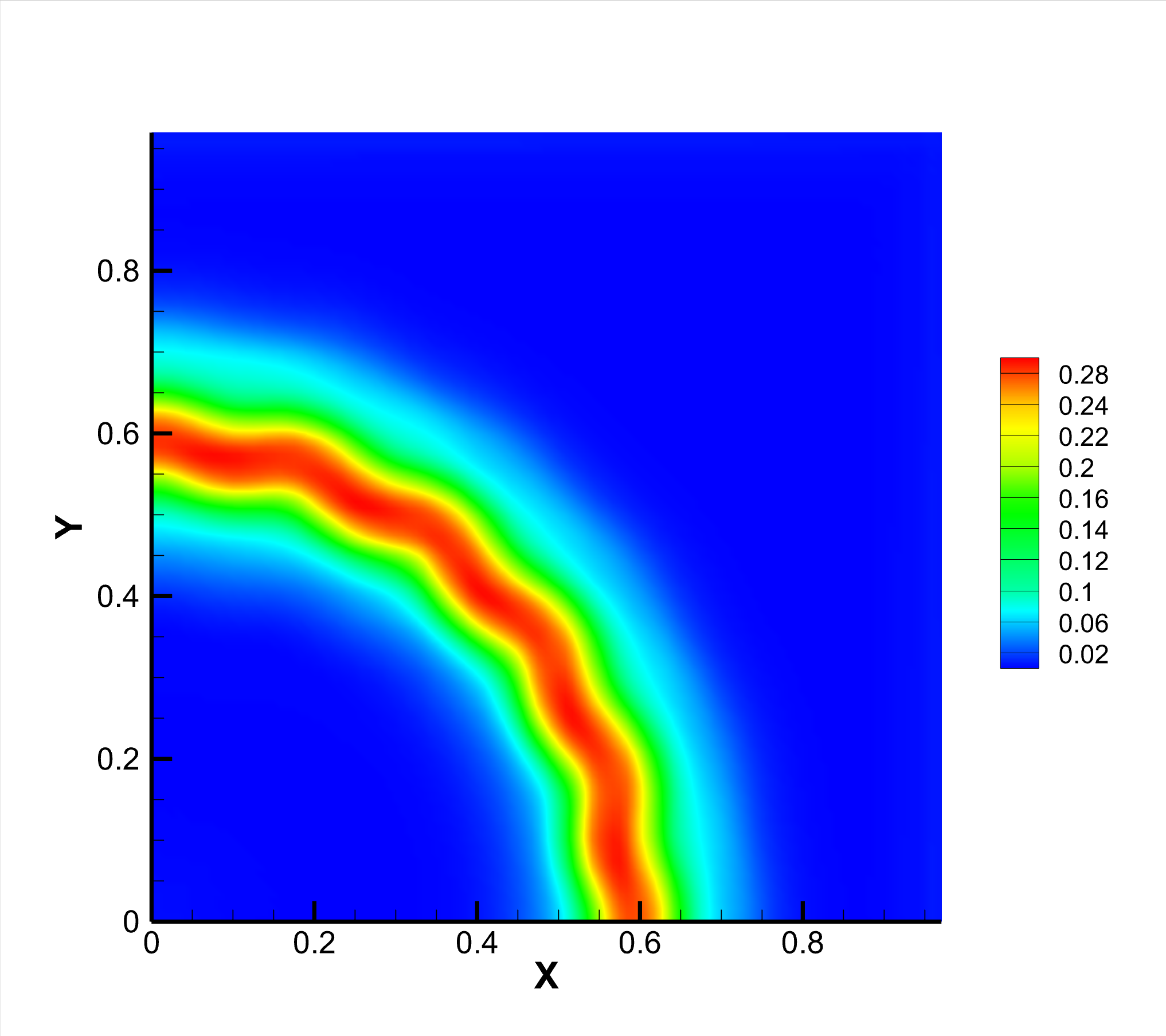}
    }
    \caption{The deviation of the particle distribution function from equilibrium at the final time instant in the Rayleigh–Taylor instability problem.}
    \label{fig:RT NE}
\end{figure}

The density contours at different time instants are shown in Figure \ref{fig:RT}.
The evolution patterns of the Rayleigh-Taylor instability across varying rarefaction regimes have been identified.
Within the continuum regime, as illustrated in the first panel of Figure \ref{fig:RT}, the frequent intermolecular collisions lead to intense fluid mixing around the interface between heavy and light fluids.
Conversely, increasing Knudsen number accelerates the fluid mixing process through enhanced particle transport phenomena, thereby significantly reducing the manifestations of interfacial instability.
Figure \ref{fig:RT line} presents the density distribution in the radial direction across all elements, which provides a quantitative characterization of the mixing zone.
Figure \ref{fig:RT NE} shows the degree of deviation of the distribution function away from equilibrium at the final time, defined as
\begin{equation}
    \Vert{f-\mathcal M}\Vert_2,
\end{equation}
where $\Vert\cdot\Vert_2$ denotes the $L^2$ norm.
It can be observed that non-equilibrium effects are intrinsically coupled with flow patterns and scale proportionally with the degree of rarefaction.
This numerical experiment again verifies the asymptotic-preserving and well-balanced properties of the solution algorithm.
In all flow regimes, the hydrostatic solution is well preserved in the computation, while the mixing process only occurs near the
Rayleigh–Taylor interface.

Table \ref{tab:RT cost} provides the computational costs of different methods.
The results demonstrate that the computational efficiency of the current method is significantly improved compared to classical discrete velocity methods while maintaining equivalent accuracy.
30- to 50-fold acceleration is achieved for the flows ranging from continuum (Kn=0.0001) to rarefied (Kn=1).
The measured data validates the scalability of the solution algorithm for multi-scale flows.

\begin{table}[htbp]
    \caption{Computational setup of the Rayleigh-Taylor instability problem (NC/GH refers to the Newton-Cotes/Gauss-Hermite quadrature).}
    \centering
    \begin{tabular}{lllllllll} 
        \hline
        Gas & $t$ & $x$ & $N_x$ & $y$ & $N_y$ & $u$ & $N_u$ \\
        Argon & $(0,2.0]$ & $[0,1]$ & 120 & $[0,1]$ & 600 & $[-5,5]$ & $[18,70]$ \\
        %\hline
        \rule{0pt}{0ex}\\
        $v$ & $N_v$ & Order & $N_m$ & $a$ & Kn & Quadrature & Integrator & \\
        $[-5,5]$ & $[18,70]$ & 2 & $[0,4]$ & -1.5 & $[0.0001,1]$ & NC/GH & IMEX & \\
        \rule{0pt}{0ex}\\
        Boundary & CFL \\
        Specular & 0.5 \\
        \hline
    \end{tabular}
    \label{tab:RT}
\end{table}

\begin{table}[h]
    \centering
    \caption{Computational costs of different methods for solving the Rayleigh-Taylor problem (NC/GH refers to the Newton-Cotes/Gauss-Hermite quadrature).}
    \begin{tabular}{llcl}
        \toprule
        Knudsen number& Numerical method & Number of collocation points & Time (s) \\
        \midrule
        0.0001& finite difference (NC) & 50& 259314.78\\
         & Hermite (NC, $N_m=0$)& 30& 43436.27\\
         & Hermite (GH, $N_m=0$)& 18& 8926.044\\
        \midrule
        0.01& finite difference (NC) & 70& 46080.32\\
         & Hermite (NC, $N_m=1$)& 40& 6925.77\\
         & Hermite (GH, $N_m=1$)& 20& 930.39\\
        \midrule
        1& finite difference (NC) & 70& 176955.60\\
         & Hermite (NC, $N_m=4$)& 40& 45807.11\\
         & Hermite (GH, $N_m=4$)& 24& 3836.53\\
         \bottomrule
    \end{tabular}
    \label{tab:RT cost}
\end{table}

\section{Conclusion}

This work presents a unified, Hermite-spectral-based solution algorithm for multi-scale flows under an external force grounded in the Boltzmann–BGK framework.
A seamless connection between computational methods and physical models is achieved by merging a high‐order Hermite spectral representation of the velocity distribution with a multi‐scale evolution model inspired by the unified gas‐kinetic scheme.
The equilibrium and non-equilibrium flow physics across continuum and rarefied regimes are robustly coupled without sacrificing computational efficiency.
The use of the spectral-collocation method not only provides spectral accuracy in the representation of non‐equilibrium distribution functions, but also, through careful selection of quadrature points and weights, yields up to 50-fold acceleration compared to classical discrete velocity methods, all while preserving key physical structures such as the well‐balanced hydrostatic equilibrium and correct asymptotics.
Through a comprehensive suite of benchmark tests, we have demonstrated the accuracy, efficiency, and robustness of the current methodology across a wide range of Knudsen numbers and external‐force magnitudes. 
Looking forward, the inherent flexibility of the solution algorithm offers promising avenues for extension to three-dimensional problems, complex geometries, and more generalized molecule models, paving the way for efficient, high‐fidelity simulations of non‐equilibrium force-driven flows in aerothermodynamics, microfluidics, astrophysics, and beyond.

\section*{Acknowledgements}

The current research is funded by the National Science Foundation of China (No. 12302381) and the Chinese Academy of Sciences Project for Young Scientists in Basic Research (YSBR107). 
The computing resources provided by Hefei Advanced Computing Center and ORISE Supercomputer are acknowledged.

\bibliographystyle{unsrt}
\bibliography{main}

\begin{thebibliography}{10}

\bibitem{toro2013riemann}
Eleuterio~F Toro.
\newblock {\em Riemann solvers and numerical methods for fluid dynamics: a
  practical introduction}.
\newblock Springer Science \& Business Media, 2013.

\bibitem{chapman1990mathematical}
Sydney Chapman and Thomas~George Cowling.
\newblock {\em The mathematical theory of non-uniform gases: an account of the
  kinetic theory of viscosity, thermal conduction and diffusion in gases}.
\newblock Cambridge university press, 1990.

\bibitem{benilov2019enskog}
ES~Benilov and MS~Benilov.
\newblock The enskog--vlasov equation: a kinetic model describing gas, liquid,
  and solid.
\newblock {\em Journal of Statistical Mechanics: Theory and Experiment},
  2019(10):103205, 2019.

\bibitem{tsien1946superaerodynamics}
Hsue-Shen Tsien.
\newblock Superaerodynamics, mechanics of rarefied gases.
\newblock {\em Journal of the Aeronautical Sciences}, 13(12):653--664, 1946.

\bibitem{tij1994perturbation}
Mohamed Tij and Andr{\'e}s Santos.
\newblock Perturbation analysis of a stationary nonequilibrium flow generated
  by an external force.
\newblock {\em Journal of statistical physics}, 76:1399--1414, 1994.

\bibitem{xiao2018investigation}
Tianbai Xiao, Kun Xu, Qingdong Cai, and Tiezheng Qian.
\newblock An investigation of non-equilibrium heat transport in a gas system
  under external force field.
\newblock {\em International Journal of Heat and Mass Transfer}, 126:362--379,
  2018.

\bibitem{jin2010asymptotic}
Shi Jin.
\newblock Asymptotic preserving (ap) schemes for multiscale kinetic and
  hyperbolic equations: a review.
\newblock {\em Lecture notes for summer school on methods and models of kinetic
  theory (M\&MKT), Porto Ercole (Grosseto, Italy)}, pages 177--216, 2010.

\bibitem{dimarco2013asymptotic}
Giacomo Dimarco and Lorenzo Pareschi.
\newblock Asymptotic preserving implicit-explicit runge--kutta methods for
  nonlinear kinetic equations.
\newblock {\em SIAM Journal on Numerical Analysis}, 51(2):1064--1087, 2013.

\bibitem{xiao2021flux}
Tianbai Xiao.
\newblock A flux reconstruction kinetic scheme for the boltzmann equation.
\newblock {\em Journal of Computational Physics}, 447:110689, 2021.

\bibitem{chacon2017multiscale}
Luis Chacon, Guangye Chen, Dana~A Knoll, C~Newman, H~Park, William Taitano,
  Jeff~A Willert, and Geoffrey Womeldorff.
\newblock Multiscale high-order/low-order (holo) algorithms and applications.
\newblock {\em Journal of Computational Physics}, 330:21--45, 2017.

\bibitem{xu2010unified}
Kun Xu and Juan-Chen Huang.
\newblock A unified gas-kinetic scheme for continuum and rarefied flows.
\newblock {\em Journal of Computational Physics}, 229(20):7747--7764, 2010.

\bibitem{guo2013discrete}
Zhaoli Guo, Kun Xu, and Ruijie Wang.
\newblock Discrete unified gas kinetic scheme for all knudsen number flows:
  Low-speed isothermal case.
\newblock {\em Physical Review E—Statistical, Nonlinear, and Soft Matter
  Physics}, 88(3):033305, 2013.

\bibitem{xiao2020velocity}
Tianbai Xiao, Chang Liu, Kun Xu, and Qingdong Cai.
\newblock A velocity-space adaptive unified gas kinetic scheme for continuum
  and rarefied flows.
\newblock {\em Journal of Computational Physics}, 415:109535, 2020.

\bibitem{zhu2021general}
Lianhua Zhu, Xingcai Pi, Wei Su, Zhi-Hui Li, Yonghao Zhang, and Lei Wu.
\newblock General synthetic iterative scheme for nonlinear gas kinetic
  simulation of multi-scale rarefied gas flows.
\newblock {\em Journal of Computational Physics}, 430:110091, 2021.

\bibitem{fei2020unified}
Fei Fei, Jun Zhang, Jing Li, and ZhaoHui Liu.
\newblock A unified stochastic particle bhatnagar-gross-krook method for
  multiscale gas flows.
\newblock {\em Journal of Computational Physics}, 400:108972, 2020.

\bibitem{liu2020unified}
Chang Liu, Yajun Zhu, and Kun Xu.
\newblock Unified gas-kinetic wave-particle methods i: Continuum and rarefied
  gas flow.
\newblock {\em Journal of Computational Physics}, 401:108977, 2020.

\bibitem{liu2020simplified}
Sha Liu, Chengwen Zhong, and Ming Fang.
\newblock Simplified unified wave-particle method with quantified
  model-competition mechanism for numerical calculation of multiscale flows.
\newblock {\em Physical Review E}, 102(1):013304, 2020.

\bibitem{yang2023discrete}
LM~Yang, ZH~Li, C~Shu, YY~Liu, W~Liu, and J~Wu.
\newblock Discrete unified gas-kinetic wave-particle method for flows in all
  flow regimes.
\newblock {\em Physical Review E}, 108(1):015302, 2023.

\bibitem{li2021simple}
Peng Li and Zhen Gao.
\newblock Simple high order well-balanced finite difference weno schemes for
  the euler equations under gravitational fields.
\newblock {\em Journal of Computational Physics}, 437:110341, 2021.

\bibitem{navas2023family}
Adri{\'a}n Navas-Montilla and Isabel Echeverribar.
\newblock A family of well-balanced weno and teno schemes for atmospheric
  flows.
\newblock {\em Journal of Computational Physics}, 489:112273, 2023.

\bibitem{mantri2021well}
Yogiraj Mantri and Sebastian Noelle.
\newblock Well-balanced discontinuous galerkin scheme for 2$\times$ 2
  hyperbolic balance law.
\newblock {\em Journal of Computational Physics}, 429:110011, 2021.

\bibitem{ghosh2016well}
Debojyoti Ghosh and Emil~M Constantinescu.
\newblock Well-balanced, conservative finite difference algorithm for
  atmospheric flows.
\newblock {\em AIAA Journal}, 54(4):1370--1385, 2016.

\bibitem{luo2011well}
Jun Luo, Kun Xu, and Na~Liu.
\newblock A well-balanced symplecticity-preserving gas-kinetic scheme for
  hydrodynamic equations under gravitational field.
\newblock {\em SIAM Journal on Scientific Computing}, 33(5):2356--2381, 2011.

\bibitem{bhatnagar1954model}
Prabhu~Lal Bhatnagar, Eugene~P Gross, and Max Krook.
\newblock A model for collision processes in gases. i. small amplitude
  processes in charged and neutral one-component systems.
\newblock {\em Physical review}, 94(3):511, 1954.

\bibitem{xiao2017well}
Tianbai Xiao, Qingdong Cai, and Kun Xu.
\newblock A well-balanced unified gas-kinetic scheme for multiscale flow
  transport under gravitational field.
\newblock {\em Journal of Computational Physics}, 332:475--491, 2017.

\bibitem{xiao2022well}
Tianbai Xiao.
\newblock A well-balanced unified gas-kinetic scheme for multicomponent flows
  under external force field.
\newblock {\em Entropy}, 24(8):1110, 2022.

\bibitem{li2020gas}
Zhi-Hui Li, Wen-Qiang Hu, Ao-Ping Peng, Jun-Lin Wu, and Chun-Hian Lee.
\newblock Gas-kinetic unified algorithm for plane external force-driven flows
  covering all flow regimes by modeling of boltzmann equation.
\newblock {\em International Journal for Numerical Methods in Fluids},
  92(8):922--949, 2020.

\bibitem{premnath2009steady}
Kannan~N Premnath, Martin~J Pattison, and Sanjoy Banerjee.
\newblock Steady state convergence acceleration of the generalized lattice
  boltzmann equation with forcing term through preconditioning.
\newblock {\em Journal of Computational Physics}, 228(3):746--769, 2009.

\bibitem{shan2006kinetic}
Xiaowen Shan, Xue-Feng Yuan, and Hudong Chen.
\newblock Kinetic theory representation of hydrodynamics: a way beyond the
  navier--stokes equation.
\newblock {\em Journal of Fluid Mechanics}, 550:413--441, 2006.

\bibitem{xu2014direct}
Kun Xu.
\newblock {\em Direct modeling for computational fluid dynamics: construction
  and application of unified gas-kinetic schemes}, volume~4.
\newblock World Scientific, 2014.

\bibitem{xu2001gas}
Kun Xu.
\newblock A gas-kinetic bgk scheme for the navier--stokes equations and its
  connection with artificial dissipation and godunov method.
\newblock {\em Journal of Computational Physics}, 171(1):289--335, 2001.

\bibitem{grad1949kinetic}
Harold Grad.
\newblock On the kinetic theory of rarefied gases.
\newblock {\em Communications on pure and applied mathematics}, 2(4):331--407,
  1949.

\bibitem{Cai2019}
Zhenning Cai and Manuel Torrilhon.
\newblock On the holway-weiss debate: Convergence of the grad-moment-expansion
  in kinetic gas theory.
\newblock {\em Physics of Fluids}, 31(12):126105, 12 2019.

\bibitem{Yan2024}
Su~Yan and Xiaowen Shan.
\newblock Lattice boltzmann simulation of nonequilibrium flows using spectral
  multiple-relaxation-time collision model.
\newblock {\em AIAA Journal}, 62, 11 2024.

\bibitem{leveque1999wave}
Randall~J LeVeque and Derek~S Bale.
\newblock Wave propagation methods for conservation laws with source terms.
\newblock In {\em Hyperbolic Problems: Theory, Numerics, Applications: Seventh
  International Conference in Z{\"u}rich, February 1998 Volume II}, pages
  609--618. Springer, 1999.

\end{thebibliography}

\end{document}